\documentclass[12pt]{article}
\usepackage{amsmath}
\usepackage{graphicx}
\usepackage{enumerate}
\usepackage{xcolor}
\usepackage{natbib}
\usepackage{url} % not crucial - just used below for the URL 
\usepackage{comment}
\usepackage{siunitx}

\newcommand{\blind}{1}

\usepackage{hyperref}
\usepackage{mathtools}
\usepackage{cleveref}
\usepackage{enumitem}

\usepackage{multirow}
\usepackage{arydshln}
\usepackage{float}

\usepackage{xr}
\usepackage[utf8]{inputenc}

\usepackage{amsmath,amssymb,graphicx,color,amsthm}
\usepackage[linesnumbered,ruled]{algorithm2e}
\usepackage{tikz}
\usepackage{dsfont}
\usepackage{mathtools}
\usetikzlibrary{shapes.geometric, arrows}
\usepackage[english]{babel}
\usepackage[utf8]{inputenc}
\usepackage{mathrsfs}  
\usepackage{afterpage}
\usepackage{amsfonts}
\usepackage[colorinlistoftodos]{todonotes}
\usepackage{enumerate}
\usepackage{hyperref}
\usepackage{cleveref}

\newcommand{\given}{\, |\,}

\newcommand\T{^\top}

\newcommand\RR{\mathbb{R}}

\newcommand\EE{\mathbb{E}}

\newcommand{\cov}{\mathrm{Cov}}

\newtheorem{theorem}{Theorem}
\newtheorem{definition}{Definition}

\DeclareMathOperator\GP{GP}

\DeclarePairedDelimiterX{\infdivx}[2]{(}{)}{%
	#1\;\delimsize\|\;#2%
}

\DeclarePairedDelimiter{\norm}{\lVert}{\rVert}

\newcommand{\RNum}[1]{\uppercase\expandafter{\romannumeral #1\relax}}

\xdefinecolor{dukeblue}{rgb}{0.004,0.129,0.412}

\newcommand{\bfu}{\mathbf{u}}
\newcommand{\bfs}{\mathbf{s}}
\newcommand{\bfv}{\mathbf{v}}
\newcommand{\bfx}{\mathbf{x}}

\usepackage{algpseudocode}
\allowdisplaybreaks

\usepackage{circledsteps}
\newcommand{\circleone}{\raisebox{.5pt}{\text{\textcircled{\raisebox{-.9pt} {1}}}}}
\newcommand{\circletwo}{\raisebox{.5pt}{\text{\textcircled{\raisebox{-.9pt} {2}}}}}
\newcommand{\circlethree}{\raisebox{.5pt}{\text{\textcircled{\raisebox{-.9pt} {3}}}}}

\begin{document}

\def\spacingset#1{\renewcommand{\baselinestretch}%
{#1}\small\normalsize} \spacingset{1}

%%%%%%%%%%%%%%%%%%%%%%%%%%%%%%%%%%%%%%%%%%%%%%%%%%%%%%%%%%%%%%%%%%%%%%%%%%%%%%
\if1\blind
{
  \title{\bf The Nearest-Neighbor Derivative Process: Modeling Spatial Rates of Change in Massive Datasets}
  %\author{
  %Jiawen Chen\\
  %   Department of Biostatistics, The University of North Carolina at Chapel Hill\\
  %   Caiwei Xiong \\
  %   Department of Biostatistics, The University of North Carolina at Chapel Hill\\
  %   Quan Sun\\
  %   Department of Biostatistics, The University of North Carolina at Chapel Hill\\
  %   Geoffery W. Wang\\
  %   Department of Statistics, North Carolina State University\\
  %   Gaorav P. Gupta\\
  %   Department of Radiation Oncology, The University of North Carolina at Chapel Hill\\
  %   Aritra Halder\\
  %   Department of Epidemiology and Biostatistics, Drexel University\\
  %   Yun Li\\
  %   Department of Genetics, The University of North Carolina at Chapel Hill \\
  %   and \\
  %   Didong Li\\
  %   Department of Biostatistics, The University of North Carolina at Chapel Hill}
\author{Jiawen Chen$^{1,2}$, Aritra Halder$^3$, Yun Li$^{4,5}$, Sudipto Banerjee$^6$, Didong Li$^{4}$\\
$^1$Department of Biomedical Data Science, Stanford University,\\
$^2$Gladstone Institutes,\\
$^3$Department of Biostatistics and Epidemiology, Drexel University,\\
$^4$Department of Biostatistics, University of North Carolina at Chapel Hill,\\
$^5$Department of Genetics, University of North Carolina at Chapel Hill,\\
$^6$Department of Biostatistics, University of California, Los Angeles\\
}
\date{}
  \maketitle
} \fi

\if0\blind
{
  \bigskip
  \bigskip
  \bigskip
  \begin{center}
    {\LARGE\bf The Nearest-Neighbor Derivative Process: Modeling Spatial Rates of Change in Massive Datasets}
\end{center}
  \medskip
} \fi

\bigskip
\begin{abstract}
Gaussian processes (GPs) are instrumental in modeling spatial processes, offering precise interpolation and prediction capabilities across fields such as environmental science and biology. Recently, there has been growing interest in extending GPs to infer spatial derivatives, which are vital for analyzing spatial dynamics and detecting subtle changes in data patterns. Despite their utility, traditional GPs suffer from computational inefficiencies, due to the cubic scaling with the number of spatial locations. Fortunately, the computational challenge has spurred extensive research on scalable GP methods. However, these scalable approaches do not directly accommodate the inference of derivative processes. A straightforward approach is to use scalable GP models followed by finite-difference methods, known as the plug-in estimator. This approach, while intuitive, suffers from sensitivity to parameter choices, and the approximate gradient may not be a valid GP, leading to compromised inference. To bridge this gap, we introduce the Nearest-Neighbor Derivative Process (NNDP), an innovative framework that models the spatial processes and their derivatives within a single scalable GP model. NNDP significantly reduces the computational time complexity from $O(n^3)$ to $O(n)$, making it feasible for large datasets. We provide various theoretical supports for NNDP and demonstrate its effectiveness through extensive simulations and real data analysis. 

\end{abstract}

\noindent%
{\it Keywords:}  Bayesian modeling; Mean-square derivatives; Nearest-Neighbor Gaussian processes; Spatial gradients.
%3 to 6 keywords, that do not appear in the title
\vfill

\newpage
\spacingset{1.9} % DON'T change the spacing!

\section{Introduction}\label{sec:intro}

Spatial processes play a critical role in environmental science, ecology, geology, and biology \citep{banerjee2014hierarchical} and are prominent in the modeling and analysis of scientific data referenced with respect to geographic space. Gaussian processes (GPs) are widely used in spatial modeling due to their flexibility in capturing complex spatial dependencies through covariance functions. They provide a probabilistic framework for spatial interpolation, commonly known as ``kriging'', prediction, and uncertainty quantification, making them indispensable tools in spatial statistics~\citep{williams2006gaussian}.

Estimation of the underlying spatial process is often followed by analysis of local features of the spatial surface. This exercise is often referred to as ``wombling'', named after a seminal paper by \cite{womble1951differential}. Spatial gradients (first-order derivatives) and curvature (second-order derivatives) provide information on the rate and direction of changes in the spatial domain \citep{banerjee2003directional,halder2024bayesian}. For example, gradients are crucial to identify boundaries of ecological differences \citep[see][]{banerjee2006wombling} or changes in hedonic house prices across administrative boundaries \citep{majumdar2006gradients}. Estimating derivatives from spatial data enables the analysis of local features such as peaks, troughs, and ridges, providing deeper insight into spatial phenomena, for example discovering the lurking effects of a natural boundary \citep[see, e.g.,][models heavy-metal concentrations in a river-valley]{halder2024bayesian}. Recent investigations into modeling how biological characteristics change in tissues within the human brain~\citep[see, e.g.,][]{chen2024investigating} provide insight into the neural microenvironment. 

Although several of the aforementioned articles have addressed probabilistic inference of spatial gradients, such inference has not scaled to analyze massive volumes of spatial data arising from rapidly emerging spatial data science technologies such as remote sensing. Some of the most effective spatial processes for scaling massive datasets are derived from parent GPs using directed acyclic graphs to induce sparsity. Although such constructions produce computationally efficient GPs such as the Nearest-Neighbor Gaussian Process (NNGP) introduced by \citet{datta2016hierarchical} and generalizations thereof, they lose the smoothness properties of the parent GP and generally do not admit spatial gradients. Inferring derivatives requires specifying the smoothness of the spatial process to ensure that spatial gradients are well-defined. Although such smoothness can be modeled using a flexible spatial covariance kernel such as the Mat\'ern \cite{stein2012interpolation} to ensure that gradient and curvature processes can be derived, such modeling is hindered if the GP needs to be scaled to analyze massive volumes of spatial data.

Our current contribution here is to develop derivative processes from an NNGP to enable spatial modelers to infer spatial gradients in surfaces estimated from massive volumes of spatial data. Typically, joint inference is performed on the parent process and its induced derivatives~\citep{banerjee2003directional}. Further advances by \citet{halder2024bayesian} extend inference to spatial curvature processes% , allowing for the joint estimation of the process, its gradient, and curvature.
. The GP framework offers attractive exact posterior predictive inference for such processes. It only requires a valid cross-covariance matrix, which depends on the smoothness specification of the parent process.  % The cross-covariance matrix required between the process and its derivatives is characterized by the derivatives of the covariance function, enabling exact inference for derivatives within the GP framework.
However, incorporating derivative processes into GPs exacerbates the computational burden. The combined process comprised of the parent and induced derivative processes introduces a complex dependence structure that does not facilitate efficient computation.
% Joint modeling of the process and its derivatives increases the dimensions and complexity of the covariance structures. 
This limitation becomes % particularly acute 
severe in applications involving massive spatial datasets, such as those arising from high-resolution environmental monitoring, remote sensing, and spatial genomics. The need for scalable methods that can efficiently estimate spatial derivatives without sacrificing accuracy is therefore pressing.

%An application domain that exemplifies the urgent need for scalable spatial derivative processes is spatial transcriptomics (ST)~\citep{staahl2016visualization}. By analyzing the spatial dynamics of gene expression, derivatives can reveal biologically meaningful patterns, such as identifying tissue boundaries or regions of rapid change. They can also help detect specific genes with significant changes across spatial boundaries, known as cliff genes, which are critical for understanding cellular differentiation and tissue organization~\citep{chen2024investigating}. However, recent technologies such as the Visium High Definition (VisiumHD) platform by 10x Genomics generate ultra-high-resolution spatial gene expression data, capturing tens of thousands of genes across half a million to millions of spatial locations. The resulting large scale of these datasets poses  significant computational challenges for traditional GP methods.

To address these challenges, \citet{chen2024investigating} proposed StarTrail, a two-step scalable gradient process, where the spatial process is modeled as an NNGP followed by the plug-in estimate of spatial gradients through finite-difference approximations \citep{liu2022optimal}. While this method offers computational scalability, it inherits limitations inherent in finite difference techniques. Specifically, finite difference approximations can be sensitive to the choice of step size and may introduce significant approximation errors, particularly when estimating higher-order derivatives or dealing with irregularly spaced data. Moreover, the finite difference approach does not provide valid probabilistic inference for the gradients; it treats the gradient estimates as deterministic quantities derived from the approximated function rather than modeling the joint distribution of the process and its derivatives.

This article introduces the Nearest Neighbor Derivative Process (NNDP) to jointly model the spatial process and its derivatives within a unified probabilistic framework. By leveraging the properties of GPs, where the derivatives of a GP are themselves GPs, we construct the derivatives of NNGP kernels and prove that NNDP is a rigorous GP. Furthermore, we establish the smoothness of the approximated process and derive the joint distribution for NNGP and NNDP while ensuring mathematical rigor and robustness. 
NNDP provides several advantages over existing methods. It eliminates the need for step size selection inherent in finite difference methods, provides accurate derivative estimates, and enables parameter inference and uncertainty quantification. Its scalability is maintained by exploiting the conditional independence structure of the NNGP, making it computationally efficient for large datasets. Our method is supported by extensive simulation studies and two real data applications in high-resolution ST data and climate science. 

The remainder of the paper is organized as follows. \Cref{sec:prelim} reviews preliminaries. \Cref{sec:NNDP} introduces NNDP. \Cref{sec:sim} presents simulations while \Cref{sec:application} applies the NNDP to spatial transcriptomics and air temperature data. 
\Cref{sec:diss} concludes with a discussion. Proofs and additional experiments are supplied in an Appendix.

\section{Preliminary} \label{sec:prelim}
\subsection{Gaussian Process (GP)}
A GP is a stochastic process such that any finite collection of its realizations follows a multivariate normal distribution \citep{stein2012interpolation}. Formally, if \(\omega\) is a GP defined over the domain \(\mathcal{D}\) with mean function \(\mu:\mathcal{D} \to \mathbb{R}\) and covariance function (also referred to as kernel) \(C: \mathcal{D} \times \mathcal{D} \to \mathbb{R}\), then for any finite set of points \(\bfx_1,\ldots,\bfx_n \in \mathcal{D}\),
\[
[\omega(\bfx_1),\ldots,\omega(\bfx_n)]^\top \sim N_n\bigl(\mathbf{v},\mathbf{C}\bigr),
\]
where \(v = [\mu(\bfx_1),\ldots,\mu(\bfx_n)]^\top\) and \(\mathbf{C} = [{C}(\mathbf{x}_i,\mathbf{x}_j)]\). We use $N$ for univariate normal distribution and $N_n$ for $n$-variate normal distribution.

For simplicity and without loss of generality, we assume \(\mu = 0\). If the mean is not zero, a suitable data transformation or an additive mean model can be used to achieve zero mean residuals, which is a practice commonly adopted in machine learning and statistical modeling \citep{williams2006gaussian,banerjee2014hierarchical}. Under this assumption, the kernel function \(C\) determines the behavior of the GP. In the setting \(\mathcal{D} = \mathbb{R}^d\), the choice of kernel is critical, and various widely used kernels are available, each imposing distinct smoothness and dependence properties on the underlying GP.

A commonly used kernel is the Radial Basis Function (RBF), also known as the squared exponential or Gaussian kernel:
\[
C(\bfx,\bfx') = \sigma^2 \exp(-\alpha \|\bfx - \bfx'\|^2),
\]
where \(\sigma^2\) controls the marginal variance at each point and \(\alpha\) controls the rate of correlation decay with distance. The RBF kernel is infinitely differentiable, ensuring that the resulting GP is mean-square differentiable (MSD) of all orders. Formally, a GP \(f\) is mean-square continuous (MSC) if \(\mathbb{E}[(f(\bfx+\bfu)-f(\bfx))^2]\to0\) as \(\bfu\to0\), and is MSD if the limit \(\lim_{h \to 0}(f(\bfx+h\bfu)-f(\bfx))/h\) exists in the mean-square sense for any $\bfu \in \RR^d$. Higher order mean-square differentiability can be defined in the same manner. In particular, a GP with RBF kernel is infinitely differentiable, which, while desirable in some contexts, may lead to over-smoothing in applications such as spatial interpolation \citep{stein2012interpolation}.

To address this, the Mat\'ern kernel family provides a more flexible class of covariance functions that includes a smoothness parameter \(\nu\):
\[
C(\bfx,\bfx') = \sigma^2 \frac{2^\nu}{\Gamma(\nu)} \left(\alpha\|\bfx - \bfx'\|\right)^\nu K_\nu(\alpha \|\bfx - \bfx'\|),
\]
where \(K_\nu\) is the modified Bessel function of the second kind. The parameter \(\nu\) directly controls the smoothness of the resulting GP; a GP with a Mat\'ern kernel is \(\lceil \nu \rceil - 1\) times MSD. The exponential kernel, often used in geostatistics, arises as a special case of the Mat\'ern kernel when \(\nu = 1/2\), which is MSC but not MSD:
$$C(\bfx,\bfx') = \sigma^2 \exp(-\alpha \|\bfx - \bfx'\|).$$

Although inferring \(\nu\) can be challenging from both theoretical and empirical standpoints \citep{zhang2004inconsistent, tangEtAl2021jrssb}, it is common practice to fix \(\nu\) to certain values (e.g., \(\nu = 1/2,3/2,5/2,\ldots\)) to simplify the Bessel function and the parameter inference~\citep{williams2006gaussian}. The parameters \(\sigma^2\) and \(\alpha\) are typically estimated via the maximum likelihood estimate (MLE), and standard optimization methods such as L-BFGS~\citep{liu1989limited}, can be employed to maximize the log-likelihood in practice.

\subsection{Nearest Neighbor Gaussian Process}
Let $\{\omega(\bfs): \bfs\in \mathcal{D} \subset\RR^d\}$ be a GP with zero mean and covariance function $C(\bfs,\bfs')= \cov(\omega(\bfs),\omega(\bfs'))$. Following \cite{datta2016hierarchical}, we define the \emph{reference set} $\mathcal{S} = \{\bfs_1,\dots,\bfs_k\}$ consisting of $k$ distinct locations that are fixed in $\mathcal{D}$. %For a (weakly) stationary process, the covariance function could be written as $C(\bfs,\bfs')=C(\bfs-\bfs')$, where $K$ is a valid covariance function on $\RR^d$. For an isotropic process, we can write $C(\bfs,\bfs')=\widetilde{K}(\norm{\bfs-\bfs'})$, where $\norm{\bfs-\bfs'}$ is the Euclidean distance between $\bfs$ and $\bfs'$. 

Model-based inference for the spatial process requires the inverse and determinant of the covariance matrix $\mathbf{C}$ involving $O(n^3)$ floating point operations (flops). By now, there is a burgeoning literature on GP models that are able to scale massive datasets. Here, we focus on the Nearest Neighbor Gaussian Process (NNGP), devised by \cite{datta2016hierarchical}, that reduces the time complexity from $O(n^3)$ to $O(n)$. The NNGP is a legitimate stochastic process that extends Vecchia's likelihood approximation \cite{vecchia1988estimation},
\[
p(\boldsymbol{\omega_\mathcal{S}})=p(\omega(\bfs_1))\prod_{i=2}^{k}p(\omega(\bfs_i)\given \omega(\bfs_{i-1}),\cdots, \omega(\bfs_1)) \approx \prod_{i=1}^k p(\omega(\bfs_i)\given \boldsymbol{\omega_{\mathcal{N}(\bfs_i)}}) = \widetilde{p}(\boldsymbol{\omega_\mathcal{S}}),
\] 
where $\boldsymbol{\omega_{\mathcal{S}}}$ is $k\times 1$ with elements $\omega(\bfs_i)$. Here, the larger conditioning sets on the right-hand side is replaced with smaller and carefully chosen conditioning sets of size $m$, where $m \ll k$. For each $\bfs_i \in \mathcal{S}$, $\mathcal{N}_\mathcal{S}=\{ \mathcal{N}(\bfs_i);i=1,2,\dots,k\}$ is the collection of all smaller conditioning set. If the directed graph $\mathcal{G}$ with $\mathcal{S}$ as nodes and $\mathcal{N}_\mathcal{S}$ as the set of directed edges is a \textit{directed acyclic graph (DAG)}, the nearest-neighbor density $\widetilde{p}(\boldsymbol{\omega_\mathcal{S}})$ is a proper multivariate joint density. This valid joint density is the key concept distinguishing NNGP from other scalable techniques such as low-rank models~\citep{cressie2008fixed, banerjeeEtAl2008jrssb} or covariance-tapering \citep{furrer2006covariance}.

\subsection{Derivative Process}
We turn to spatial derivatives at any $\bfs_0 \in \RR^d$. The process $\omega(\bfs)$ is mean-square differentiable at $\bfs_0$ if there exists a vector $\nabla \omega(\bfs_0)$, such that, for any scalar $h$ and any unit vector $\bfu$, $\omega(\bfs_0+h\bfu) = \omega(\bfs_0)+h\bfu\T \nabla\omega(\bfs_0)+r(\bfs_0,h\bfu)$, where $r(\bfs_0,h\bfu)/h \rightarrow 0$ in the $L_2$ sense as $h\rightarrow 0$. Equivalently, the mean-square limit $D_\bfu \omega (\bfs) := \lim_{h \rightarrow 0}\omega_{\bfu,h}(\bfs) = \bfu \T \nabla \omega(\bfs)$ is a well-defined process in $\RR^d$, where $\omega_{\bfu,h}(\bfs) = (\omega(\bfs+h\bfu)-\omega(\bfs))/h$ is the finite difference process \citep{banerjee2003directional}. We refer to $D_\bfu \omega (\bfs)$ as the \textit{derivative process} of $\omega$ in the direction $\bfu$. Here, we develop a Nearest-Neighbor Derivative Process (NNDP) to scale inference on directional spatial gradients for massive datasets.

%NNGP represents a scalable adaptation of the traditional GP, maintaining the essence of GP modeling while addressing its computational limitations. Recognized as a valid process, the NNGP allows for the effective handling of large datasets by inducing sparsity in the covariance matrix computations through localized neighborhood approximations. This scalability is crucial for extending the applicability of GPs to more extensive and complex data scenarios, where calculating derivatives of the spatial process becomes essential for detailed analysis. 

\section{Nearest-Neighbor Derivative Process} \label{sec:NNDP}
\subsection{Distribution Theory}
%We assume $\{\omega(\bfs): \bfs\in \mathcal{D} \subset \RR^d\}$ is a univariate spatial process with zero mean and finite second moment. 
If $\omega(\bfs) \sim \GP(0, C(\cdot, \cdot))$ is a GP with zero mean and covariance function $C(\bfs,\bfs')$ on the domain $\mathcal{D} \subset \RR^d$, then we investigate inference for $\nabla \omega(\bfs) = \left(\partial \omega(\bfs)/\partial s_1,\ldots, \partial \omega(\bfs)/\partial s_d \right)^{\T}$, assuming that it exists. Smoothness of the spatial process is easily expressed in terms of the covariance function \citep{stein2012interpolation, banerjee2003smoothness} and produces the multivariate $(1+d)\times 1$ process $(\omega(\bfs), \nabla\omega(\bfs)^{\T})^{\T}$ with the $(d+1)\times (d+1)$ matrix-valued cross-covariance function
\begin{equation}\label{eq: gradients_cross_covariance}
\begin{bmatrix}
\text{cov}\left\{\omega(\bfs), \omega(\bfs')\right\} & \text{cov}\left\{\omega(\bfs), \nabla\omega(\bfs')\right\} \\
\text{cov}\left\{\nabla\omega(\bfs), \omega(\bfs')\right\} & \text{cov}\left\{\nabla\omega(\bfs), \nabla\omega(\bfs')\right\}
\end{bmatrix} = \begin{bmatrix} 
C(\bfs,\bfs') & -(\nabla_{\bfs'} C(\bfs,\bfs'))^\top  \\ 
\nabla_{\bfs} C(\bfs,\bfs') & \nabla_{\bfs}\circ\nabla_{\bfs'}C(\bfs,\bfs')
\end{bmatrix}\;,
\end{equation}
where $\nabla_{\bfs}$ and $\nabla_{\bfs'}$ indicate partial derivative vectors with respect to the first and second arguments of $C(\bfs,\bfs')$, and $\nabla_{\bfs}\circ\nabla_{\bfs'} := \partial^2/\partial s_i\partial s_j'$ is the operator producing the matrix of mixed partial derivatives $\partial^2 C(\bfs,\bfs')/\partial s_i\partial s_j'$, which is the negative of the hessian of $C$. 

The existence of the cross-covariance function in \eqref{eq: gradients_cross_covariance} depends on $\partial^2 C(\bfs,\bfs')/\partial s_i\partial s_j'$ for each $j,j'=1,\ldots,d$. If $C(\bfs,\bfs')$ is modeled using a Mat\'ern covariance functions \citep{stein2012interpolation}, then the smoothness of $\omega(\bfs)$ in the mean-square sense is fully determined by the value of a smoothness parameter $C(\bfs,\bfs')$. \cite{banerjee2003directional} and \cite{banerjee2006wombling} make use of this parameter to construct a multivariate process involving $\omega(\bfs)$ and $\nabla \omega(\bfs)$, while \cite{halder2024bayesian} and \cite{halder2024Spatiotemporal} use the Mat\'ern family to construct higher-dimensional curvature processes for spatial and space-time random fields. Because the NNGP is built from a discrete topological ordering of the nodes of a graph, there are inherent discontinuities in the covariance function of an NNGP and, hence, an analogue of \eqref{eq: gradients_cross_covariance} does not exist at all locations in the domain. Thus, our first task is to identify the set of locations where a spatial process for the gradients is well-defined.   

Recall the construction of the NNGP using a topologically ordered DAG, a reference set of $k$ locations $\mathcal{S}$ and a collection $\mathcal{N}_\mathcal{S}$ of neighbor sets ${\cal N}(\bfs)$ consisting of a fixed number, say $m$, of neighbors. The joint density derived from the DAG over ${\cal S}$ is
\begin{equation}\label{eqn:NNGP_density}
    \widetilde{p}(\boldsymbol{\omega_\mathcal{S}}) = \prod_{i=1}^k N(\omega(\bfs_i) \given \mathbf{b}_{\bfs_i}^{\T}\boldsymbol{\omega_{\mathcal{N}(\bfs_i)}},f_{\bfs_i}) = N_k(\boldsymbol{\omega_\mathcal{S}} \given 0,\widetilde{\mathbf{C}}_\mathcal{S}),
\end{equation}
where $\mathbf{b}_{\bfs_i}^{\T} = \mathbf{C}_{\bfs_i, \mathcal{N}(\bfs_i)}^{\T}\mathbf{C}_{\mathcal{N}(\bfs_i)}^{-1}$ is $1\times m$, $f_{\bfs_i} = C(\bfs_i,\bfs_i) - \mathbf{C}_{\bfs_i, \mathcal{N}(\bfs_i)}^{\T}\mathbf{C}_{\mathcal{N}(\bfs_i)}^{-1}\mathbf{C}_{\mathcal{N}(\bfs_i),\bfs_i}$ and $\widetilde{\mathbf{C}}_{\mathcal{S}}^{-1} = (\mathbf{I} -\widetilde{\mathbf{B}})^{\T}\widetilde{\mathbf{F}}^{-1}(\mathbf{I} -\widetilde{\mathbf{B}})$, where $\widetilde{\mathbf{B}}$ is $k\times k$ such that $\boldsymbol{ \widetilde{\omega}_{\mathcal{S}}} = \widetilde{\mathbf{B}}\boldsymbol{\omega_{\mathcal{S}}} + \boldsymbol{\mathbf{\epsilon}_{\mathcal{S}}}$, $\boldsymbol{\mathbf{\epsilon}_{\mathcal{S}}}$ is $k\times 1$ with elements $\epsilon(\bfs_i) \overset{ind}{\sim} N(0, f_{\bfs_i})$, and $\widetilde{\mathbf{F}}^{-1}$ is diagonal with $1/f_{\bfs_i}$ as its diagonal elements. The inverse covariance matrix $\widetilde{\mathbf{C}}_{\mathcal{S}}^{-1}$ is sparse as a consequence of $\mathbf{\widetilde{B}}$. The $i$-th row of $\mathbf{\widetilde{B}}$ has nonzero elements, given by the elements of $\mathbf{b}_{\bfs_i}^{\T}$, only in the $m$ positions corresponding to $\mathcal{N}(\bfs_i)$ and the remaining $k-m$ positions are zero. We extend the finite approximation in \eqref{eqn:NNGP_density} to
\begin{equation}
    \label{eq: nngp_process}
    \widetilde{\omega}(\bfv) = \mathbf{b}(\bfv)^{\T}\boldsymbol{\omega_{\mathcal{S}}} + \widetilde{\epsilon}(\bfv)\;;\quad \widetilde{\epsilon}(\bfv) {\sim} N(0, f_{\bfv})\;, 
\end{equation}
where $\bfv$ is any point in $\mathcal{D}$, $\mathbf{b}(\bfv)^{\T}$ is $1\times k$ with exactly $m$ nonzero elements in positions in $\mathcal{N}(\bfv)$, and $\widetilde{\epsilon}(\bfv)$ are independently distributed over any finite set $\mathcal{V} = \{ \bfv_1, \bfv_2, \ldots, \bfv_r\}$ of spatial locations with variance $f_{\bfv_i} = C(\bfv_i,\bfv_i) - \mathbf{C}_{\bfv_i, \mathcal{N}(\bfv_i)}^{\T}\mathbf{C}_{\mathcal{N}(\bfv_i)}^{-1}\mathbf{C}_{\bfv_i, \mathcal{N}(\bfv_i)}$. The nonzero elements of $\mathbf{b}(\bfv)^{\T}$ comprise the $m$ elements of $\mathbf{b}_\bfv\T \coloneqq \mathbf{C}_{\bfv, \mathcal{N}(\bfv)}^{\T}\mathbf{C}_{\mathcal{N}(\bfv)}^{-1}$.

%We define $\mathcal{N}_0(\bfv)$ for any location $\bfv\in \mathcal{D}$ as follows: (i) if $\bfv \in \mathcal{S}$ then $\mathcal{N}_0(\bfv)$ includes itself and its $m-1$ nearest neighbors from locations preceding it in the topological ordering of $\mathcal{S}$; and (ii) if $\bfv \in \mathcal{D}\setminus\mathcal{S}$, then $\mathcal{N}_0(\bfv)$ is the set of its $m$ nearest neighbors from locations in $\mathcal{S}$. The quantity $\omega(\bfv)$ in \eqref{eq: nngp_process} is a linear model representation of the NNGP. Based upon the construction of $\mathcal{N}_0(\bfv)$, it is straightforward to see that $\widetilde{\omega}(\bfs_i) = \omega(\bfs_i)$ for all $\bfs_i \in \mathcal{S}$. Because $\mathcal{N}_0(\bfv)$ includes $\bfv$ when $\bfv\in \mathcal{S}$, $\mathbf{c}_{\bfv, \mathcal{N}(\bfv)}^{\T}$ is, in fact, the $i$-th row of $\mathbf{C}_{\mathcal{N}(\bfv)}$, which implies that $\mathbf{c}_{\bfv, \mathcal{N}(\bfv)}^{\T}\mathbf{C}_{\mathcal{N}(\bfv)}^{-1} = \mathbf{e}_i^{\T}$ is the $i$-th canonical basis vector in $\mathbb{R}^m$. This implies that $\mathbf{b}(\bfv)^{\T}\omega_{\mathcal{S}} = \omega(\bfs_i)$ and $f_{\bfv} = 0$ so that $\widetilde{\omega}(\bfv) = \omega(\bfv)$ almost surely for $\bfv \in \mathcal{S}$.

We remark that the linear representation of the NNGP in \eqref{eq: nngp_process} differs from the hierarchical construction of \cite{datta2016hierarchical}, where the current development is better suited for investigating the smoothness and subsequently building the NNDP. In particular, we have a concise representation of covariance function for the process $\widetilde{\omega}(\bfv)$ as  
\begin{equation} \label{eqn:C_tilde}
\widetilde{C}(\bfv,\bfv') = \mathbf{b}(\bfv)^{\T}\widetilde{\mathbf{C}}_{\mathcal{S}}\mathbf{b}(\bfv') + \delta_{\bfv = \bfv'} f_{\bfv} = \langle\mathbf{b}(\bfv'),\mathbf{b}(\bfv)\rangle_{\widetilde{\mathbf{C}}_{\mathcal{S}}} + \delta_{\bfv = \bfv'} f_{\bfv}
%\begin{cases}
%\widetilde{\mathbf{C}}_{\bfs_i,\bfs_j} & \bfv_1=\bfs_i,\bfv_2=\bfs_j,~\bfs_i, \bfs_j\in\mathcal{S},\\
%\mathbf{B}_{\bfv_1}\widetilde{\mathbf{C}}_{\mathcal{N}(\bfv_1),\bfs_j} & \bfv_2=\bfs_j\in\mathcal{S},~\bfv_1\notin\mathcal{S},\\
%\mathbf{B}_{\bfv_1}\widetilde{\mathbf{C}}_{\mathcal{N}(\bfv_1),\mathcal{N}(\bfv_2)}\mathbf{B}_{\bfv_2}\T+\delta_{(\bfv_1=\bfv_2)}\mathbf{F}_{\bfv_1} & \bfv_1,\bfv_2\notin \mathcal{S}.
%\end{cases}
\end{equation}
for any pair $\bfv,\bfv'\in\mathcal{D}$, where $\langle\cdot,\cdot\rangle_{\widetilde{\mathbf{C}}_{\mathcal{S}}}$ is the Euclidean inner-product with respect to the positive definite matrix $\widetilde{\mathbf{C}}_{\mathcal{S}}$ defined in \eqref{eqn:NNGP_density} and $\delta_{\bfv = \bfv'}$ is the Kronecker delta. %Smoothness of $\widetilde{C}(\bfv,\bfv')$ is determined by $\mathbf{b}(\bfv)$. 
If both $\bfv, \bfv' \in \mathcal{S}$, say the $i$-th and $j$-th locations, then $\widetilde{C}(\bfv,\bfv') = \mathbf{e}_i^{\T}\widetilde{\mathbf{C}}_{\mathcal{S}}\mathbf{e}_j = \widetilde{\mathbf{C}}_{\mathcal{S}}[i,j]$ is the $(i,j)$-th element of $\widetilde{\mathbf{C}}_{\mathcal{S}}$. If $\bfv$ is the $i$-th member in the list $\mathcal{S}$ and $\bfv' \notin \mathcal{S}$, then \eqref{eqn:C_tilde} simplifies to $\widetilde{C}(\bfv,\bfv')$ being the dot product between row $i$ of $\widetilde{\mathbf{C}}$ and $\mathbf{b}(\bfv')$.       

%{\color{red} I think the above development of the NNGP will be much cleaner for our current purposes to transition to the NNDP. Please note that the above construction is slightly different from the original construction in which we would have obtained $\widetilde{\omega}(\bfv) = \delta_{\bfv \in \mathcal{S}}\omega(\bfv) + (1-\delta_{\bfv\in\mathcal{S}})\left(\mathbf{b}(\bfv)^{\T}\omega_{\mathcal{S}} + \widetilde{\epsilon}(\bfv)\right)$. The stochastic continuity properties for this model and that of \eqref{eq: nngp_process} will likely differ on a set of Lebesque measure zero, but \eqref{eq: nngp_process} is easier to work with for characterizing the existence of gradients and curvature.}

Let $\mathcal{Z}_1$ be the set of all points $\bfv$ that are equidistant from any two points in $\mathcal{S}$ and $\mathcal{Z}_2 = (\mathcal{Z}_1 \times \mathcal{Z}_1) \cup \{(\bfv,\bfv) | \bfv \in \mathcal{D}\backslash\mathcal{S}\}$, then $\mathcal{Z}_2$ has Lebesgue measure zero in the Euclidean domain $\RR^d \times \RR^d$. $\widetilde{C}(\bfv_1,\bfv_2;\theta)$ is proved to be continuous for all pairs $(\bfv_1,\bfv_2)$ outside $\mathcal{Z}_2$~\citep{datta2016hierarchical}. We denote this NNGP derived from the parent GP $\omega(\bfs) \sim \GP (0,\mathbf{C})$ as $\widetilde{\omega}(\bfs) \sim \GP (0,\widetilde{\mathbf{C}})$, which is still a valid GP.

\begin{theorem} \label{theorem:msd}
If $\omega(\bfs)$ is MSD of first-order, then $\widetilde{\omega}(\bfs)$ is also MSD of first-order outside a Lebesgue measure zero set $\mathcal{Z}_1$, that is, $\displaystyle{\lim_{h\to0}\frac{\widetilde{\omega}(\bfs+h\bfu)-\widetilde{\omega}(\bfs)}{h}}$ exists in the mean-square topology for any $\bfs \notin \mathcal{Z}_1$ and any $\bfu\in\mathcal{D}$.
\end{theorem}

\begin{definition}[NNDP]
To formulate a scalable gradient process using NNGP, we first formulate the \textit{nearest-neighbor finite difference derivative process} (NNFDP) at step size $h$ in direction $\bfu$ as
\(
\widetilde{\omega}_{\bfu,h}(\bfs) = \frac{\widetilde{\omega}(\bfs+h\bfu)-\widetilde{\omega}(\bfs)}{h}.
\) Assuming $\omega$ is MSD of first-order, then the NNDP, denoted by $D_\bfu \widetilde{\omega}(\bfs)$, is defined as 
\[D_\bfu \widetilde{\omega}(\bfs)\coloneqq \lim_{h\rightarrow 0} \widetilde{\omega}_{\bfu,h}(\bfs), \bfs \notin \mathcal{Z}_1,\]
where the limit is in the mean-square topology and the existence is guaranteed by \Cref{theorem:msd}. %we can calculate the cross-covariance 
%\[
%\cov(\widetilde{\omega}(\bfv_1),\widetilde{\omega}_{\bfu,h}(\bfv_2)) = \frac{\widetilde{C}(\bfv_1,\bfv_2+h\bfu)-\widetilde{C}(\bfv_1,\bfv_2)}{h},
%\] and NNDP $D_\bfu \widetilde{\omega}(\bfs) \coloneqq \lim_{h\rightarrow 0} \widetilde{\omega}_{\bfu,h}(\bfs)$, 

%\[
%\cov(D_\bfu \widetilde{\omega}(\bfv_1),\widetilde{\omega}(\bfv_2)) = \lim_{h \rightarrow 0}\frac{\widetilde{C}(\bfv_1+h\bfu,\bfv_2)-\widetilde{C}(\bfv_1,\bfv_2)}{h},
%\]

%\[
%\cov(D_\bfu \widetilde{\omega}(\bfv_1),D_\bfu \widetilde{\omega}(\bfv_2)) = \lim_{h \rightarrow 0}\frac{1}{h^2}(\widetilde{C}(\bfv_1+h\bfu,\bfv_2+h\bfu)-\widetilde{C}(\bfv_1+h\bfu,\bfv_2)-\widetilde{C}(\bfv_1,\bfv_2+h\bfu)+\widetilde{C}(\bfv_1,\bfv_2)).
%\]
\end{definition}
Following NNGP \citep{datta2016hierarchical}, we choose $m$-nearest neighbor to ensures a DAG $\mathcal{G}$. That is, for each $\bfs_i \in \mathcal{S}$, the neighbor set $\mathcal{N}(\bfs_i)$ is constructed using $m$ nearest neighbors of $\bfs_i$ with respect to its Euclidean distance with members of $\{\bfs_1,\bfs_2,\dots,\bfs_{i-1}\}$. We define some notations for neighbor sets as follows:

\begin{definition}
Suppose $\bfs_i \in \mathcal{S}$ is the closest point of $\bfv \notin \mathcal{S}$,  we define 
\[
\mathcal{N}(\bfv)=\begin{cases}
    \{\bfs_i,\mathcal{N}(\bfs_i)\} & i < m \\
    \{\bfs_i,\mathcal{N}(\bfs_i)_{1:(m-1)}\} & i \geq m \\
\end{cases};~~
\mathcal{N}_0(\bfs_i)=\begin{cases}
    \{\bfs_i,\mathcal{N}(\bfs_i)\} & i < m \\
    \{\bfs_i,\mathcal{N}(\bfs_i)_{1:(m-1)}\} & i \geq m \\
\end{cases}.
\] 
\end{definition}
\noindent For step size $h$, direction $\bfu$ and $\bfs_i \in \mathcal{S}$, when $h$ is sufficiently small, then $\bfs_i$ itself is the closest point of $\bfs_i+h\bfu$ in $\mathcal{S}$, that is, $\mathcal{N}(\bfs_i+h\bfu) = \mathcal{N}_0(\bfs_i)$. We derive the cross-covariance between $D_\bfu \widetilde{\omega}(\bfv_1)$ and $\widetilde{\omega}(\bfv_2)$, given by \Cref{eqn:C_duC_cov} and \Cref{thm:C_duC_cov} as below,
\begin{equation}
\begin{split} \label{eqn:C_duC_cov}
&\cov(D_\bfu \widetilde{\omega}(\bfv_1), \widetilde{\omega}(\bfv_2))= \\
&\begin{cases}
D_\bfu C(\bfs_i,\mathcal{N}_0(\bfs_i))\mathbf{C}_{\mathcal{N}_0(\bfs_i)}^{-1}\widetilde{\mathbf{C}}_{\mathcal{N}_0(\bfs_i),\bfs_j}
&\bfv_1=\bfs_i,~\bfv_2=\bfs_j \in \mathcal{S}\\
D_\bfu C(\bfs_i,\mathcal{N}_0(\bfs_i))\mathbf{C}_{\mathcal{N}_0(\bfs_i)}^{-1}\widetilde{\mathbf{C}}_{\mathcal{N}_0(\bfs_i),\mathcal{N}(\bfv_2)} \mathbf{b}_{\bfv_2} 
&\bfv_1=\bfs_i,~\bfv_2 \notin \mathcal{S}\\
D_\bfu C(\bfv_1,\mathcal{N}(\bfv_1))\mathbf{C}_{\mathcal{N}(\bfv_1)}^{-1} \widetilde{\mathbf{C}}_{\mathcal{N}(\bfv_1),\bfs_j}
& \bfv_1\notin \mathcal{S},~\bfv_2=\bfs_j\\
D_\bfu C(\bfv_1,\mathcal{N}(\bfv_1))\mathbf{C}_{\mathcal{N}(\bfv_1)}^{-1} \widetilde{\mathbf{C}}_{\mathcal{N}(\bfv_1),\mathcal{N}(\bfv_2)}\mathbf{b}_{\bfv_2}
&\bfv_1,\bfv_2\notin \mathcal{S}, ~\bfv_1 \neq \bfv_2, ~\bfv_1+h\bfu \neq \bfv_2 %\\
%\inft &\bfv_1,\bfv_2\notin \mathcal{S},~\bfv_1 = \bfv_2, ~\bfv_1+\Delta \neq \bfv_2
\end{cases},
\end{split}
\end{equation}
where $D_\bfu C(\bfv_1,\bfv_2) = \lim_{h\rightarrow 0} (C(\bfv_1+h\bfu,\bfv_2)-C(\bfv_1,\bfv_2))/h$, which exists because $\omega$ is MSD~\citep{stein2012interpolation}. Also, let $D_\bfu C(\bfv_1,\mathcal{N}(\bfv_1)) = [D_\bfu C(\bfv_1,\mathcal{N}(\bfv_1)_1),\dots,D_\bfu C(\bfv_1,\mathcal{N}(\bfv_1)_{m_{\bfv_1}})]$ be the row vector, where $\mathcal{N}(\bfv_1)_j$ is the $j$-th entry of $\mathcal{N}(\bfv_1)$ and $m_{\bfv_1}$ is the number of neighbors of $\bfv_1$. For example, $C$ could be set as the Mat\'ern kernel allowing us to obtain a closed form for $D_\bfu C(\bfv_1,\bfv_2)$ as in \citep{halder2024bayesian}. To ease notation, we write $\mathbf{C}_{\mathcal{A},\mathcal{B}}$ to denote a submatrix of $\mathbf{C}$ indexed by integers in sets $\mathcal{A}$ and $\mathcal{B}$ and $\mathbf{C}_{\mathcal{A}}\coloneqq\mathbf{C}_{\mathcal{A},\mathcal{A}}$. 

\begin{theorem} \label{thm:C_duC_cov}
Let $\omega(\bfs)\sim \GP(0,C)$ and $\widetilde{\omega}(\bfs)$ be its NNGP, $\bfs \in \mathcal{S}$. Then $\cov(D_\bfu \widetilde{\omega}(\bfv_1),\widetilde{\omega}(\bfv_2))$, the covariance  between NNGP $\widetilde{\omega}(\bfs)$ and its NNDP $D_\bfu \widetilde{\omega}(\bfs)$, is given by \Cref{eqn:C_duC_cov}.
Moreover, the covariance between $\widetilde{\omega}$ and $D_\bfu (\widetilde{\omega})$ is continuous for any pair of $(\bfv_1,\bfv_2) \in \mathcal{D}\times \mathcal{D} \backslash \mathcal{Z}_2$, provided that the covariance between ${\omega}$ and $D_\bfu ({\omega})$ is continuous.
\end{theorem}

We are also able to express the NNDP covariance function, i.e., the covariance between $D_\bfu \widetilde{\omega}(\bfv_1)$ and $D_\bfu \widetilde{\omega}(\bfv_2)$ as given in \Cref{eqn:duC_var} and \Cref{thm:duC_var}.
\begin{equation}
\begin{split}
\label{eqn:duC_var}
&\cov(D_\bfu \widetilde{\omega}(\bfv_1),D_\bfu \widetilde{\omega}(\bfv_2))= \\
&\begin{cases}
D_\bfu C(\bfs_i,\mathcal{N}_0(\bfs_i))\mathbf{C}_{\mathcal{N}_0(\bfs_i)}^{-1}\widetilde{\mathbf{C}}_{\mathcal{N}_0(\bfs_i),\mathcal{N}_0(\bfs_j)}\mathbf{C}_{\mathcal{N}_0(\bfs_j)}^{-1} D_\bfu C(\mathcal{N}_0(\bfs_j),\bfs_j) & \bfv_1,\bfv_2 \in \mathbf{S}, \bfv_1\neq \bfv_2 \\
 D_\bfu C(\bfs_i,\mathcal{N}_0(\bfs_i))\mathbf{C}_{\mathcal{N}_0(\bfs_i)}^{-1}\widetilde{\mathbf{C}}_{\mathcal{N}_0(\bfs_i)}\mathbf{C}_{\mathcal{N}_0(\bfs_i)}^{-1} D_\bfu C(\mathcal{N}_0(\bfs_i),\bfs_i) \\
~~+D_{\bfu,\bfu}^{(2)}C(\bfs_i,\bfs_i) - D_\bfu C(\bfs_i,\mathcal{N}_0(\bfs_i))\mathbf{C}_{\mathcal{N}_0(\bfs_i)}^{-1}D_\bfu C(\mathcal{N}_0(\bfs_i),\bfs_i)& \bfv_1= \bfv_2=\bfs_i\in \mathbf{S}\\%\bfv_1+h\bfu=\bfv_2+h\bfu \\
D_\bfu C(\bfs_i,\mathcal{N}_0(\bfs_i))\mathbf{C}_{\mathcal{N}_0(\bfs_i)}^{-1}\widetilde{\mathbf{C}}_{\mathcal{N}_0(\bfs_i),\mathcal{N}(\bfv_2)}\mathbf{C}_{\mathcal{N}(\bfv_2)}^{-1} D_\bfu C(\mathcal{N}(\bfv_2),\bfv_2) & \bfv_1=\bfs_i,~\bfv_2 \notin \mathcal{S} \\
D_\bfu C(\bfv_1,\mathcal{N}(\bfv_1))\mathbf{C}_{\mathcal{N}(\bfv_1)}^{-1}\widetilde{\mathbf{C}}_{\mathcal{N}(\bfv_1),\mathcal{N}_0(\bfs_j)}\mathbf{C}_{\mathcal{N}_0(\bfs_j)}^{-1} D_\bfu C(\mathcal{N}_0(\bfs_j),\bfs_j) & \bfv_1\notin \mathcal{S},\bfv_2=\bfs_j \\
D_\bfu C(\bfv_1,\mathcal{N}(\bfv_1))\mathbf{C}_{\mathcal{N}(\bfv_1)}^{-1}\widetilde{\mathbf{C}}_{\mathcal{N}(\bfv_1),\mathcal{N}(\bfv_2)}\mathbf{C}_{\mathcal{N}(\bfv_2)}^{-1} D_\bfu C(\mathcal{N}(\bfv_2),\bfv_2) & \bfv_1,\bfv_2\notin \mathcal{S},\bfv_1 \neq \bfv_2,\\&\bfv_1+h\bfu \neq \bfv_2 \\
D_{\bfu} C(\bfv_1,\mathcal{N}(\bfv_1))\mathbf{C}_{\mathcal{N}(\bfv_1)}^{-1}\widetilde{\mathbf{C}}_{\mathcal{N}(\bfv_1)}\mathbf{C}_{\mathcal{N}(\bfv_1)}^{-1} D_{\bfu} C(\mathcal{N}(\bfv_1),\bfv_1)  \\
~~+D^{(2)}_{\bfu}C(\bfv_1,\bfv_1)- D_{\bfu} C(\bfv_1,\mathcal{N}(\bfv_1))\mathbf{C}_{\mathcal{N}(\bfv_1)}^{-1}D_{\bfu} C(\mathcal{N}(\bfv_1),\bfv_1) & \bfv_1,\bfv_2\notin \mathcal{S},\bfv_1 =\bfv_2
\end{cases},
\end{split}
\end{equation}
where $D_{\bfu,\bfu}^{(2)}C(\bfv_1,\bfv_2) = \lim_{h\rightarrow 0 } (D_\bfu C(\bfv_1+h\bfu,\bfv_2)-D_\bfu C(\bfv_1,\bfv_2))/h$, which exists because $\omega$ is MSD~\citep{stein2012interpolation}.

\begin{theorem} \label{thm:duC_var}
Let $\omega(\bfs)\sim \GP(0,C)$ and $\widetilde{\omega}(\bfs)$ be the NNGP derived from $\omega(\bfs)$. Then $\cov(D_\bfu \widetilde{\omega}(\bfv_1),D_\bfu \widetilde{\omega}(\bfv_2))$, the covariance of the NNDP $D_\bfu \widetilde{\omega}(\bfs)$, is given by \Cref{eqn:duC_var}.
 Moreover, the covariance of $D_\bfu \widetilde{\omega}$ is continuous for any pair of $(\bfv_1,\bfv_2) \in \mathcal{D}\times \mathcal{D} \backslash \mathcal{Z}_2$, provided that the covariance of $D_\bfu {\omega}$ is continuous.
\end{theorem}

This completes the construction of the cross-covariance matrix. In practice, for $\mathcal{S} \subset \RR^d$, we only work with the derivative for an orthonormal basis of $\RR^d$, for example, the Euclidean canonical unit vectors along each axis $\{\mathbf{e}_1,\mathbf{e}_2,\dots,\mathbf{e}_d\}$. We could express $\bfu = \sum_{i=1}^d u_i \mathbf{e}_i$ and $\mathbf{v} = \sum_{i=1}^d v_i \mathbf{e}_i$. Thus, we can compute $D_\bfu \omega(\bfs)=\sum_{i=1}^d u_i D_{\mathbf{e}_i}\omega(\bfs)$, and $D_{\bfu,\bfu} \omega(\bfs)=\sum_{i=1}^d \sum_{j=1}^d u_i D_{\mathbf{e}_i}\omega(\bfs)u_j$. We investigate the NNGP and NNDP using a differential operator $\mathcal{L}: \RR^1 \rightarrow \RR^{1+d}$, where $\mathcal{L}\widetilde{\omega} = (\widetilde{\omega},\nabla \widetilde{\omega}\T)$. The resulting process $\mathcal{L}\widetilde{\omega}$ is still a GP with a zero-mean and a cross-covariance matrix
\[
   \begin{bmatrix} 
C & (\nabla \widetilde{C})^\top  \\ 
\nabla \widetilde{C} & \nabla^2 \widetilde{C} %& \nabla^3 K^\top %\\ 
%\nabla^2 K & -\nabla^3 K & \nabla^4 K 
\end{bmatrix},
\]
where $C$ and $\nabla\widetilde{C}$ are both function of $C$, with detailed formula listed in \Cref{eqn:duC_var} and \Cref{eqn:duC_var}. This extends the results in \cite{banerjee2003directional} to a scalable version. Joint inference for the gradients is achieved using this cross-covariance matrix, as in \cite{banerjee2003directional} and \cite{halder2024bayesian}.

%We could use any valid isotropic cross-covariance function to construct $C$ in the parent GP (see, e.g., \cite{banerjee2003directional,halder2024bayesian}). 
The smoothness requirement remains the same for the NNDP kernel as the original GP kernel \citep{banerjee2003directional, halder2024bayesian}. For example, for the Mat\'ern kernel, we require $\nu > 1$ so that the NNDP exists. Despite the choice of an isotropic kernel in the parent GP, the resulting NNDP kernel is not isotropic due to its dependence on the neighborhood set. See Algorithm~S1%\Cref{alg:NNDP} 
for further details on computing with NNDP.

\subsection{Bayesian Hierarchical Model}

We construct a spatial regression model for a spatially indexed response $Y(\bfs)$
\[
Y(\bfs) = \mu(\bfs, \boldsymbol{\beta}) + \widetilde{\omega}(\bfs) + \epsilon(\bfs),
\]
where $\mu(\bfs,\boldsymbol{\beta} )$ is a mean function parametrized by $\boldsymbol{\beta}$, \(\widetilde{\omega} (\bfs) \sim \text{NNGP}(0, \widetilde{C}(\cdot, \cdot; \boldsymbol{\theta})) \) is a latent NNGP derived from the parent $\GP(0,C(\cdot, \cdot; \boldsymbol{\theta}))$, $\boldsymbol{\theta}=\{\sigma^2,\phi\}$ and \(\epsilon(\bfs) \sim N(0, \tau^2) \) is an independent white noise process. The mean function is customarily modeled as \( \mu(\bfs, \boldsymbol{\beta}) = \mathbf{x}(\bfs)^\top \boldsymbol{\beta} \), where $\mathbf{x}(\bfs)$ is a $p\times 1$ vector of covariates and $\boldsymbol{\beta}$ is a $p\times 1$ vector of coefficients. 

Let ${\cal S} = \{\bfs_1,\ldots,\bfs_n\}$ be a set of $n$ spatial locations where we have observed the response and the covariates. Let $\mathbf{y} = (Y(\bfs_1),\ldots, Y(\bfs_n))^{\T}$ be the $n\times 1$ vector of observed responses and let $\mathbf{X}$ be a fixed and known $n\times p$ matrix of predictors with rows $\bfx(\bfs_i)^{\T}$. A Bayesian specification specifies prior distributions on $\boldsymbol{\beta}$, $\boldsymbol{\theta}$ and $\tau^2$. The posterior distribution $p(\boldsymbol{\beta},\boldsymbol{\theta},\tau,\boldsymbol{\widetilde{\omega}} \given \mathbf{y})$ is proportional to the joint model 
\begin{equation}\label{eq:posterior}
\begin{split}
p(\boldsymbol{\theta}) \times {\rm IG}(\tau^2 \given a_\tau, b_\tau) 
\times N_p(\boldsymbol{\beta}\given \mu_\beta, \mathbf{V}_\beta) \times N_n(\boldsymbol{\widetilde{\omega}}\given 0, \widetilde{\mathbf{C}}) \times N_n(\mathbf{y}\given \mathbf{X}\boldsymbol{\beta}+\boldsymbol{\widetilde{\omega}},\tau^2\mathbf{I}),
\end{split}
\end{equation}
where $p(\boldsymbol{\theta}) = {\rm IG}(\sigma^2\mid a_\sigma,b_\sigma)\times U(\phi\mid a_\phi,b_\phi)$. %The NNDP is also a valid GP, as proved in \Cref{thm:NNDP_proper}. This property ensures that the NNDP retains the foundational strengths of GP, including accurate inference and prediction, and statistical rigor.

\begin{theorem} \label{thm:NNDP_proper}
    Let $\omega$ be a GP on $\mathcal{D}$. NNDP is a valid GP on $\mathcal{D} \setminus \mathcal{Z}_2$, with joint density, $\widetilde{p}(\boldsymbol{\widetilde{\omega}_\mathcal{S}},D_\bfu \boldsymbol{\widetilde{\omega}_\mathcal{S}})$, given by
    \[
    p(\omega(\bfs_1))\prod_{i=2}^k p(\omega(\bfs_i)\given\boldsymbol{\omega_{\mathcal{N}(\bfs_i)}}) p(D_\bfu\omega( \bfs_1)\given \boldsymbol{\omega_\mathcal{S}}) \prod_{i=2} p(D_\bfu \omega(\bfs_i)\given D_\bfu \boldsymbol{\omega_{\mathcal{N}_0(\bfs_i)\setminus \{s_i\}}}, \boldsymbol{\omega_\mathcal{S}}),
    \]which is a proper joint density. 
\end{theorem}

For each posterior sample of $\{\boldsymbol{\widetilde{\omega}},\boldsymbol{\theta}\}$ drawn from \Cref{eq:posterior}, we draw $D_\bfu \boldsymbol{\widetilde{\omega}}\given \boldsymbol{\widetilde{\omega}},\boldsymbol{\theta} \sim N_n(-(\nabla \widetilde{\mathbf{C}})\T\widetilde{\mathbf{C}}^{-1} \widetilde{\boldsymbol{\omega}},\nabla^2 \widetilde{\mathbf{C}}-(\nabla \widetilde{\mathbf{C}})\T\widetilde{\mathbf{C}}^{-1}\nabla \widetilde{\mathbf{C}})$, where $\nabla^2 \widetilde{\mathbf{C}}$ and $\nabla \widetilde{\mathbf{C}}$ are computed from expressions housed in \Cref{eqn:C_duC_cov} and \Cref{eqn:duC_var}. For a new location $\bfs' \in {\cal D}\setminus ({\cal Z}_2 \cup {\cal S})$, where $Y(\bfs')$ is unknown, we draw one instance of 
$$D_\bfu\widetilde{\omega}(\bfs')\given \boldsymbol{\widetilde{\omega}},\boldsymbol{\theta} \sim N(-(\nabla \widetilde{\mathbf{C}}_{\mathcal{S},\bfs'})\T\widetilde{\mathbf{C}}^{-1}\widetilde{\boldsymbol{\omega}},\cov(D_\bfu \widetilde{\omega}(\bfs'))-(\nabla \widetilde{\mathbf{C}})_{\mathcal{S},\bfs'}\T\widetilde{\mathbf{C}}^{-1}\nabla \widetilde{\mathbf{C}}_{\mathcal{S},\bfs'})$$ 
for each posterior draw of $\{\boldsymbol{\widetilde{\omega}},\boldsymbol{\theta}\}$, where $\nabla \widetilde{\mathbf{C}}_{\mathcal{S},\bfs'} = \cov(\widetilde{\boldsymbol{\omega}},D_\bfu\widetilde{\omega}(\bfs')) $ is computed from \Cref{eqn:C_duC_cov} and $\cov(D_\bfu \widetilde{\omega}(\bfs'))$ is computed from \Cref{eqn:duC_var}.

%In \Cref{apdx:cont_cov} and \Cref{apdx:cont_var}, we also shows that $\cov(\widetilde{\omega}(\bfv_1),D_\bfu \widetilde{\omega}(\bfv_2))$ and $\cov(D_\bfu \widetilde{\omega}(\bfv_1),D_\bfu \widetilde{\omega}(\bfv_2))$ are continuous for all pairs of $(\bfv_1,\bfv_2)$ outside $\mathcal{Z}_2$, which is a (Lebesgue) measure zero set, under the assumption that the covariance function of the original GP is continuous.

\subsection{Simplified NNDP covariance}

The computational efficiency of NNDP could be further improved by investigating the structure of the covariance matrix of NNDP. To do so, we first investigate the relationship between the covariance function of the NNGP, \(\widetilde{C}\), and that of its parent GP, \(C\). Consider \(\bfs_i\) and its set of neighboring locations \(\mathcal{N}(s_i)\), then covariance matrices \(\mathbf{C}_{\mathcal{N}(\bfs_i)}\) and \(\widetilde{\mathbf{C}}_{\mathcal{N}(s_i)}\) consist of covariances between neighbors in the GP and NNGP, respectively. Here, we consider a special case, that when the covariance among the neighbors in the NNGP equal to the parent GP, that is, \(\widetilde{\mathbf{C}}_{\mathcal{N}(s_i)} = \mathbf{C}_{\mathcal{N}(s_i)}\), what are the consequences? In fact, as asserted by the following \Cref{thm:NNGP_GP_equal}, the covariance between a location \(s_i\) and its neighbors in the NNGP will be identical to that in the parent GP. Similarly in this case, we examine the variance at location \(s_i\) in the NNGP. If \(\widetilde{\mathbf{C}}_{\mathcal{N}(\bfs_i)} = \mathbf{C}_{\mathcal{N}(\bfs_i)}\), then the variance of the NNGP equals that of the parent GP, meaning \(\widetilde{\mathbf{C}}_{s_i, s_i} = \mathbf{C}_{s_i, s_i}\).

\begin{theorem} \label{thm:NNGP_GP_equal}
    If $\mathbf{C}_{\mathcal{N}(\bfs_i)}=\widetilde{\mathbf{C}}_{\mathcal{N}(\bfs_i)}$, then $\mathbf{C}_{\bfs_i,\mathcal{N}(\bfs_i)}=\widetilde{\mathbf{C}}_{\bfs_i,\mathcal{N}(\bfs_i)}$ and $\mathbf{C}_{\bfs_i,\bfs_i}=\widetilde{\mathbf{C}}_{\bfs_i,\bfs_i}$.
\end{theorem}

Theorem~\ref{thm:NNGP_GP_equal} states where the neighbor covariances match, the NNGP preserves the variance at each location, aligning it with the variance in the parent GP. %These findings highlight that when the covariance among neighboring locations in the NNGP is consistent with that of the parent GP, the NNGP effectively retains the local covariance structure of the GP. This has important implications for the fidelity of the NNGP approximation. It suggests that the NNGP can capture essential characteristics of the GP, particularly in terms of local covariance relationships, while potentially offering computational advantages due to its reliance on a subset of neighboring locations.
Moreover, in practical implementations \citep{finley2019efficient}, prediction at a new location essentially uses Bayesian ``kriging'' based on the neighborhood set with the original kernel function \(\mathbf{C}\), rather than the NNGP's approximated kernel function \(\widetilde{\mathbf{C}}\). Topological ordering of coordinates is flexible and, for the initial \(m\) points, the covariance structure is constructed using the true kernel function. %This can be viewed as an iterative application of the above results. 
Consequently, the cross-covariance between a new point (indexed as \(m+1\)) and the first \(m\) points, along with its variance, matches that derived from the original GP. Hence, inference for a new point is achieved by reordering the data, placing its neighbors in positions \(1\) through \(m\) and the new point at position \(m+1\). Leveraging these results, yields a simplified NNDP kernel. It involves only the original covariance function \(\mathbf{C}\), rather than its NNGP kernel \(\widetilde{\mathbf{C}}\). Specifically, for $\bfv_1=\bfs_i,~\bfv_2=\bfs_j \in \mathcal{S}$, and when $\bfs_j \in \mathcal{N}(\bfs_i)$
\begin{equation}\label{eqn:dC_NN}
    \cov(\widetilde{\omega}(\bfv_1),D_\bfu \widetilde{\omega}(\bfv_2))= D_\bfu C(\bfs_i,\mathcal{N}_0(\bfs_i))\mathbf{C}_{\mathcal{N}_0(\bfs_i)}^{-1}\widetilde{\mathbf{C}}_{\mathcal{N}_0(\bfs_i),\bfs_j} = D_\bfu C(\bfs_i,\bfs_j).
\end{equation}
For $\bfv_1\in \mathcal{S}$, we easily obtain $\cov(D_\bfu \widetilde{\omega}(\bfv_1),D_\bfu \widetilde{\omega}(\bfv_1)) = D_{\bfu,\bfu}^{(2)}C(\bfs_i,\bfs_i)~\refstepcounter{equation}(\theequation)\label{eqn:varC_NN}$.
\begin{comment}
\begin{equation}\label{eqn:varC_NN}
\begin{aligned}
    \cov(D_\bfu \widetilde{\omega}(\bfv_1),D_\bfu \widetilde{\omega}(\bfv_1)) &= D_\bfu C(\bfs_i,\mathcal{N}_0(\bfs_i))\mathbf{C}_{\mathcal{N}_0(\bfs_i)}^{-1}\widetilde{\mathbf{C}}_{\mathcal{N}_0(\bfs_i)}\mathbf{C}_{\mathcal{N}_0(\bfs_i)}^{-1} D_\bfu C(\mathcal{N}_0(\bfs_i),\bfs_i) \\
    & \quad +D_{\bfu,\bfu}^{(2)}C(\bfs_i,\bfs_i) - D_\bfu C(\bfs_i,\mathcal{N}_0(\bfs_i))\mathbf{C}_{\mathcal{N}_0(\bfs_i)}^{-1}D_\bfu C(\mathcal{N}_0(\bfs_i),\bfs_i) \\
    & = D_{\bfu,\bfu}^{(2)}C(\bfs_i,\bfs_i).
\end{aligned}
\end{equation}
\end{comment}
These simplifications result in the NN-version NNDP algorithm summarized in Algorithm S1. Algorithm S2 elucidates predictions in new locations. \Cref{thm:NNGP_GP_equal} ensures a valid covariance matrix, although the NNGP itself does not supply an explicit covariance matrix \citep[][]{wang2018rejoinder}.

\subsection{Computational complexity and choice of $\mathcal{S}$}

The NNDP, as developed above, shares the same computational complexity as NNGP. Implementing NNDP involves first fitting an NNGP to the data, after which the joint covariance matrix is used to sample posterior distributions. Flop counts for generating posterior samples are dominated by operations involving the neighbor sets, each of size \(m\). Computing \(\mathbf{B}_{s_i}\) and \(\mathbf{F}_{s_i}\) requires inversion of an $m\times m$ matrix, which takes \(O(m^3)\) flops per location. The total computational cost for parameter updates and sampling is \(\sim O((n + k)m^3)\) flops, where \(n\) is the number of observed locations and \(k\) is the size of the reference set. This linearity with respect to \(n\) and \(k\) ensures that the NNDP model remains computationally feasible even for large \(n\) and \(k\), especially since \(m\) is typically small (e.g., \(m \approx 10\)). Therefore, the NNDP maintains the same computational efficiency as NNGP, allowing for practical implementation in large-scale spatial modeling tasks.% This is particularly advantageous when modeling spatial gradients or processes that require detailed local covariance structures, as the NNDP can provide accurate approximations without incurring prohibitive computational costs.

%In the context of calculating gradients, the choice of the set \(\mathcal{S}\) is crucial. For points in \(\mathcal{Z}_2\) where the variable \(v_1\) equals \(v_1\) outside of \(\mathcal{S}\), the function may exhibit discontinuities. These discontinuities make it challenging to compute second-order gradients at those points. To address this issue, we select \(\mathcal{S} = T\), where \(T\) is the set of observed locations. This ensures that the gradients are calculated only at locations where the function is continuous, which allows an accurate estimation of the second-order gradients and avoids the complications arising from discontinuities. Moreover, this choice of \(\mathcal{S}\) aligns with findings in the NNGP literature, where using the set of observed locations for gradient calculations has demonstrated superior performance compared to other choices of \(\mathcal{S}\). %By focusing on the observed data points, we leverage the available information most effectively, enhancing the model's predictive capabilities and ensuring the reliability of the gradient estimates.

\subsection{Comparison with finite difference}

An alternative approach to estimating spatial gradients is to first fit a GP and then approximate the gradients using finite difference methods~\citep{chen2024investigating}. Denoting the posterior mean of the NNGP as \(\widehat{f}\), the gradients are approximated using finite differences,
\[
D_{\bfu} \widetilde{\omega}(\bfs) \approx \frac{\widehat{f}(\bfs + h \cdot \bfu) - \widehat{f}(\bfs)}{h},
\]
where \(h\) is a sufficiently small step size specified by the user. For example, we could set $h$ as \(\iota = \min_{i\neq j} \|\bfs_i - \bfs_j \|\), which is the minimal separation among observed locations \(\{ \bfs_i \}_{i=1}^n\).

Finite difference methods, while straightforward to implement, have limitations that are obviated by the NNDP. First, the NNDP provides a valid joint distribution of the process and its derivatives, ensuring that the gradient estimates are sampled from a valid posterior distribution. In contrast, finite difference approximations do not capture this joint probabilistic structure, potentially leading to inconsistencies in inference. Moreover, the errors introduced by finite difference approximations can become more pronounced when estimating higher-order gradients. As the order of the derivative increases, the approximation error tends to grow, which can lead to significant inaccuracies in the gradient estimates. This can adversely affect downstream analyses that rely on precise gradient information.

Additionally, finite difference methods require specifying a step size \(h\), which is data-dependent and sensitizes inference. This choice can compromise the accuracy of the gradient estimation or amplify numerical errors due to the underlying spatial configuration \citep{banerjee2003directional}. Larger values of \(h\) may fail to capture local behavior of the random field and yield poor gradient estimates. Conversely, small values of \(h\) can lead to significant numerical errors, particularly when the minimal separation \(\iota\) between data points is large. Choosing a step size \(h\) that balances accuracy with numerical stability is challenging. This underscores the importance of closed-form distribution theory for spatial random fields.

Adopting the NNDP circumvents these issues by accounting for spatial dependence and providing direct estimates of the gradients without resorting to numerical approximations. This yields more accurate and theoretically consistent gradient estimates, free from the additional error sources and tuning parameters associated with finite difference methods. Next, we validate these observations with experiments in the following section.%The NNDP thus offers a robust and efficient alternative for gradient estimation in spatial modeling, enhancing both the reliability and interpretability of the results.

\section{Simulation Studies}\label{sec:sim}

In this section, we conducted a series of simulation studies to evaluate the performance of our proposed NNDP in estimating spatial gradients. We compared NNDP with exact inference (only for small sample sizes when exact inference is computable) and finite difference methods. The following are a set of simulations designed to assess the accuracy and robustness of the NNDP under various functional forms and sampling schemes, particularly examining scenarios where finite difference methods might struggle due to their reliance on step size selection. Due to space constraints, details about priors and posterior samplings are presented in the Appendix Section H. We note here that for simplicity, all the experiments are done on the $\omega$ level instead of $Y$ level, that is, we consider the simple model $Y(\bfs) = \omega(\bfs)$. We use $\phi \sim \text{unif}(0.01,300)$, $\sigma^2 \sim {\rm IG}(0.01,10)$ for priors. Gradient estimates are the batch median of the posterior samples. Batch size is set as $50$ for the simulation of pattern 2 equal spacing case (\Cref{tab:pattern2_scale}, \Cref{fig:pattern2_scale1}), and $100$ for all other experiments.

\subsection{Gradient Estimation}

We first consider a smooth sinusoidal function, $\omega_1(\bfs) = 10 \left( \sin(3\pi s_1) + \cos(3\pi s_2) \right)$ (Pattern 1), where $\bfs = (s_1,s_2)\T \in [0,1]^2$. Data are generated on a regular grid with mesh size $0.01$, which also serves as the minimal separation \( \iota \), i.e., the minimal distance between distinct points. We estimate spatial gradients using three approaches: (i) exact inference; (ii) the NDP; and (iii) the finite difference method (shown as FD in all tables) with \( h = \iota \).

\begin{figure}[!h]
    \centering
    \includegraphics[width=\linewidth]{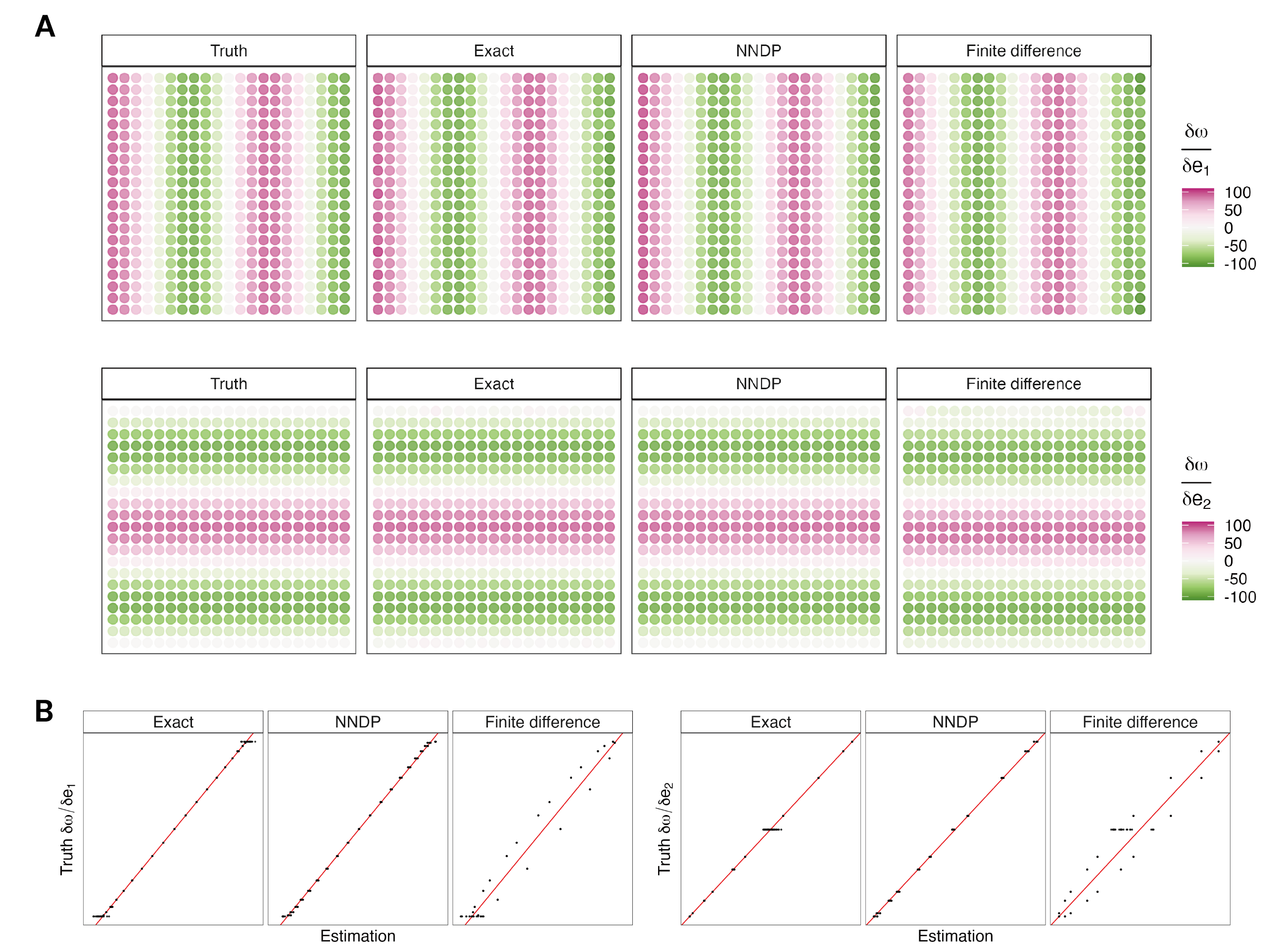}
    \caption{Spatial gradient estimates for Pattern 1. (A) Spatial gradient for $\frac{\delta \omega}{\delta \mathbf{e}_1}$ (top) and $\frac{\delta \omega}{\delta \mathbf{e}_2}$ (bottom). (B) Scatter plot comparing truth and estimated spatial gradient for exact inference, NNDP and finite difference. The red line is the $y=x$ line.}
    \label{fig:pattern1_scale1_combined}
\end{figure}

Spatial gradients are evaluated along canonical directions $\mathbf{e}_1=(1,0)\T$ and $\mathbf{e}_2 = (0,1)\T$, where the true spatial gradients are $\nabla \omega_1(\bfs) = 30 \pi(\cos (3\pi s_1),-\sin (3\pi s_2))\T$ (\Cref{fig:pattern1_scale1_combined}A first column). For all simulations, we considered using the Mat\'ern kernel as $C$ when fitting GP in exact inference and fitting NNGP in NNDP and finite difference, where $\nu=5/2$ so that $\nabla^2\omega_1$ exists. %The Mat\'ern kernel is given by
%$C(x,x')=\frac{\sigma^2 2\left(\frac{\alpha}{2}\left\|x-x'\right\|\right)^\nu}{\Gamma(\nu)}K_\nu (\alpha \left\|x-x'\right\|)$, where $K_\nu$ is the modified Bessel function of the second kind and $\Gamma$ is the Gamma function. When $\nu=1/2$, the Mat\'ern kernel becomes the exponential kernel:$C(x,x')=\sigma^2 \exp(-\alpha \left\|x-x'\right\|)$. The parameter $\nu$ is called the smoothness, since the smoothness of a Mat\'ern GP is exactly $\ceil{\nu}-1$. 
%We fix $\nu=5/2$ to ensure that $\nabla^2\omega_1$ exists. 
\Cref{fig:pattern1_scale1_combined} shows that NNDP accurately approximates true gradients and is comparable with exact inference. NNDP outperforms the finite difference method in terms of correlation and mean squared error (MSE) metrics~(\Cref{tab:mse_cor_scale1}). The finite difference method yields markedly less stable estimates due to its sensitivity to the step size $h$.

% \usepackage{multirow}
\begin{table}[]
\caption{Gradient estimation correlation and MSE.}
\vspace{0.5cm}
\centering
\label{tab:mse_cor_scale1}
\begin{tabular}{llllll}
\hline
Pattern                    & Method            & Cor $\frac{\delta \omega}{\delta \mathbf{e}_1}$ & Cor $\frac{\delta \omega}{\delta \mathbf{e}_2}$ & MSE $\frac{\delta \omega}{\delta \mathbf{e}_1}$ & MSE $\frac{\delta \omega}{\delta \mathbf{e}_2}$ \\ \hline
\multirow{3}{*}{Pattern 1} & Exact  & 0.99968                           & 0.99953                           & 3.0105                            & 3.6847                            \\
& {\bf NNDP}              & {\bf 0.99941  }                         & {\bf 0.99975}                           & {\bf 5.5283}                            & {\bf 2.9736}                            \\
& FD & 0.97653                           & 0.96675                           & 236.41                            & 256.71                            \\ \hline
\multirow{3}{*}{Pattern 2} & Exact  & 0.94955                           & -                                 & 177.98                            & -                                 \\
& {\bf NNDP}              & {\bf 0.99981}                           & -                                 & {\bf 0.74745}                           & -                                 \\ 
                           & FD & 0.92898                           & -                                 & 244.3                             & -                                 \\ \hline
\end{tabular}
\end{table}

\begin{figure}[!h]
    \centering
    \includegraphics[width=\linewidth]{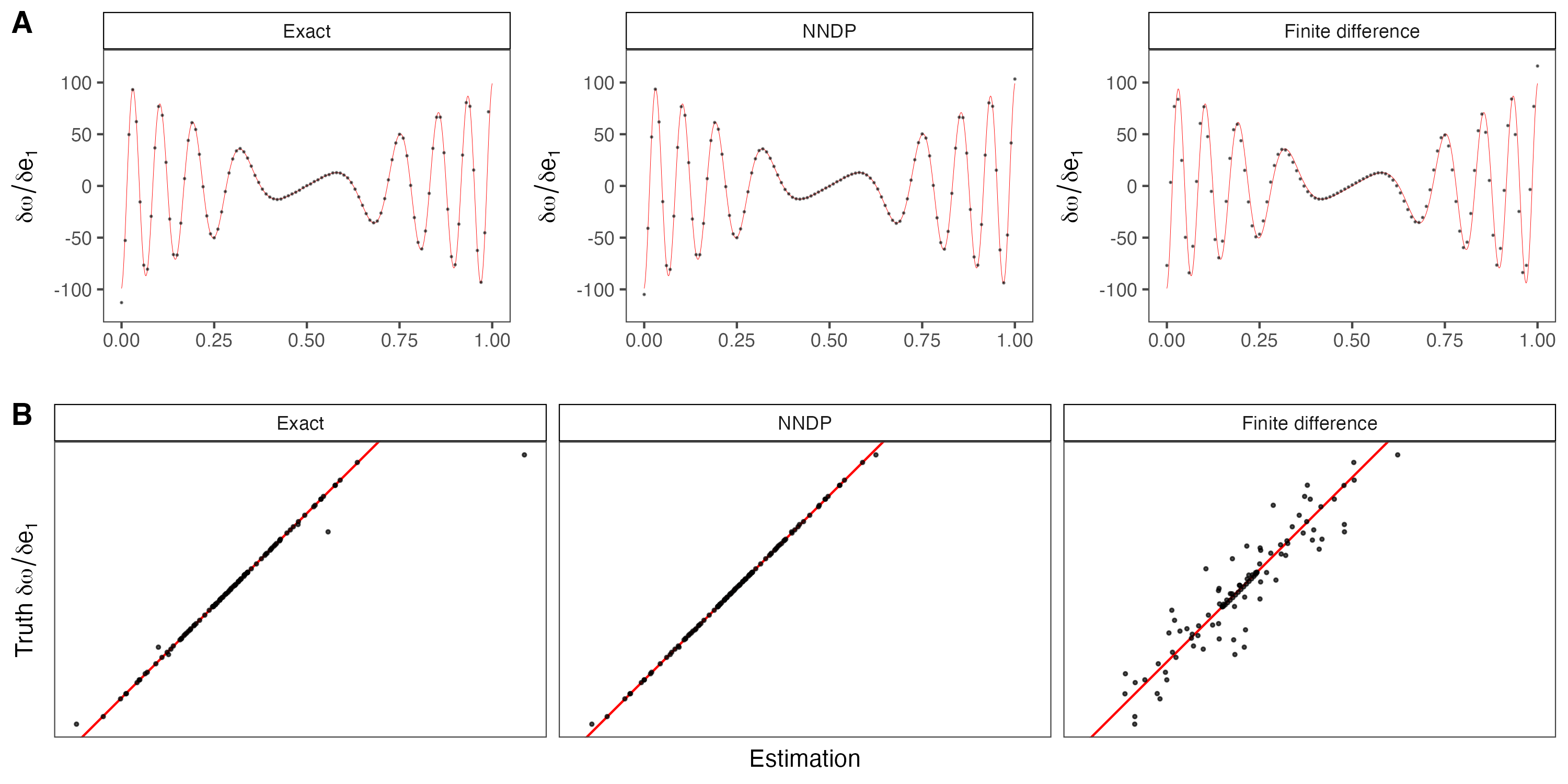}
    \caption{Spatial gradient estimates for Pattern 2. Spatial gradient estimates for Pattern 1. (A) Estimated spatial gradient (black dots) for $\frac{\delta \omega}{\delta \mathbf{e}_1}$. The red line indicates the truth. (B) Scatter plot comparing truth and estimated spatial gradient for exact inference, NNDP and finite difference. The red line is the $y=x$ line.}
    \label{fig:pattern2_scale1}
\end{figure}

We also examine a highly oscillatory function, Pattern 2: $\omega_2(\bfs) = \sin\left( 100 (s_1 - 0.5)^2 \right)$, with $\bfs = (s_1) \in [0,1]$ on a regular grid with mesh size \( 0.01 \). This function is more challenging due to rapid changes and non-stationary behavior. Spatial gradients are evaluated along the canonical direction $\mathbf{e}_1=(1,0)\T$, where the true spatial gradient is given by $\nabla \omega_2(\bfs) = 200 \cos ((s_1-0.5)^2) (s_1-0.5)$ (\Cref{fig:pattern2_scale1} red line). NNDP again excels over the finite difference method as evinced by higher correlation with the exact gradients and lower MSE (\Cref{fig:pattern2_scale1},~\Cref{tab:mse_cor_scale1}). This suggests finite difference methods may struggle more for functions with high-frequency components or non-stationary behavior.

\subsection{Effect of Step Size in Finite Difference Method}

A key advantage of NNDP over finite difference approximation is the challenges of step size (\( h \)) selection in the finite difference method. Here, we conducted several simulations using various step sizes to investigate this problem. For Pattern 1, we simulate \( s_1 \) and \( s_2 \) as equidistant sequences of points ranging from 0 to 1 with minimal separation of \( \iota = 0.005,\, 0.01,\, \text{and}\ 0.02 \). For each \( \iota \), we compared the performance of NNDP with the finite difference method applied using different scaling factors of the minimal separation \( \iota \) between data points, specifically setting \( h = \text{scale} \times \iota \) with \( \text{scale} = 0.005,\, 0.05,\, 0.5,\, 1,\, 5,\, 10 \). 

As shown in \Cref{tab:pattern1_scale}, the finite difference method achieved its best performance in terms of correlation and MSE when \( \text{scale} = 0.05 \). %In Pattern 2, the optimal scale for the finite difference method varied with different \( \iota \) values, reflecting the sensitivity of the method to both the function's characteristics and the data resolution.
Despite extensive tuning, even the best-performing finite difference approximations could not outperform the accuracy of the NNDP in estimating gradients (\Cref{tab:pattern1_scale}).

\begin{table}[!h]
\caption{Pattern 1: spatial gradient estimation with varying $\iota$ and scale}
\label{tab:pattern1_scale}
\vspace{0.5cm}
\centering
\begin{tabular}{lllllll}
\hline
$\iota$ & Method & Scale & Cor $\delta \omega/\delta \mathbf{e}_1$ & Cor $\delta \omega/\delta \mathbf{e}_2$ & MSE $\delta \omega/\delta \mathbf{e}_1$ & MSE $\delta \omega/\delta \mathbf{e}_2$ \\
\hline 
\multirow{7}{*}{0.02}
& NNDP & - & 0.9999916& 0.9999957& 0.091& 0.058 \\\cdashline{2-7}
& \multirow{6}{*}{FD} & 0.005 & 0.9999512 & 0.9999385 & 0.445 & 0.489 \\
& & \textbf{0.05} & \textbf{0.9999904} & \textbf{0.9999911} & \textbf{0.091} & \textbf{0.071} \\
& & 0.5 & 0.999053 & 0.9987022 & 9.397 & 10.312 \\
& & 1 & 0.996081 & 0.9948974 & 38.817 & 40.507 \\
& & 5 & 0.9036247 & 0.8902348 & 923.080 & 845.255 \\
& & 10 & 0.6358632 & 0.624448 & 3171.776 & 2644.353 \\
\hline
\multirow{7}{*}{0.01}
& NNDP & - & 0.9999997& 0.9999997& 0.003& 0.003 \\\cdashline{2-7}
& \multirow{6}{*}{FD} & 0.005 & 0.9999679 & 0.9999633 & 0.288 & 0.295 \\
&  & \textbf{0.05} & \textbf{0.9999982} & \textbf{0.9999979} & \textbf{0.017} & \textbf{0.017} \\
&  & 0.5 & 0.9997618 & 0.9997018 & 2.348 & 2.389 \\
&  & 1 & 0.9990041 & 0.998753 & 9.796 & 9.989 \\
&  & 5 & 0.9751834 & 0.9709178 & 241.686 & 231.172 \\
&  & 10 & 0.9013132 & 0.8909226 & 931.865 & 848.240 \\
\hline
\multirow{7}{*}{0.005}
& NNDP & - & 1 & 0.9999999 & 0 & 0.001 \\\cdashline{2-7}
& \multirow{6}{*}{FD} & 0.005 & 0.9999984 & 0.999998 & 0.014 & 0.016 \\
&  & \textbf{0.05} & \textbf{0.9999996} & \textbf{0.9999995} & \textbf{0.004} & \textbf{0.004} \\
&  & 0.5 & 0.9999395 & 0.9999254 & 0.593 & 0.6 \\
&  & 1 & 0.9997485 & 0.9996922 & 2.464 & 2.477 \\
&  & 5 & 0.9937166 & 0.9925235 & 61.336& 60.048 \\
&  & 10 & 0.9747562 & 0.9708506 & 243.638 & 232.918 \\
\hline
\end{tabular}
\end{table}

We subsequently extend our simulations to include data sampled from a uniform distribution rather than a regular grid to assess the robustness of the methods under irregular sampling schemes. We simulated \( n = 1{,}000 \) points with \( s_1 \sim \text{Uniform}(0, 1) \) and \( s_2 \sim \text{Uniform}(0, 3) \), resulting in a minimal separation of approximately \( \iota = 0.0014 \) (\Cref{tab:pattern1_unif}). In this case, the finite difference method excelled with \( \text{scale} = 0.5 \), which differed from the optimal scale in the regular grid scenario (\Cref{tab:pattern1_unif}). 

This finding is particularly noteworthy because in practice gradients are not directly observed, which makes it impossible to tune the optimal step size in a finite difference method. Selecting an inappropriate step size can be extremely dangerous, as our simulations showed that both too small and too large step sizes led to significantly increased MSEs, making gradient estimation unreliable.

\begin{table}[!t]
\caption{Pattern 1: spatial gradient estimation performance with randomly sampled points.}
\centering
\vspace{0.5cm}
\label{tab:pattern1_unif}
\begin{tabular}{llllll}
\hline
Method   & Cor $\delta \omega/\delta \mathbf{e}_1$ & Cor $\delta \omega/\delta \mathbf{e}_2$ & MSE $\delta \omega/\delta \mathbf{e}_1$ & MSE $\delta \omega/\delta \mathbf{e}_2$ & Scale\\ \hline
NNDP     & 0.9997931    & 0.9998362   & 1.83    & 1.49    & -    \\ \hdashline
\multirow{8}{*}{FD} & 0.9674478    & 0.9164227   & 296.40  & 788.44  & 0.005\\
 %& 0.9330558    & 0.990939    & 668.38  & 77.19   & 0.01 \\
 & 0.9992453    & 0.9993614   & 6.66    & 5.51     & 0.05 \\
% & 0.9997 & 0.9997337   & 2.65    & 2.36     & 0.1  \\
 & \textbf{0.9997809}   & \textbf{0.999829}   & \textbf{1.93 }    & \textbf{1.54 }   & \textbf{0.5} \\
 & 0.9997705    & 0.9998099   & 2.03     & 1.71     & 1    \\
 & 0.9993027    & 0.9992765   & 6.57     & 6.28     & 5    \\
 & 0.9978645    & 0.9975708   & 20.67   & 20.79   & 10   \\ \hline
\end{tabular}
\end{table}

For Pattern 2, we repeated the same experiments as in Pattern 1, considering both regular grid data and uniformly sampled points. For the regular grid case, we explored different mesh sizes, \( \iota = 0.0005,\, 0.001,\, \text{and}\ 0.005 \), and scales, \( \text{scale} = 0.001,\, 0.01,\, 0.1,\, 0.5,\, 1,\, 3,\, 5 \), for the finite different method (\Cref{tab:pattern2_scale}). 

\begin{table}[!htpb]
\caption{Pattern 2: spatial gradient estimation with varying $\iota$ and scale.}
\label{tab:pattern2_scale}
\vspace{0.5cm}
\centering
\begin{tabular}{lllll}
\hline
$\iota$ & Method & Scale & Cor $\delta \omega/\delta \mathbf{e}_1$ & MSE $\delta \omega/\delta \mathbf{e}_1$ \\
\hline
\multirow{6}{*}{0.0005}
& NNDP & - & 1 & 0 \\\cdashline{2-5}
& \multirow{6}{*}{FD} & 0.001 & 0.0791 & 102306.04 \\
&  & 0.01  & 0.65468 & 2035.41 \\
&  & 0.1   & 0.98998 & 33.32 \\
%&  & 0.5   & 0.99978 & 0.75 \\
&  & \textbf{1} & \textbf{0.99982} & \textbf{0.60} \\
&  & 3     & 0.9983 & 5.63 \\
&  & 5     & 0.99526 & 15.72 \\
\hline
\multirow{6}{*}{0.001}
& NNDP & - & 1 & 0 \\\cdashline{2-5}
& \multirow{6}{*}{FD} & 0.001 & 0.64145 & 2524.98 \\
&  & 0.01  & 0.99452 & 18.57 \\
&  & \textbf{0.1} & \textbf{0.99997} & \textbf{0.11} \\
%&  & 0.5   & 0.99981 & 0.64 \\
&  & 1     & 0.99924 & 2.54 \\
&  & 3     & 0.99315 & 22.76 \\
&  & 5     & 0.98104 & 62.81 \\
\hline
\multirow{6}{*}{0.005}
& NNDP & - & 0.99998 & 0.08 \\\cdashline{2-5}
& \multirow{6}{*}{FD} & \textbf{0.001} & \textbf{0.99997} & \textbf{0.09} \\
&  & 0.01  & 0.99997 & 0.09 \\
&  & 0.1   & 0.9998 & 0.69 \\
%&  & 0.5   & 0.99532 & 15.90 \\
&  & 1     & 0.98138 & 63.10 \\
&  & 3     & 0.83972 & 526.17 \\
&  & 5     & 0.59538 & 1253.18 \\
\hline
\end{tabular}
\end{table}

The results show that the finite difference method does not have a consistent best scale across mesh sizes. For example, with $\iota=0.0005$ the best scale was $1$, while for $\iota=0.001$ it was $0.1$, and for $\iota=0.005$ it dropped to $0.001$ (\Cref{tab:pattern2_scale}). This variability illustrates the difficulty of tuning the step size even under regular grids. To further test robustness, we repeated the same experiment under irregular sampling by drawing points uniformly. The results (\Cref{tab:pattern2_unif}) again showed that the best finite difference scale shifted unpredictably (here scale $=3$ gave the best performance), confirming that the method’s accuracy depends strongly on sampling design.

\begin{table}[H]
\caption{Pattern 2: spatial gradient estimation performance with randomly sampled points.}
\label{tab:pattern2_unif}
\centering
\vspace{0.5cm}
\begin{tabular}{llll}
\hline
Method & Scale & Cor $\delta \omega/\delta \mathbf{e}_1$ & MSE $\delta \omega/\delta \mathbf{e}_1$ \\ \hline
NNDP  & - & 1 & 0 \\ \cdashline{1-4}
\multirow{6}{*}{FD} 
& 0.001 & -0.049407 & 1587233.2 \\
& 0.01  & 0.24659 & 16579.5 \\
%& 0.05  & 0.90937 & 339.4 \\
& 0.1   & 0.97203 & 97.4 \\
%& 0.5   & 0.99833 & 5.7 \\
& 1     & 0.99969 & 1.1 \\
& \textbf{3} & \textbf{0.99984} & \textbf{0.6} \\
& 5     & 0.99976 & 0.8 \\
%& 10    & 0.99913 & 3 \\
%& 20    & 0.99653 & 12 \\
%& 100   & 0.90995 & 297.1 \\ 
\hline
\end{tabular}
\end{table}

These experiments reveal that the optimal scaling factor for the step size in the finite difference method varies not only between different spatial patterns, but also depends on the sampling scheme and the scales of the input variables. If the spatial coordinates \( s_1 \) and \( s_2 \) have different ranges or scales, the optimal step size for finite difference approximations may become inconsistent across dimensions making it challenging to tune the step size. This complicates practical application of finite difference methods, as the lack of a universally optimal step size necessitates case-by-case adjustments---a process that is impractical when gradients are unobserved and the true optimal step size is unknown. The NNDP, in contrast, avoids tuning the step size, yielding numerically robust inference regardless of inherent complexities in spatial patterns, sampling scheme, or variable scales. %This robustness underscores the NNDP's suitability for spatial gradient estimation in diverse scenarios, offering a reliable alternative to finite difference methods.

\subsection{Computation complexity}
We report computation times for the gradient estimation in \Cref{tab:time}. We simulated data using Pattern 1 with varying sample sizes from 100 to 40,000 to assess the scalability of NNDP, finite difference, and exact inference methods. The computations were performed using 9 cores and a neighborhood size of $m = 10$ for the NNDP and finite difference.

\Cref{tab:time} presents the total time for computing gradients along $s_1$ and $s_2$ for each method and sample size. We observe that the NNDP and finite difference methods exhibit linear time complexity with respect to the sample size. In contrast, the exact inference method shows a significantly higher computational time, which is cubic in sample size. %This demonstrates the scalability advantage of the NNDP and finite difference methods over exact inference in large datasets.

Notably, the finite difference method requires more computational time than the NNDP, despite both methods having same linear time complexity. Specifically, for NNDP, when estimating the gradient on $\mathbf{e}_1$ and $\mathbf{e}_2$, we only calculate the covariance matrix for its neighbors once. However, the finite difference method requires independent sampling of $f(\bfs + h\bfu)$ for different directions at each location $\bfs$, which leads to additional computational cost.

\iffalse
\begin{figure}[!h]
    \centering
    \includegraphics[width=0.8\linewidth]{figure/time.png}
    \caption{Computation time (seconds) for NNDP, Finite Difference and Exact Inference.}
    \label{fig:time}
\end{figure}

\begin{table}[!h]
\caption{Computation time (seconds) for NNDP, Finite Difference and Exact Inference.}
\label{tab:time}
\begin{center}
\begin{tabular}{llll}
\hline
Sample size & NNDP  & Finite Difference & Exact Inference \\ \hline
100 & 0.1  & 0.3& 5.2    \\
%400 & 0.5  & 1.1& 60.3   \\
1,600& 1.7  & 4.3& 3108.2  \\
%4,900& 5.1  & 12.4     & - \\
10,000 & 9.9 & 14.2     & - \\
%14,400 & 14.7 & 37.9     & - \\
40,000 & 40.0 & 55.9    & - \\ \hline
\end{tabular}
\end{center}
\end{table}
\fi

\begin{table}[!h]
\caption{Computation time (seconds) for Exact Inference, NNDP, and Finite Difference.}
\label{tab:time}
\begin{center}
\begin{tabular}{llll}
\hline
Sample size & Exact Inference & NNDP  & FD \\ \hline
100 & 5.2 & \bf 0.1  & 0.3 \\
%400 & 60.3 & 0.5  & 1.1 \\
1,600 & 3108.2 & \bf 1.7  & 4.3 \\
%4,900 & - & 5.1  & 12.4 \\
10,000 & - & \bf9.9 & 14.2 \\
%14,400 & - & 14.7 & 37.9 \\
40,000 & - & \bf40.0 & 55.9 \\ \hline
\end{tabular}
\end{center}
\end{table}

%In contrast, the NNDP does not require the specification of a step size parameter and consistently provided accurate curvature estimates across all tested conditions. Its ability to capture complex curvature structures without parameter tuning demonstrates its robustness and suitability for spatial curvature estimation in diverse scenarios, offering a significant advantage over finite difference methods.

%\newpage

%\bigskip

\section{Applications} \label{sec:application}

In this section, we apply NNDP on real-world massive spatial data. Same as simulation, we use $\phi \sim \text{unif}(0.01,300)$, $\sigma^2 \sim {\rm IG}(0.01,10)$ for priors. Gradient estimates are the batch median of the posterior samples. Batch size is set as $100$ for all experiments.

\subsection{Mouse brain spatial transcriptomics}

%Spatial data are ubiquitous across various scientific disciplines, and the analysis of spatial gradients has proven to be instrumental in understanding complex phenomena in fields such as housing economics, environmental studies, and biology. Spatial gradients capture the rate and direction of change in variables across space, providing insights into local dynamics and interactions. 
In biology, significant spatial gradients can help identify tissue heterogeneity, cellular interactions, and developmental processes \citep{chen2024investigating}. Spatial Transcriptomics (ST) is an emerging technology that measures gene expression while preserving its spatial location, allowing researchers to study the spatial organization of gene activity within tissues \citep{staahl2016visualization}. This technology offers unprecedented insights into cellular function, tissue architecture, and disease mechanisms by preserving the spatial context of gene expression. One of the latest advancements in this field is the Visium High Definition (VisiumHD) platform developed by 10x Genomics. VisiumHD captures gene expression at near-single-cell resolution across entire tissue sections with a 2\si{\micro\meter} $\times$ 2\si{\micro\meter} window, producing datasets with about $100,000$ to 1 million spatial locations (spots). This richness presents both opportunities and computational challenges for analysis.%, necessitating novel statistical methods capable of handling large-scale spatial data.

The mouse brain serves as an excellent model for studying spatial gene expression due to its complex structure and well-characterized anatomy \citep{yao2021taxonomy}. In particular, the cerebral cortex executes multiple functions, for example, generation of voluntary behavior, emotion, cognition, learning, and memory. It is further partitioned into multiple areas with specific input and output connections to many subcortical and other cortical regions~\citep{yao2021taxonomy}. Here, we apply the NNDP to VisiumHD data, leveraging its computational speed and effectiveness to analyze gene expression gradients in the mouse brain.

We obtained VisiumHD data from a coronal section of the mouse brain (See Appendix H for data availability). The dataset consists of expression measurements of 19,059 genes at 98,917 spatial locations within the tissue section, each associated with coordinates in the two-dimensional plane of the tissue slice, providing a highly detailed map of gene activity across the brain section.

Following the convention in the literature, we preprocessed the gene expression using the widely used Seurat package \citep{hao2024dictionary}, including normalization, scaling, smoothing, and clustering to identify distinct cellular populations. From the clustering results, we identified top marker genes related to each cluster. Specifically, we selected two marker genes as the outcome variable for spatial gradient analysis: \textit{Wipf3}, which is enriched in the hippocampal formation (HPF) region, specifically in the cornu ammonis(CA)1, CA2, CA3 pyramidal layers and the dentate gyrus granule cell layer; and \textit{Slc17a7}, which is enriched in both the isocortex and the HPF region.

\begin{figure}[!h]
\centering
\includegraphics[width=\linewidth]{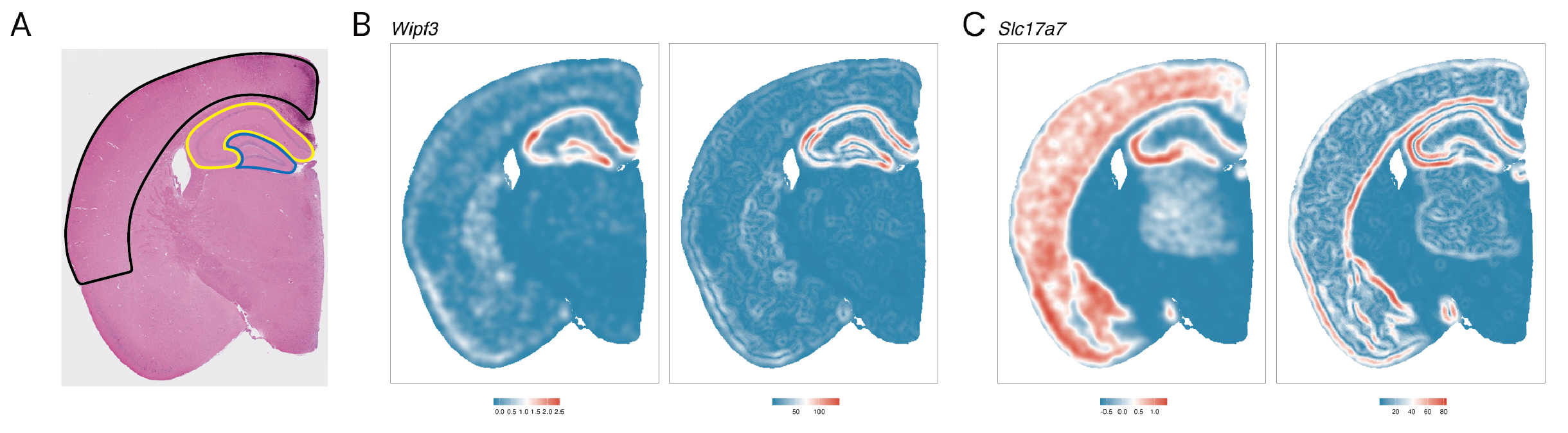}
\caption{Spatial gradient estimates for mouse brain VisiumHD data. (A) H\&E image of a slide of VisiumHD data. The annotation is based on Allen Brain Atlas as a reference. Black area: isocortex. Yellow area: CA. Blue area: dentate gyrus. (B) Gene expression (left) and its estimated spatial gradients (right) of gene \textit{Wipf3}. (C) Gene expression (left) and its estimated spatial gradients (right) of gene \textit{Slc17a7}. The left panels of B and C are colored by gene expression, while the right panels are colored by \( \| \nabla \widetilde{\omega}(\mathbf{s}) \| \). For both B and C, warmer colors signify higher values, while colder colors indicate lower values.}
\label{fig:visium}
\end{figure}

%To enhance spatial continuity and reduce noise, we smoothed the gene expression data using a weighted mean over the \( k \) nearest neighbors, with \( k = 200 \). The weights were determined by a Gaussian kernel based on spatial distance:
%\[
%w_{ij} = \exp\left( -\frac{\norm{\mathbf{s}_i - \mathbf{s}_j}^2}{2\beta^2} \right),
%\]
%where \( \beta = 0.01 \), and \( \mathbf{s}_i \) and \( \mathbf{s}_j \) represent the spatial coordinates of locations \( i \) and \( j \), respectively.

We then applied the NNDP to estimate spatial gradients of expressions of the two aforementioned genes, \textit{Wipf3} and \textit{Slc17a7}, using a Mat\'ern covariance function with smoothness parameter \(\nu = 5/2\). The gradients were computed along the two canonical spatial directions, \( \mathbf{e}_1 = (1, 0)^{\T} \) and \( \mathbf{e}_2 = (0, 1)^{\T} \), corresponding to the horizontal and vertical axes of the tissue section, respectively. At each spatial location \( \mathbf{s} \), the magnitude of the gradient was calculated using the \( L^2 \) norm:
\[
\norm{\nabla \widetilde{\omega}(\mathbf{s})} = \sqrt{ \nabla\widetilde{\omega}_{\mathbf{e}_1}(\mathbf{s})^2 + \nabla\widetilde{\omega}_{\mathbf{e}_2}(\mathbf{s})^2 },
\]
where \( \widetilde{\omega}_{\mathbf{e}_1}(\mathbf{s}) \) and \( \widetilde{\omega}_{\mathbf{e}_2}(\mathbf{s}) \) are the estimated partial derivatives of the gene expression \( \widetilde{\omega}(\mathbf{s}) \) along \( \mathbf{e}_1 \) and \( \mathbf{e}_2 \), respectively.

Modeling the mouse brain Visium HD data using NNDP yielded detailed maps of gene expression gradients for each selected gene. \Cref{fig:visium} presents the estimated gradient for \textit{Wipf3} and \textit{Slc17a7}, which took approximately 81 seconds. Gradient maps evince regions of rapid change in gene expression. We see high gradients along the borders between subregions, such as the dentate gyrus (B right panel) and the CA fields (C right panel), reflecting sharp transitions in gene expression profiles. The NNDP is able to estimate gradients at high spatial resolution, which leads to detecting rapidly changing regions related to biological functions such as disease propagation and the formation of functional structures.

%Moreover, the computational efficiency of the NNDP makes it feasible to apply spatial gradient estimation to high-dimensional datasets with large sample sizes and numerous features. This efficiency is crucial for modern spatial data, where traditional methods may be computationally prohibitive.

\subsection{Air temperature}

\begin{figure}[H]
\centering
\includegraphics[width=\linewidth]{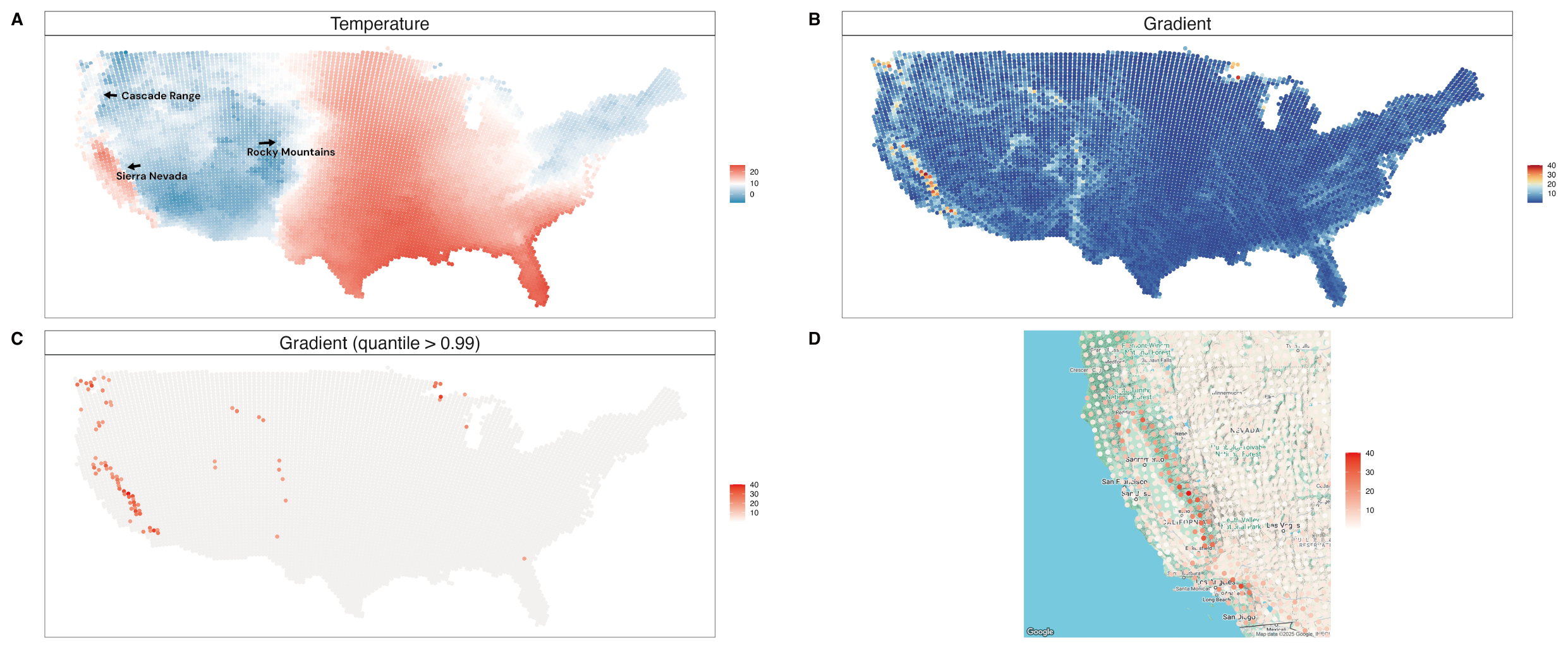}
\caption{Spatial gradient estimates for NARR temperature data. (A) The air temperature heatmap. (B) Estimated spatial gradients for air temperature, colored by the magnitude of the gradient \( \| \nabla \widetilde{\omega}(\mathbf{s}) \| \), with warmer colors indicating regions of higher gradient magnitude. (C) Estimated spatial gradients for air temperature with magnitude greater than the 0.99 quantile. (D) A map of Sierra Nevada mountain area with estimated spatial gradient overlaying on the map. The map was obtained using Google Maps. }
\label{fig:tempe}
\end{figure}

Spatial gradient analysis also plays a crucial role in geospatial studies, for example, in understanding how geographical features influence environmental variables. In this application, we focus on meteorological data sourced from the NCEP North American Regional Reanalysis (NARR) product~\citep{chen2022deepkrigingspatiallydependentdeep}, see Appendix I for the data. Specifically, we examine air temperature data at 2 meters above ground level, recorded on June 05, 2019 (\Cref{fig:tempe}A). This dataset comprises gridded data covering the continental United States with a spatial resolution of approximately 32$\times$32 km, encompassing a total of 7,706 gridded cells.

We applied the NNDP to estimate spatial gradients of temperature, using the Mat\'ern covariance function with smoothness parameter \( \nu = 5/2 \). The gradients were computed along the two canonical spatial directions, and the magnitude of the gradient was calculated using the \( L^2 \) norm (\Cref{fig:tempe}B). The NNDP analysis reveals large spatial variability in temperature gradients across the region. Notably, we observe large gradients primarily in the Sierra Nevada mountain area indicating sharp changes in temperature across this geographical feature (\Cref{fig:tempe}C,D). Other notable regions displaying substantial temperature gradients include the Cascade Range and the Rocky Mountains (\Cref{fig:tempe}C). These findings underscore the impact of topography on temperature, illustrating how mountains cause rapid changes in environmental conditions over relatively short spatial distances.

\section{Discussion} \label{sec:diss}

In this paper, we introduce NNDP, a scalable GP framework for estimating spatial derivatives for large-scale spatial data. NNDP develops derivative processes from an NNGP, providing a fully Bayesian framework. Our theoretical results establish the validity of NNDP as a valid GP, characterize its smoothness properties, and derive the joint distribution between the parent process and its derivatives.

%by offering a valid process and overcoming the difficulty that the NNGP does not admit an explicit covariance function. %Additionally, NNDP enables scalable inference for spatial derivatives without the need for numerical approximations or hyperparameter tuning associated with traditional finite difference methods.

Our simulation studies demonstrate the comparable performance of NNDP and exact inference as well as the superior performance of NNDP over finite difference methods in estimating gradients across various spatial patterns and sampling schemes. %Specifically, NNDP consistently provided more accurate estimates, with higher correlation to true values and lower MSEs. 
In contrast, the finite difference method's performance was sensitive to the choice of step size $h$, which varied depending on the function, sampling density, and spatial dimensions. This sensitivity makes finite difference methods impractical for applications where the optimal $h$ is unknown and hard to tune since the gradients are unobserved.

The application of NNDP to high-resolution ST data from the mouse brain illustrates the method's practical utility in analyzing modern, large-scale spatial datasets. By estimating spatial gradients of gene expression, we identified regions of rapid change corresponding to anatomical structures within the brain, such as hippocampal formation. %NNDP efficiently handled the computational demands of the dataset, which included nearly 100,000 spatial locations, demonstrating its scalability and effectiveness in real-world applications. 
We also applied NNDP to analyze air temperature data, in which we found the rapid changing region matches the geographical features like mountain areas including the Sierra Nevada mountain area. This alignment further highlights the capability of NNDP in geospatial applications and underscores the broader importance of studying spatial derivatives across diverse scientific domains.

For future research, the technique we used to construct NNDP could be extended to higher-order derivatives, such as the curvature process, offering a practical and reliable solution for analyzing complex higher-order spatial phenomena across diverse scientific domains. Furthermore, extending the NNDP framework to spatio-temporal derivative processes proposed by \cite{halder2024Spatiotemporal} is a promising direction. Many real-world phenomena are dynamic over time, and incorporating temporal dynamics alongside spatial dependencies would enable the analysis of the time point when spatial gradients evolve. Developing spatiotemporal NNDP models could be meaningful for application in large-scale environmental monitoring, where pollutant levels change over time, or in epidemiology, where disease spread has both spatial and temporal components. Another promising direction for future work is the extension of the NNDP to multi-output GP. In many applications, multiple related spatial variables are observed simultaneously, such as different pollutants in environmental studies or multiple gene expressions in genomics. Joint modeling of such variables can borrow strength from their cross-covariance structures and, hence, potentially lead to improved estimation of variable relationships.

%\bibliographystyle{chicago}
%\bibliography{ref}

\clearpage
\newpage
\appendix

{\LARGE \noindent \textbf{Appendix}}

\noindent In this appendix, we present proofs for Theorem 1-5 in Section A-E, simplified version of NNDP prediction on new locations in Section F, detailed algorithms for NNDP in Section G, data availability and experimental details for VisiumHD in Section H and air temperature in Section I.

\section{Proof of \Cref{theorem:msd}: Smoothness of the NNGP}
\begin{proof}
$\widetilde{\omega}$ is MSD if $\lim_{h\to0}\frac{\widetilde{\omega}(\bfv+h\bfu)-\widetilde{\omega}(\bfv)}{h}$ exists in the mean-square topology, that is,

\[\lim_{h\to0}\EE\left(\frac{\widetilde{\omega}(\bfv+h\bfu)-\widetilde{\omega}(\bfv)}{h}\right)^2~\text{exists}.\]
\begin{align*}
\text{First observe that }&\lim_{h\to0}\EE\left(\frac{\widetilde{\omega}(\bfv+h\bfu)-\widetilde{\omega}(\bfv)}{h}\right)^2\\& = \lim_{h\to0}\frac{1}{h^2}\left(\widetilde{C}(\bfv+h\bfu,\bfv+h\bfu)-\widetilde{C}(\bfv+h\bfu,\bfv)-\widetilde{C}(\bfv,\bfv+h\bfu)+\widetilde{C}(\bfv,\bfv)\right).
\end{align*}

Define $\mathbf{e} = (1,0,\dots,0)\T$ a $m\times 1$ vector, $m$ is the neighborhood size of ${\bf s}$. We now calculate $\lim_{h\to0}\EE\left(\frac{\widetilde{\omega}(\bfv+h\bfu)-\widetilde{\omega}(\bfv)}{h}\right)^2$ by considering the following two cases. 

\begin{enumerate}
    \item $\mathbf{v} = \bfs \in \mathcal{S}$, $\bfv + h\bfu \notin \mathcal{S}$.
\begin{align*}
\lim_{h\to0}&\frac{1}{h^2}(\widetilde{C}(\bfv+h\bfu,\bfv+h\bfu)-\widetilde{C}(\bfv+h\bfu,\bfv)-\widetilde{C}(\bfv,\bfv+h\bfu)+\widetilde{C}(\bfv,\bfv))  \\
& = \lim_{h\to0} \frac{1}{h}\left(\frac{(\mathbf{b}_{\bfs+h\bfu}\T \widetilde{\mathbf{C}}_{\mathcal{N}(\bfs+h\bfu)} \mathbf{b}_{\bfs+h\bfu} - \mathbf{b}_{\bfs+h\bfu}\T \widetilde{\mathbf{C}}_{\mathcal{N}(\bfs+h\bfu),\bfs}}{h}\right) \\&\quad-\lim_{h\to0}\frac{1}{h}\left( \frac{\widetilde{\mathbf{C}}_{\bfs,\mathcal{N}(\bfs+h\bfu)}\mathbf{b}_{\bfs+h\bfu} - \widetilde{\mathbf{C}}_{\bfs,\bfs})}{h}\right) + \lim_{h\to0}\frac{1}{h^2}\mathbf{F}_{\bfs+h\bfu} \\
& = \circleone - \circletwo + \circlethree.
\\
\circleone &= \lim_{h\to0} \frac{1}{h}\left(\frac{\mathbf{b}_{\bfs+h\bfu}\T \widetilde{\mathbf{C}}_{\mathcal{N}(\bfs+h\bfu)} \mathbf{b}_{\bfs+h\bfu} - \mathbf{b}_{\bfs+h\bfu}\T \widetilde{\mathbf{C}}_{\mathcal{N}(\bfs+h\bfu),\bfs}}{h}\right)\\
&=\lim_{h\to0} \frac{\mathbf{b}_{\bfs+h\bfu}\T}{h}\left(\frac{ \widetilde{\mathbf{C}}_{\mathcal{N}(\bfs+h\bfu)} \mathbf{C}_{\mathcal{N}(\bfs+h\bfu)}^{-1}\mathbf{C}_{\mathcal{N}(\bfs+h\bfu),\bfs+h\bfu} - \widetilde{\mathbf{C}}_{\mathcal{N}(\bfs+h\bfu),\bfs}}{h}\right)
\end{align*}
The main difficulty in calculating the above limit is that the neighbor set $\mathcal{N}(\bfs+h\bfu)$ is a discrete object that depends on $h$. However, when taking the limit, it suffices to consider $h$ that is sufficiently small, so that the neighbor set remains unchanged. 

For a rigorous argument, we let $\iota\coloneqq \min_{\mathbf{s},\mathbf{s}'\in \mathcal{S}, \mathbf{s}\neq \mathbf{s}'} \|\mathbf{s}-\mathbf{s}'\|$ be the minimal separation of the reference set $\mathcal{S}$, then when $0<|h|<\frac{1}{4}\iota$, for any $\bfs,\bfs'\in\mathcal{S}$,
\begin{align*}
    \norm{\bfs'-\bfs+h\bfu}\geq \norm{\bfs'-\bfs}-\norm{\bfs-\bfs+h\bfu}\geq \iota-|h|>|h|=\norm{\bfs-\bfs+h\bfu}.
\end{align*}
Thus, $\bfs$ is the closest point in $\mathcal{S}$ for $\bfs+h\bfu$. That is, when $0<|h|<\frac{1}{4}\iota$, under our construction of neighbor set for $\bfs+h\bfu \notin \mathcal{S}$, $\mathcal{N}(\bfs+h\bfu)=\mathcal{N}_0(\bfs)$, which does not depend on $h$ anymore. In all the following proof, when taking the limit $\lim_{h\to0}$, we always assume $0<|h|<\frac{1}{4}\iota$, which does not affect the limit but helps to simplify the calculation. As a result,
\begin{align*}
\circleone&=\lim_{h\to0} \frac{\mathbf{b}_{\bfs+h\bfu}\T}{h}\left(\frac{ \widetilde{\mathbf{C}}_{\mathcal{N}_0(\bfs)} \mathbf{C}_{\mathcal{N}_0(\bfs)}^{-1}\mathbf{C}_{\mathcal{N}_0(\bfs),\bfs+h\bfu} - \widetilde{\mathbf{C}}_{\mathcal{N}_0(\bfs),\bfs}}{h}\right)\\
&=\lim_{h\to0} \frac{\mathbf{b}_{\bfs+h\bfu}\T}{h}\left(\frac{ \widetilde{\mathbf{C}}_{\mathcal{N}_0(\bfs)} \mathbf{C}_{\mathcal{N}_0(\bfs)}^{-1}\mathbf{C}_{\mathcal{N}_0(\bfs),\bfs+h\bfu} - \widetilde{\mathbf{C}}_{\mathcal{N}_0(\bfs)}\mathbf{e}}{h}\right)\\
&=\lim_{h\to0} \frac{\mathbf{b}_{\bfs+h\bfu}\T\widetilde{\mathbf{C}}_{\mathcal{N}_0(\bfs)} \mathbf{C}_{\mathcal{N}_0(\bfs)}^{-1}}{h}\left(\frac{\mathbf{C}_{\mathcal{N}_0(\bfs),\bfs+h\bfu} - \mathbf{C}_{\mathcal{N}_0(\bfs)}\mathbf{e}}{h}\right)\\
&=\lim_{h\to0} \frac{\mathbf{b}_{\bfs+h\bfu}\T\widetilde{\mathbf{C}}_{\mathcal{N}_0(\bfs)} \mathbf{C}_{\mathcal{N}_0(\bfs)}^{-1}}{h}\left(\frac{\mathbf{C}_{\mathcal{N}_0(\bfs),\bfs+h\bfu} - \mathbf{C}_{\mathcal{N}_0(\bfs),\bfs}}{h}\right).\\
\\
\circletwo &= \lim_{h\to0}\frac{1}{h}\left( \frac{\widetilde{\mathbf{C}}_{\bfs,\mathcal{N}(\bfs+h\bfu)}\mathbf{b}_{\bfs+h\bfu} - \widetilde{\mathbf{C}}_{\bfs,\bfs})}{h}\right) \\
&=\lim_{h\to0}\frac{1}{h}\left( \frac{\widetilde{\mathbf{C}}_{\bfs,\mathcal{N}(\bfs+h\bfu)}\mathbf{C}_{\mathcal{N}(\bfs+h\bfu)}^{-1}\mathbf{C}_{\mathcal{N}(\bfs+h\bfu),\bfs+h\bfu} - \widetilde{\mathbf{C}}_{\bfs,\bfs})}{h}\right)\\
&=\lim_{h\to0}\frac{1}{h}\left( \frac{\widetilde{\mathbf{C}}_{\bfs,\mathcal{N}_0(\bfs)}\mathbf{C}_{\mathcal{N}_0(\bfs)}^{-1}\mathbf{C}_{\mathcal{N}_0(\bfs),\bfs+h\bfu} - \widetilde{\mathbf{C}}_{\bfs,\mathcal{N}_0(\bfs)}\mathbf{e}}{h}\right)\\
&=\lim_{h\to0}\frac{\widetilde{\mathbf{C}}_{\bfs,\mathcal{N}_0(\bfs)}\mathbf{C}_{\mathcal{N}_0(\bfs)}^{-1}}{h}\left( \frac{\mathbf{C}_{\mathcal{N}_0(\bfs),\bfs+h\bfu} - \mathbf{C}_{\mathcal{N}_0(\bfs)}\mathbf{e}}{h}\right)\\
&=\lim_{h\to0}\frac{\widetilde{\mathbf{C}}_{\bfs,\mathcal{N}_0(\bfs)}\mathbf{C}_{\mathcal{N}_0(\bfs)}^{-1}}{h}\left( \frac{\mathbf{C}_{\mathcal{N}_0(\bfs),\bfs+h\bfu} - \mathbf{C}_{\mathcal{N}_0(\bfs),\bfs}}{h}\right).\\
\\
\circleone-\circletwo & = \lim_{h\to0} \frac{\mathbf{b}_{\bfs+h\bfu}\T\widetilde{\mathbf{C}}_{\mathcal{N}_0(\bfs)} \mathbf{C}_{\mathcal{N}_0(\bfs)}^{-1}}{h}\left(\frac{\mathbf{C}_{\mathcal{N}_0(\bfs),\bfs+h\bfu} - \mathbf{C}_{\mathcal{N}_0(\bfs),\bfs}}{h}\right) \\&\quad- \lim_{h\to0}\frac{\widetilde{\mathbf{C}}_{\bfs,\mathcal{N}_0(\bfs)}\mathbf{C}_{\mathcal{N}_0(\bfs)}^{-1}}{h}\left( \frac{\mathbf{C}_{\mathcal{N}_0(\bfs),\bfs+h\bfu} - \mathbf{C}_{\mathcal{N}_0(\bfs),\bfs}}{h}\right)\\
& = \lim_{h\to0} \frac{\mathbf{b}_{\bfs+h\bfu}\T\widetilde{\mathbf{C}}_{\mathcal{N}_0(\bfs)} -\widetilde{\mathbf{C}}_{\bfs,\mathcal{N}_0(\bfs)}}{h}\mathbf{C}_{\mathcal{N}_0(\bfs)}^{-1}\left( \frac{\mathbf{C}_{\mathcal{N}_0(\bfs),\bfs+h\bfu} - \mathbf{C}_{\mathcal{N}_0(\bfs),\bfs}}{h}\right)
\\
& = \lim_{h\to0} \left( \frac{\mathbf{C}_{\bfs+h\bfu,\mathcal{N}_0(\bfs)} - \mathbf{C}_{\bfs,\mathcal{N}_0(\bfs)}}{h}\right)\mathbf{C}_{\mathcal{N}_0(\bfs)}^{-1}\widetilde{\mathbf{C}}_{\mathcal{N}_0(\bfs)}\mathbf{C}_{\mathcal{N}_0(\bfs)}^{-1}\left( \frac{\mathbf{C}_{\mathcal{N}_0(\bfs),\bfs+h\bfu} - \mathbf{C}_{\mathcal{N}_0(\bfs),\bfs}}{h}\right).
\end{align*}
Thus, if $\widetilde{\omega}$ MSD of first order, $\lim_{h\to0}  \frac{\mathbf{C}_{\bfs+h\bfu,\mathcal{N}_0(\bfs)} - \mathbf{C}_{\bfs,\mathcal{N}_0(\bfs)}}{h}$ and $ \lim_{h\to0}\frac{\mathbf{C}_{\mathcal{N}_0(\bfs),\bfs+h\bfu} - \mathbf{C}_{\mathcal{N}_0(\bfs),\bfs}}{h}$ exists.
Then, $\circleone-\circletwo$ also exists and $$\circleone-\circletwo = D_{\bfu} C(\bfs,\mathcal{N}_0(\bfs))\mathbf{C}_{\mathcal{N}_0(\bfs)}^{-1}\widetilde{\mathbf{C}}_{\mathcal{N}_0(\bfs)}\mathbf{C}_{\mathcal{N}_0(\bfs)}^{-1} D_{\bfu} C(\mathcal{N}_0(\bfs),\bfs).$$
\begin{align*}
\circlethree &=   \lim_{h\to0}\frac{1}{h^2}(\mathbf{C}_{\bfs+h\bfu,\bfs+h\bfu}-\mathbf{C}_{\bfs+h\bfu,\mathcal{N}(\bfs+h\bfu)}\mathbf{C}_{\mathcal{N}(\bfs+h\bfu)}^{-1}\mathbf{C}_{\mathcal{N}(\bfs+h\bfu),\bfs+h\bfu})\\
&= \lim_{h\to0}\frac{1}{h^2}(\mathbf{C}_{\bfs+h\bfu,\bfs+h\bfu} - \mathbf{C}_{\bfs+h\bfu,\bfs} + \mathbf{C}_{\bfs+h\bfu,\bfs} - \mathbf{C}_{\bfs+h\bfu,\mathcal{N}(\bfs+h\bfu)}\mathbf{C}_{\mathcal{N}(\bfs+h\bfu)}^{-1}\mathbf{C}_{\mathcal{N}(\bfs+h\bfu),\bfs+h\bfu})\\
&= \lim_{h\to0} \frac{1}{h} D_{\bfu} \mathbf{C} (\bfs+h\bfu,\bfs) + \lim_{h\to0}\frac{1}{h^2} ( \mathbf{C}_{\bfs+h\bfu,\mathcal{N}_0(\bfs)}\mathbf{e}_1 - \mathbf{C}_{\bfs+h\bfu,\mathcal{N}_0(\bfs)}\mathbf{C}_{\mathcal{N}_0(\bfs)}^{-1}\mathbf{C}_{\mathcal{N}_0(\bfs),\mathcal{N}_\Delta(\bfs)}\mathbf{e}_1) \\
&=\lim_{h\to0} \frac{1}{h} D_{\bfu} \mathbf{C} (\bfs+h\bfu,\bfs) - \frac{1}{h} \mathbf{C}_{\bfs+h\bfu,\mathcal{N}_0(\bfs)}\mathbf{C}_{\mathcal{N}_0(\bfs)}^{-1}D_{\bfu} C(\mathcal{N}_0(\bfs),\bfs) \\
&=\lim_{h\to0} \frac{1}{h}(D_{\bfu} \mathbf{C} (\bfs+h\bfu,\bfs) - D_{\bfu} \mathbf{C} (\bfs,\bfs)) \\&\quad+ \frac{1}{h} (D_{\bfu} \mathbf{C} (\bfs,\bfs) -\mathbf{C}_{\bfs+h\bfu,\mathcal{N}_0(\bfs)}\mathbf{C}_{\mathcal{N}_0(\bfs)}^{-1}D_{\bfu} C(\mathcal{N}_0(\bfs),\bfs)).
\end{align*}
Thus, if $\widetilde{\omega}$ MSD of first order, $\lim_{h\to0} \frac{1}{h}(D_{\bfu} \mathbf{C} (\bfs+h\bfu,\bfs) - D_{\bfu} \mathbf{C} (\bfs,\bfs))$ exists. Then, $\circlethree$ also exists, and 
\begin{align*} 
\circlethree&=D^{(2)}_{\bfu}C(\bfs,\bfs) - D_{\bfu} C(\bfs,\mathcal{N}_0(\bfs))\mathbf{C}_{\mathcal{N}_0(\bfs)}^{-1}D_{\bfu} C(\mathcal{N}_0(\bfs),\bfs),\\
\circleone - \circletwo + \circlethree&=D^{(2)}_{\bfu}C(\bfs,\bfs) + D_{\bfu} C(\bfs,\mathcal{N}_0(\bfs))\mathbf{C}_{\mathcal{N}_0(\bfs)}^{-1}\widetilde{\mathbf{C}}_{\mathcal{N}_0(\bfs)}\mathbf{C}_{\mathcal{N}_0(\bfs)}^{-1} D_{\bfu} C(\mathcal{N}_0(\bfs),\bfs)  \\
&\quad - D_{\bfu} C(\bfs,\mathcal{N}_0(\bfs))\mathbf{C}_{\mathcal{N}_0(\bfs)}^{-1}D_{\bfu} C(\mathcal{N}_0(\bfs),\bfs).
\end{align*}

\item $\bfv, \bfv+h\bfu \notin \mathcal{S}$.
%Let $\partial^2 \widetilde{\mathbf{C}}_{\bfv_i} = D_{\bfu} C(\bfv,\mathcal{N}(\bfv))\mathbf{C}_{\mathcal{N}(\bfv)}^{-1}\widetilde{\mathbf{C}}_{\mathcal{N}(\bfv),\mathcal{N}(\bfv)}\mathbf{C}_{\mathcal{N}(\bfv)}^{-1} D_{\bfu} C(\mathcal{N}(\bfv),\bfv)$,
\begin{align*}
\lim_{h\to0}&\frac{1}{h^2}(\widetilde{C}(\bfv+h\bfu,\bfv+h\bfu)-\widetilde{C}(\bfv+h\bfu,\bfv)-\widetilde{C}(\bfv,\bfv+h\bfu)+\widetilde{C}(\bfv,\bfv))  \\
& = \lim_{h\to0}\frac{1}{h}\left(\frac{(\mathbf{b}_{\bfv+h\bfu}\T \widetilde{\mathbf{C}}_{\mathcal{N}(\bfv+h\bfu),\mathcal{N}(\bfv+h\bfu)} \mathbf{b}_{\bfv+h\bfu} - \mathbf{b}_{\bfv+h\bfu}\T \widetilde{\mathbf{C}}_{\mathcal{N}(\bfv+h\bfu),\bfv}}{h}\right) \\&\quad-\lim_{h\to0}\frac{1}{h}\left( \frac{\widetilde{\mathbf{C}}_{\bfv,\mathcal{N}(\bfv+h\bfu)}\mathbf{b}_{\bfv+h\bfu} - \widetilde{\mathbf{C}}_{\bfv,\bfv})}{h}\right) + \lim_{h\to0}\frac{1}{h^2}(\mathbf{F}_{\bfv+h\bfu}-\mathbf{F}_{\bfv}) \\
& = \circleone - \circletwo + \circlethree.
\end{align*}
Similar as the previous case, if $\widetilde{\omega}$ MSD of first order, $\circleone-\circletwo$ also exists and $$\circleone-\circletwo = D_{\bfu} C(\bfv,\mathcal{N}(\bfv))\mathbf{C}_{\mathcal{N}(\bfv)}^{-1}\widetilde{\mathbf{C}}_{\mathcal{N}(\bfv)}\mathbf{C}_{\mathcal{N}(\bfv)}^{-1} D_{\bfu} C(\mathcal{N}(\bfv),\bfv).$$
\begin{align*}
\circlethree&= \lim_{h\to0}\frac{1}{h^2}(\mathbf{C}_{\bfv+h\bfu,\bfv+h\bfu}-\mathbf{C}_{\bfv+h\bfu,\mathcal{N}(\bfv+h\bfu)}\mathbf{C}_{\mathcal{N}(\bfv+h\bfu)}^{-1}\mathbf{C}_{\mathcal{N}(\bfv+h\bfu),\bfv+h\bfu}\\&\quad - \mathbf{C}_{\bfv,\bfv}+\mathbf{C}_{\bfv,\mathcal{N}(\bfv)}\mathbf{C}_{\mathcal{N}(\bfv)}^{-1}\mathbf{C}_{\mathcal{N}(\bfv),\bfv})\\
&= \lim_{h\to0}\frac{1}{h^2}(\mathbf{C}_{\bfv+h\bfu,\bfv+h\bfu} - \mathbf{C}_{\bfv+h\bfu,\bfv} + \mathbf{C}_{\bfv+h\bfu,\bfv} -\mathbf{C}_{\bfv,\bfv}) \\&\quad - D_{\bfu} C(\bfv,\mathcal{N}(\bfv))\mathbf{C}_{\mathcal{N}(\bfv)}^{-1}D_{\bfu} C(\mathcal{N}(\bfv),\bfv).
\end{align*}
Thus, if $\widetilde{\omega}$ MSD of first order, $\lim_{h\to0}\frac{1}{h^2}(\mathbf{C}_{\bfv+h\bfu,\bfv+h\bfu} - \mathbf{C}_{\bfv+h\bfu,\bfv} + \mathbf{C}_{\bfv+h\bfu,\bfv} -\mathbf{C}_{\bfv,\bfv})$ exists. Then, $\circlethree$ also exists:
\begin{align*}
\circlethree & = D^{(2)}_{\bfu}C(\bfv,\bfv) - D_{\bfu} C(\bfv,\mathcal{N}(\bfv))\mathbf{C}_{\mathcal{N}(\bfv)}^{-1}D_{\bfu} C(\mathcal{N}(\bfv),\bfv),\\
\circleone - \circletwo + \circlethree &=D^{(2)}_{\bfu}C(\bfv,\bfv) + D_{\bfu} C(\bfv,\mathcal{N}(\bfv))\mathbf{C}_{\mathcal{N}(\bfv)}^{-1}\widetilde{\mathbf{C}}_{\mathcal{N}(\bfv)}\mathbf{C}_{\mathcal{N}(\bfv)}^{-1} D_{\bfu} C(\mathcal{N}(\bfv),\bfv)  \\
&\quad - D_{\bfu} C(\bfv,\mathcal{N}(\bfv))\mathbf{C}_{\mathcal{N}(\bfv)}^{-1}D_{\bfu} C(\mathcal{N}(\bfv),\bfv).
\end{align*}
\end{enumerate}
\end{proof}

\section{Proof of \Cref{thm:C_duC_cov}: The cross-covariance between NNGP and NNDP}
\begin{proof}[Proof of \Cref{thm:C_duC_cov}]. 
We first calculate the covariance between the NNGP $\widetilde{\omega}$ and the NNDP $D_\bfu \widetilde{\omega}$. Let $\mathbf{e}_i = (0,0,\dots, 1,\dots,0)\T$ be the i-th canonical basis, and
\[
\mathcal{N}_{\bfu,h}(\bfs_i)=\begin{cases}
    \{\bfs_i+h\bfu,\mathcal{N}(\bfs_i)\} & i < m \\
    \{\bfs_i+h\bfu,\mathcal{N}(\bfs_i)_{1:(m-1)}\} & i \geq m ,\\
\end{cases}.
\]We consider the following four cases:

\begin{enumerate}
\item $\mathbf{v}_1=\bfs_i,\mathbf{v}_2=\bfs_j \in \mathcal{S}$, $\mathbf{v}_1+h\bfu \notin \mathcal{S}$. $\mathcal{N}(\bfs_i+h\bfu)_1=\bfs_i$ if the neighbors are ordered by distance, so
\begin{align*}
&\lim_{h \rightarrow 0}\frac{\widetilde{C}(\mathbf{v}_1+h\bfu,\mathbf{v}_2)-\widetilde{C}(\mathbf{v}_1,\mathbf{v}_2)}{h} \\
& = \lim_{h \rightarrow 0}\frac{\mathbf{b}_{\bfs_i+h\bfu}\T\widetilde{\mathbf{C}}_{\mathcal{N}(\bfs_i+h\bfu),\bfs_j}-\widetilde{\mathbf{C}}_{\bfs_i,\bfs_j}}{h}\\
& = \lim_{h \rightarrow 0} \frac{\mathbf{C}_{\bfs_i+h\bfu, \mathcal{N}_0(\bfs_i)}\mathbf{C}_{\mathcal{N}_0(\bfs_i)}^{-1}\widetilde{\mathbf{C}}_{\mathcal{N}_0(\bfs_i),\bfs_j} - \widetilde{\mathbf{C}}_{\bfs_i,\bfs_j}}{h} \\
& = \lim_{h \rightarrow 0} \frac{\mathbf{e}_1\T\mathbf{C}_{\mathcal{N}_{\bfu,h}(\bfs_i), \mathcal{N}_0(\bfs_i)}\mathbf{C}_{\mathcal{N}_0(\bfs_i)}^{-1}\widetilde{\mathbf{C}}_{\mathcal{N}_0(\bfs_i),\bfs_j} - \mathbf{e}_1\T\widetilde{\mathbf{C}}_{\mathcal{N}_0(\bfs_i),\bfs_j}}{h} \\
& = \lim_{h \rightarrow 0} \frac{\mathbf{e}_1\T(\mathbf{C}_{\mathcal{N}_{\bfu,h}(\bfs_i), \mathcal{N}_0(\bfs_i)}-\mathbf{C}_{\mathcal{N}_0(\bfs_i)})\mathbf{C}_{\mathcal{N}_0(\bfs_i)}^{-1}\widetilde{\mathbf{C}}_{\mathcal{N}_0(\bfs_i),\bfs_j}}{h} \\
& = [D_{\bfu} C(\bfs_i,\bfs_i),D_{\bfu} C(\bfs_i,\mathcal{N}(\bfs_i)_1),\dots,D_{\bfu} C(\bfs_i,\mathcal{N}(\bfs_i)_{\min(m-1,m_i)})]\mathbf{C}_{\mathcal{N}_0(\bfs_i)}^{-1}\widetilde{\mathbf{C}}_{\mathcal{N}_0(\bfs_i),\bfs_j} \\
& = D_{\bfu} C(\bfs_i,\mathcal{N}_0(\bfs_i))\mathbf{C}_{\mathcal{N}_0(\bfs_i)}^{-1}\widetilde{\mathbf{C}}_{\mathcal{N}_0(\bfs_i),\bfs_j}.
\end{align*}

\item $\mathbf{v}_1\notin \mathcal{S},\mathbf{v}_2=\bfs_j.$
\begin{align*}
&\lim_{h \rightarrow 0}\frac{\widetilde{C}(\mathbf{v}_1+h\bfu,\mathbf{v}_2)-\widetilde{C}(\mathbf{v}_1,\mathbf{v}_2)}{h} \\
& = \lim_{h \rightarrow 0}\frac{\mathbf{b}_{\mathbf{v}_1+h\bfu}\T\widetilde{\mathbf{C}}_{\mathcal{N}(\bfv_1+h\bfu),\bfs_j}-\mathbf{b}_{\mathbf{v}_1}\T\widetilde{\mathbf{C}}_{\mathcal{N}(\bfv_1),\bfs_j}}{h}\\
& = \lim_{h \rightarrow 0} \frac{\mathbf{b}_{\mathbf{v}_1+h\bfu}\T-\mathbf{b}_{\mathbf{v}_1}\T}{h} \widetilde{\mathbf{C}}_{\mathcal{N}(\bfv_1),\bfs_j}\\
& = \lim_{h \rightarrow 0} \frac{\mathbf{C}_{\mathbf{v}_1+h\bfu, \mathcal{N}(\bfv_1)}-\mathbf{C}_{\mathbf{v}_1, \mathcal{N}(\bfv_1)}}{h}\mathbf{C}_{\mathcal{N}(\bfv_1)}^{-1} \widetilde{\mathbf{C}}_{\mathcal{N}(\bfv_1),\bfs_j} \\
& = D_{\bfu} C(\mathbf{v}_1,\mathcal{N}(\bfv_1))\mathbf{C}_{\mathcal{N}(\bfv_1)}^{-1} \widetilde{\mathbf{C}}_{\mathcal{N}(\bfv_1),\bfs_j}.
\end{align*}
\item $\mathbf{v}_1=\bfs_i,\mathbf{v}_2\notin \mathcal{S}.$
\begin{align*}
&\lim_{h \rightarrow 0}\frac{\widetilde{C}(\mathbf{v}_1+h\bfu,\mathbf{v}_2)-\widetilde{C}(\mathbf{v}_1,\mathbf{v}_2)}{h} \\
& = \lim_{h \rightarrow 0}\frac{\mathbf{b}_{\bfs_i+h\bfu}\T\widetilde{\mathbf{C}}_{\mathcal{N}(\bfs_i+h\bfu),\mathcal{N}(\bfv_2)}\mathbf{b}_{\mathbf{v}_2}-\widetilde{\mathbf{C}}_{\bfs_i,\mathcal{N}(\bfv_2)}\mathbf{b}_{\mathbf{v}_2}}{h}\\
& = \lim_{h \rightarrow 0} \frac{\mathbf{e}_1\T\mathbf{C}_{\mathcal{N}_{\bfu,h}(\bfs_i), \mathcal{N}_0(\bfs_i)}\mathbf{C}_{\mathcal{N}_0(\bfs_i)}^{-1}\widetilde{\mathbf{C}}_{\mathcal{N}_0(\bfs_i),\mathcal{N}(\bfv_2)} - \mathbf{e}_1\T\widetilde{\mathbf{C}}_{\mathcal{N}_0(\bfs_i),\mathcal{N}(\bfv_2)}}{h} \mathbf{b}_{\mathbf{v}_2}\\
& = \lim_{h \rightarrow 0} \frac{\mathbf{e}_1\T(\mathbf{C}_{\mathcal{N}_{\bfu,h}(\bfs_i), \mathcal{N}_0(\bfs_i)}-\mathbf{C}_{\mathcal{N}_0(\bfs_i)})\mathbf{C}_{\mathcal{N}_0(\bfs_i)}^{-1}\widetilde{\mathbf{C}}_{\mathcal{N}_0(\bfs_i),\mathcal{N}(\bfv_2)}}{h} \mathbf{b}_{\mathbf{v}_2}\\
& = [D_{\bfu} C(\bfs_i,\bfs_i),D_{\bfu} C(\bfs_i,\mathcal{N}(\bfs_i)_1),\dots,D_{\bfu} C(\bfs_i,\mathcal{N}(\bfs_i)_{\min(m-1,m_i)})]\mathbf{C}_{\mathcal{N}_0(\bfs_i)}^{-1}\widetilde{\mathbf{C}}_{\mathcal{N}_0(\bfs_i),\mathcal{N}(\bfv_2)} \mathbf{b}_{\mathbf{v}_2}\\
& = D_{\bfu} C(\bfs_i,\mathcal{N}_0(\bfs_i))\mathbf{C}_{\mathcal{N}_0(\bfs_i)}^{-1}\widetilde{\mathbf{C}}_{\mathcal{N}_0(\bfs_i),\mathcal{N}(\bfv_2)} \mathbf{b}_{\mathbf{v}_2}.
\end{align*}
\item $\mathbf{v}_1,\mathbf{v}_2\notin \mathcal{S}$, $\mathbf{v}_1 \neq \mathbf{v}_2$, $\mathbf{v}_1+h\bfu \neq \mathbf{v}_2$. 
\begin{align*}
&\lim_{h \rightarrow 0}\frac{\widetilde{C}(\mathbf{v}_1+h\bfu,\mathbf{v}_2)-\widetilde{C}(\mathbf{v}_1,\mathbf{v}_2)}{h} \\
& = \lim_{h \rightarrow 0}\frac{\mathbf{b}_{\mathbf{v}_1+h\bfu}\T\widetilde{\mathbf{C}}_{\mathcal{N}(\bfv_1+h\bfu),\mathcal{N}(\bfv_2)}\mathbf{b}_{\mathbf{v}_2}+h\bfu_{(\mathbf{v}_1+h\bfu=\mathbf{v}_2)}\mathbf{F}_{\mathbf{v}_1+h\bfu}-\mathbf{b}_{\mathbf{v}_1}\T\widetilde{\mathbf{C}}_{\mathcal{N}(\bfv_1),\mathcal{N}(\bfv_2)}\mathbf{b}_{\mathbf{v}_2}-\delta_{(\mathbf{v}_1=\mathbf{v}_2)}\mathbf{F}_{\mathbf{v}_1}}{h}\\
& = \lim_{h \rightarrow 0} \frac{\mathbf{b}_{\mathbf{v}_1+h\bfu}\T\widetilde{\mathbf{C}}_{\mathcal{N}(\bfv_1),\mathcal{N}(\bfv_2)}\mathbf{b}_{\mathbf{v}_2}+h\bfu_{(\mathbf{v}_1+h\bfu=\mathbf{v}_2)}\mathbf{F}_{\mathbf{v}_1+h\bfu}-\mathbf{b}_{\mathbf{v}_1}\T\widetilde{\mathbf{C}}_{\mathcal{N}(\bfv_1),\mathcal{N}(\bfv_2)}\mathbf{b}_{\mathbf{v}_2}-\delta_{(\mathbf{v}_1=\mathbf{v}_2)}\mathbf{F}_{\mathbf{v}_1}}{h} \\
& = \lim_{h \rightarrow 0} \frac{\mathbf{b}_{\mathbf{v}_1+h\bfu}\T-\mathbf{b}_{\mathbf{v}_1}\T}{h}\widetilde{\mathbf{C}}_{\mathcal{N}(\bfv_1),\mathcal{N}(\bfv_2)}\mathbf{b}_{\mathbf{v}_2} \\
& = \lim_{h \rightarrow 0} \frac{\mathbf{C}_{\mathbf{v}_1+h\bfu, \mathcal{N}(\bfv_1)}-\mathbf{C}_{\mathbf{v}_1, \mathcal{N}(\bfv_1)}}{h}\mathbf{C}_{\mathcal{N}(\bfv_1)}^{-1} \widetilde{\mathbf{C}}_{\mathcal{N}(\bfv_1),\mathcal{N}(\bfv_2)}\mathbf{b}_{\mathbf{v}_2} \\
& = D_{\bfu} C(\mathbf{v}_1,\mathcal{N}(\bfv_1))\mathbf{C}_{\mathcal{N}(\bfv_1)}^{-1} \widetilde{\mathbf{C}}_{\mathcal{N}(\bfv_1),\mathcal{N}(\bfv_2)}\mathbf{b}_{\mathbf{v}_2}.
\end{align*}

\end{enumerate}

In summary, $\cov(D_\bfu \widetilde{\omega}(\bfv_1),\widetilde{\omega}(\bfv_2))$ is given by
\[ \begin{cases}
D_{\bfu} C(\bfs_i,\mathcal{N}_0(\bfs_i))\mathbf{C}_{\mathcal{N}_0(\bfs_i)}^{-1}\widetilde{\mathbf{C}}_{\mathcal{N}_0(\bfs_i),\bfs_j}
&\mathbf{v}_1=\bfs_i,~\mathbf{v}_2=\bfs_j \in \mathcal{S}\\
D_{\bfu} C(\bfs_i,\mathcal{N}_0(\bfs_i))\mathbf{C}_{\mathcal{N}_0(\bfs_i)}^{-1}\widetilde{\mathbf{C}}_{\mathcal{N}_0(\bfs_i),\mathcal{N}(\bfv_2)} \mathbf{b}_{\mathbf{v}_2} 
&\mathbf{v}_1=\bfs_i,~\mathbf{v}_2 \notin \mathcal{S}\\
D_{\bfu} C(\mathbf{v}_1,\mathcal{N}(\bfv_1))\mathbf{C}_{\mathcal{N}(\bfv_1)}^{-1} \widetilde{\mathbf{C}}_{\mathcal{N}(\bfv_1),\bfs_j}
& \mathbf{v}_1\notin \mathcal{S},~\mathbf{v}_2=\bfs_j\\
D_{\bfu} C(\mathbf{v}_1,\mathcal{N}(\bfv_1))\mathbf{C}_{\mathcal{N}(\bfv_1)}^{-1} \widetilde{\mathbf{C}}_{\mathcal{N}(\bfv_1),\mathcal{N}(\bfv_2)}\mathbf{b}_{\mathbf{v}_2}
&\mathbf{v}_1,\mathbf{v}_2\notin \mathcal{S}, ~\mathbf{v}_1 \neq \mathbf{v}_2, ~\mathbf{v}_1+h\bfu \neq \mathbf{v}_2 \\
%\infty
%&\mathbf{v}_1,\mathbf{v}_2\notin \mathcal{S},~\mathbf{v}_1 = \mathbf{v}_2, ~\mathbf{v}_1+h\bfu \neq \mathbf{v}_2
\end{cases}.
\]

Note that there is missing case: $\mathbf{v}_1,\mathbf{v}_2\notin \mathcal{S}$, $\mathbf{v}_1 = \mathbf{v}_2$, $\mathbf{v}_1+h\bfu \neq \mathbf{v}_2$.

\begin{align*}
&\lim_{h \rightarrow 0}\frac{\widetilde{C}(\mathbf{v}_1+h\bfu,\mathbf{v}_2)-\widetilde{C}(\mathbf{v}_1,\mathbf{v}_2)}{h} \\
& = \lim_{h \rightarrow 0}\frac{\mathbf{b}_{\mathbf{v}_1+h\bfu}\T\widetilde{\mathbf{C}}_{\mathcal{N}(\bfv_1+h\bfu),\mathcal{N}(\bfv_2)}\mathbf{b}_{\mathbf{v}_2}+h\bfu_{(\mathbf{v}_1+h\bfu=\mathbf{v}_2)}\mathbf{F}_{\mathbf{v}_1+h\bfu}-\mathbf{b}_{\mathbf{v}_1}\T\widetilde{\mathbf{C}}_{\mathcal{N}(\bfv_1),\mathcal{N}(\bfv_2)}\mathbf{b}_{\mathbf{v}_2}-\delta_{(\mathbf{v}_1=\mathbf{v}_2)}\mathbf{F}_{\mathbf{v}_1}}{h}\\
& = \lim_{h \rightarrow 0} \frac{\mathbf{b}_{\mathbf{v}_1+h\bfu}\T\widetilde{\mathbf{C}}_{\mathcal{N}(\bfv_1),\mathcal{N}(\bfv_2)}\mathbf{b}_{\mathbf{v}_2}-\mathbf{b}_{\mathbf{v}_1}\T\widetilde{\mathbf{C}}_{\mathcal{N}(\bfv_1),\mathcal{N}(\bfv_2)}\mathbf{b}_{\mathbf{v}_2}-\mathbf{F}_{\mathbf{v}_1}}{h} .
\end{align*}

In this case, the limit does not exist. Thus, the inference on new observations for $D_\bfu\widetilde{\omega}(\bfv_1)$ and $\widetilde{\omega}(\bfv_2)$, $\bfv_1,\bfv_2 \notin \mathcal{S}$ at the same time is now allowed.

Next, we show the continuity of $\cov(D_\bfu \widetilde{\omega}(\bfv_1),\widetilde{\omega}(\bfv_2))$, that is, 
$$\cov(D_\bfu \widetilde{\omega}(\bfv_1+\bfu_1),\widetilde{\omega}(\bfv_2+\bfu_2))\xrightarrow{\bfu_1,\bfu_2\to0} \cov(D_\bfu \widetilde{\omega}(\bfv_1),\widetilde{\omega}(\bfv_2)).$$ Excluding a measure zero set, we prove by the same four cases as above. 

%WTS $\cov(D_\bfu \widetilde{\omega}(\bfv_1+\bfu_1),\widetilde{\omega}(\bfv_2+\bfu_2))\to \cov(D_\bfu \widetilde{\omega}(\bfv_1),\widetilde{\omega}(\bfv_2))$ when $\bfu_1\to0,\bfu_2\to 0$.

\begin{enumerate}[label=\roman*.]
\item $\mathbf{v}_1=\bfs_i,~\mathbf{v}_2=\bfs_j \in \mathcal{S}$, $\mathbf{v}_1\neq \mathbf{v}_2$.
    
    \[
    \cov(D_\bfu \widetilde{\omega}(\bfv_1),\widetilde{\omega}(\bfv_2))=
    D_{\bfu} C(\bfs_i,\mathcal{N}_0(\bfs_i))\mathbf{C}_{\mathcal{N}_0(\bfs_i)}^{-1}\widetilde{\mathbf{C}}_{\mathcal{N}_0(\bfs_i),\bfs_j}.
    \]
    
    Since $\mathbf{v}_1+\bfu_1 \notin \mathcal{S}$, $\mathbf{v}_2 + \bfu_2 \notin \mathcal{S}$, $\mathbf{v}_1+\bfu_1 \neq \mathbf{v}_2+\bfu_2$, then
\begin{align*}
&\cov(D_\bfu \widetilde{\omega}(\bfv_1 +\bfu_1),\widetilde{\omega}(\bfv_2 +\bfu_2)) =  \\
&D_{\bfu} C(\mathbf{v}_1+\bfu_1,\mathcal{N}(\bfv_1+\bfu_1))\mathbf{C}_{\mathcal{N}(\bfv_1+\bfu_1)}^{-1} \widetilde{\mathbf{C}}_{\mathcal{N}(\bfv_1+\bfu_1),\mathcal{N}(\bfv_2+\bfu_2)}\mathbf{b}_{\mathbf{v}_2+\bfu_2}.
\end{align*}

$\mathbf{v}_1=\bfs_i$, thus $\mathcal{N}(\bfv_1+\bfu_1)=\mathcal{N}_0(\bfs_i)$ for small enough $\bfu_1$. Similarly, $\mathcal{N}(\bfv_2+\bfu_2)=\mathcal{N}_0(\bfs_j)$.

By the continuity of $D_{\bfu} \mathbf{C}$,
\[
\lim_{\bfu_1  \rightarrow 0} D_{\bfu} C(\mathbf{v}_1+\bfu_1,\mathcal{N}(\bfv_1+\bfu_1)) = \lim_{\bfu_1  \rightarrow 0} D_{\bfu} C(\mathbf{v}_1+\bfu_1,\mathcal{N}_0(\bfs_i)) = D_{\bfu} C(\mathbf{v}_1,\mathcal{N}_0(\bfs_i)).
\]
\begin{align*}
    \mathbf{b}_{\mathbf{v}_2+\bfu_2}\T &= \mathbf{C}_{\mathbf{v}_2+\bfu_2,\mathcal{N}(\bfv_2+\bfu_2)}\mathbf{C}_{\mathcal{N}(\bfv_2+\bfu_2)}^{-1} = \mathbf{C}_{\mathbf{v}_2+\bfu_2,\mathcal{N}_0(\bfs_j)}\mathbf{C}_{\mathcal{N}_0(\bfs_j)}^{-1} \\
    & \rightarrow \mathbf{C}_{\mathbf{v}_2,\mathcal{N}_0(\bfs_j)}\mathbf{C}_{\mathcal{N}_0(\bfs_j)}^{-1} = \mathbf{e}_1\T \mathbf{C}_{\mathcal{N}_0(\bfs_j)}\mathbf{C}_{\mathcal{N}_0(\bfs_j)}^{-1} = \mathbf{e}_1\T = (1,0,...,0).
\end{align*}

\[
\widetilde{\mathbf{C}}_{\mathcal{N}(\bfv_1+\bfu_1),\mathcal{N}(\bfv_2+\bfu_2)}\mathbf{b}_{\mathbf{v}_2+\bfu_2} \ = \widetilde{\mathbf{C}}_{\mathcal{N}_0(\bfs_i),\mathcal{N}_0(\bfs_j)}\mathbf{b}_{\mathbf{v}_2+\bfu_2} \rightarrow \widetilde{\mathbf{C}}_{\mathcal{N}_0(\bfs_i),\mathcal{N}_0(\bfs_j)} \mathbf{e}_1 = \widetilde{\mathbf{C}}_{\mathcal{N}_0(\bfs_i),\bfs_j}.
\]

Thus,
$\lim_{\bfu_1, \bfu_2 \rightarrow 0}\cov(D_\bfu \widetilde{\omega}(\bfv_1 +\bfu_1),\widetilde{\omega}(\bfv_2 +\bfu_2))=\cov(D_\bfu \widetilde{\omega}(\bfv_1),\widetilde{\omega}(\bfv_2)).$

\item $\mathbf{v}_1=\bfs_i,~\mathbf{v}_2 \notin \mathcal{S}$.

\[
\cov(D_\bfu \widetilde{\omega}(\bfv_1),\widetilde{\omega}(\bfv_2)) = D_{\bfu} C(\bfs_i,\mathcal{N}_0(\bfs_i))\mathbf{C}_{\mathcal{N}_0(\bfs_i)}^{-1}\widetilde{\mathbf{C}}_{\mathcal{N}_0(\bfs_i),\mathcal{N}(\bfv_2)} \mathbf{b}_{\mathbf{v}_2} .
\]

By assumption, $\mathbf{v}_1+\bfu_1 \notin \mathcal{S}$, $\mathbf{v}_2 + \bfu_2 \notin \mathcal{S}$, $\mathbf{v}_1+\bfu_1 \neq \mathbf{v}_2+\bfu_2$, then we can calculate
\begin{align*}
&D_{\bfu} C(\mathbf{v}_1+\bfu_1,\mathbf{v}_2+\bfu_2) =  \\
&D_{\bfu} C(\mathbf{v}_1+\bfu_1,\mathcal{N}(\bfv_1+\bfu_1))\mathbf{C}_{\mathcal{N}(\bfv_1+\bfu_1)}^{-1} \widetilde{\mathbf{C}}_{\mathcal{N}(\bfv_1+\bfu_1),\mathcal{N}(\bfv_2+\bfu_2)}\mathbf{b}_{\mathbf{v}_2+\bfu_2}.
\end{align*}
Since $\mathbf{v}_1=\bfs_i$, $\mathcal{N}(\bfv_1+\bfu_1)=\mathcal{N}_0(\bfs_i)$ for small enough $\bfu_1$.

By $D_{\bfu} C$ continuity,
\[
\lim_{\bfu_1  \rightarrow 0} D_{\bfu} C(\mathbf{v}_1+\bfu_1,\mathcal{N}(\bfv_1+\bfu_1)) = \lim_{\bfu_1  \rightarrow 0} D_{\bfu} C(\mathbf{v}_1+\bfu_1,\mathcal{N}_0(\bfs_i)) = D_{\bfu} C(\mathbf{v}_1,\mathcal{N}_0(\bfs_i)).
\]

We have $\mathcal{N}(\bfv_2)=\mathcal{N}(\bfv_2+\bfu_2)$ for small enough $\bfu_2$, then
\[
\lim_{\bfu_2 \rightarrow 0}\norm{\mathbf{v}_2+\bfu_2,\mathcal{N}(\bfv_2+\bfu_2)} = \norm{\mathbf{v}_2,\mathcal{N}(\bfv_2)},
\]
\[
\mathbf{b}_{\mathbf{v}_2+\bfu_2}\T = \mathbf{C}_{\mathbf{v}_2+\bfu_2,\mathcal{N}(\bfv_2+\bfu_2)}\mathbf{C}_{\mathcal{N}(\bfv_2+\bfu_2)}^{-1} \rightarrow \mathbf{C}_{\mathbf{v}_2,\mathcal{N}(\bfv_2)}\mathbf{C}_{\mathcal{N}(\bfv_2)}^{-1} = \mathbf{b}_{\mathbf{v}_2}\T,
\]

\[
\widetilde{\mathbf{C}}_{\mathcal{N}(\bfv_1+\bfu_1),\mathcal{N}(\bfv_2+\bfu_2)} = \widetilde{\mathbf{C}}_{\mathcal{N}_0(\bfs_i),\mathcal{N}(\bfv_2+\bfu_2)} \rightarrow \widetilde{\mathbf{C}}_{\mathcal{N}_0(\bfs_i),\mathcal{N}(\bfv_2)}.
\]

Thus,

\begin{align*}
    &\lim_{\bfu_1, \bfu_2 \rightarrow 0}\cov(D_\bfu \widetilde{\omega}(\bfv_1 +\bfu_1),\widetilde{\omega}(\bfv_2 +\bfu_2)) \\
    &=D_{\bfu} C(\bfs_i,\mathcal{N}_0(\bfs_i))\mathbf{C}_{\mathcal{N}_0(\bfs_i)}^{-1}\widetilde{\mathbf{C}}_{\mathcal{N}_0(\bfs_i),\mathcal{N}(\bfv_2)} \mathbf{b}_{\mathbf{v}_2} \\
    &=\cov(D_\bfu \widetilde{\omega}(\bfv_1),\widetilde{\omega}(\bfv_2)).
\end{align*}

\item  $\mathbf{v}_1\notin \mathcal{S},~\mathbf{v}_2=\bfs_j$

We first write out
\[
    \cov(D_\bfu \widetilde{\omega}(\bfv_1),\widetilde{\omega}(\bfv_2))=
D_{\bfu} C(\mathbf{v}_1,\mathcal{N}(\bfv_1))\mathbf{C}_{\mathcal{N}(\bfv_1)}^{-1} \widetilde{\mathbf{C}}_{\mathcal{N}(\bfv_1),\bfs_j}.
\]

We have $\mathbf{v}_1+\bfu_1 \notin \mathcal{S}$, $\mathbf{v}_2 + \bfu_2 \notin \mathcal{S}$, $\mathbf{v}_1+\bfu_1 \neq \mathbf{v}_2+\bfu_2$, then
\begin{align*}
&D_{\bfu} C(\mathbf{v}_1+\bfu_1,\mathbf{v}_2+\bfu_2) =\\
&D_{\bfu} C(\mathbf{v}_1+\bfu_1,\mathcal{N}(\bfv_1+\bfu_1))\mathbf{C}_{\mathcal{N}(\bfv_1+\bfu_1)}^{-1} \widetilde{\mathbf{C}}_{\mathcal{N}(\bfv_1+\bfu_1),\mathcal{N}(\bfv_2+\bfu_2)}\mathbf{b}_{\mathbf{v}_2+\bfu_2}.
\end{align*}

For small enough $\bfu_1$, $\mathcal{N}(\bfv_1)=\mathcal{N}(\bfv_1+\bfu_1)$ and
\[
\lim_{\bfu_1  \rightarrow 0}\norm{\mathbf{v}_1+\bfu_2,\mathcal{N}(\bfv_1+\bfu_1)} = \norm{\mathbf{v}_1,\mathcal{N}(\bfv_1)}.
\]

Thus,
\[
\lim_{\bfu_1  \rightarrow 0} D_{\bfu} C(\mathbf{v}_1+\bfu_1,\mathcal{N}(\bfv_1+\bfu_1))=  D_{\bfu} C(\mathbf{v}_1,\mathcal{N}(\bfv_1)).
\]

Same as in case 1,
\begin{align*}
    \mathbf{b}_{\mathbf{v}_2+\bfu_2}\T &= \mathbf{C}_{\mathbf{v}_2+\bfu_2,\mathcal{N}(\bfv_2+\bfu_2)}\mathbf{C}_{\mathcal{N}(\bfv_2+\bfu_2)}^{-1} = \mathbf{C}_{\mathbf{v}_2+\bfu_2,\mathcal{N}_0(\bfs_j)}\mathbf{C}_{\mathcal{N}_0(\bfs_j)}^{-1} \\
    & \rightarrow \mathbf{C}_{\mathbf{v}_2,\mathcal{N}_0(\bfs_j)}\mathbf{C}_{\mathcal{N}_0(\bfs_j)}^{-1} = \mathbf{e}_1\T \mathbf{C}_{\mathcal{N}_0(\bfs_j)}\mathbf{C}_{\mathcal{N}_0(\bfs_j)}^{-1} = \mathbf{e}_1\T = (1,0,...,0).
\end{align*}
Then,
\begin{align*}
&\widetilde{\mathbf{C}}_{\mathcal{N}(\bfv_1+\bfu_1),\mathcal{N}(\bfv_2+\bfu_2)}\mathbf{b}_{\mathbf{v}_2+\bfu_2} = \widetilde{\mathbf{C}}_{\mathcal{N}(\bfv_1+\bfu_1),\mathcal{N}_0(\bfs_j)}\mathbf{b}_{\mathbf{v}_2+\bfu_2} \\&~\rightarrow \widetilde{\mathbf{C}}_{\mathcal{N}(\bfv_1+\bfu_1),\mathcal{N}_0(\bfs_j)} \mathbf{e}_1 = \widetilde{\mathbf{C}}_{\mathcal{N}(\bfv_1+\bfu_1),\bfs_j}.
\end{align*}

Thus,
$\lim_{\bfu_1, \bfu_2 \rightarrow 0}\cov(D_\bfu \widetilde{\omega}(\bfv_1 +\bfu_1),\widetilde{\omega}(\bfv_2 +\bfu_2))=\cov(D_\bfu \widetilde{\omega}(\bfv_1),\widetilde{\omega}(\bfv_2))$.

\item $\mathbf{v}_1,\mathbf{v}_2\notin \mathcal{S}, ~\mathbf{v}_1 \neq \mathbf{v}_2, ~\mathbf{v}_1+\bfu_1 \neq \mathbf{v}_2, ~\mathbf{v}_1\neq \mathbf{v}_2+\bfu_2,~\mathbf{v}_1+\bfu_1 \neq \mathbf{v}_2 + \bfu_2$

We have 
\[
    \cov(D_\bfu \widetilde{\omega}(\bfv_1),\widetilde{\omega}(\bfv_2))=
D_{\bfu} C(\mathbf{v}_1,\mathcal{N}(\bfv_1))\mathbf{C}_{\mathcal{N}(\bfv_1)}^{-1} \widetilde{\mathbf{C}}_{\mathcal{N}(\bfv_1),\mathcal{N}(\bfv_2)}\mathbf{b}_{\mathbf{v}_2} .
\]

For small enough $\bfu_1$ and $\bfu_2$, $\mathcal{N}(\bfv_1)=\mathcal{N}(\bfv_1+\bfu_1)$, $\mathcal{N}(\bfv_2)=\mathcal{N}(\bfv_2+\bfu_2)$ and
\begin{align*}
    \lim_{\bfu_1  \rightarrow 0}\norm{\mathbf{v}_1+\bfu_2,\mathcal{N}(\bfv_1+\bfu_1)} = \norm{\mathbf{v}_1,\mathcal{N}(\bfv_1)},\\
\lim_{\bfu_2 \rightarrow 0}\norm{\mathbf{v}_2+\bfu_2,\mathcal{N}(\bfv_2+\bfu_2)} = \norm{\mathbf{v}_2,\mathcal{N}(\bfv_2)},
\end{align*}

\begin{align*}
    &\cov(D_\bfu \widetilde{\omega}(\bfv_1 +\bfu_1),\widetilde{\omega}(\bfv_2 +\bfu_2)) \\
    &= D_{\bfu} C(\mathbf{v}_1+\bfu_1,\mathcal{N}(\bfv_1+\bfu_1))\mathbf{C}_{\mathcal{N}(\bfv_1+\bfu_1)}^{-1} \widetilde{\mathbf{C}}_{\mathcal{N}(\bfv_1+\bfu_1),\mathcal{N}(\bfv_2+\bfu_2)}\mathbf{b}_{\mathbf{v}_2+\bfu_2} \\
    &\rightarrow D_{\bfu} C(\mathbf{v}_1,\mathcal{N}(\bfv_1))\mathbf{C}_{\mathcal{N}(\bfv_1)}^{-1} \widetilde{\mathbf{C}}_{\mathcal{N}(\bfv_1),\mathcal{N}(\bfv_2)}\mathbf{b}_{\mathbf{v}_2} \\
    &=\cov(D_\bfu \widetilde{\omega}(\bfv_1),\widetilde{\omega}(\bfv_2)).
\end{align*}
\end{enumerate}
\end{proof}

\section{Proof of \Cref{thm:duC_var}: The covariance of the NNDP}
\begin{proof}[Proof of \Cref{thm:duC_var}] We first calculate the covariance of $D_\bfu \widetilde{\omega}$. Recall that
\[
\cov (\widetilde{\omega}_{\bfu,h}(\bfv_1),\widetilde{\omega}_{\bfu,h}(\bfv_2)) = \frac{1}{h^2}(\widetilde{\mathbf{C}}(\bfv_1+h\bfu,\bfv_2+h\bfu)-\widetilde{\mathbf{C}}(\bfv_1+h\bfu,\bfv_2)-\widetilde{\mathbf{C}}(\bfv_1,\bfv_2+h\bfu)+\widetilde{\mathbf{C}}(\bfv_1,\bfv_2)).
\]
We then calculate the covariance of $D_\bfu \widetilde{\omega}$ by the following six cases.

\begin{enumerate}
    \item $\mathbf{v}_1,\mathbf{v}_2 \in \mathcal{S}$, $\mathbf{v}_1\neq \mathbf{v}_2$.
\begin{align*}
&\frac{1}{h^2}(\widetilde{\mathbf{C}}(\mathbf{v}_1+h\bfu,\mathbf{v}_2+h\bfu)-\widetilde{\mathbf{C}}(\mathbf{v}_1+h\bfu,\mathbf{v}_2)-\widetilde{\mathbf{C}}(\mathbf{v}_1,\mathbf{v}_2+h\bfu)+\widetilde{\mathbf{C}}(\mathbf{v}_1,\mathbf{v}_2)) \\
& = \frac{1}{h}\left(\frac{(\mathbf{b}_{\bfs_i+h\bfu}\T \widetilde{\mathbf{C}}_{\mathcal{N}(\bfs_i+h\bfu),\mathcal{N}(\bfs_j+h\bfu)} \mathbf{b}_{\bfs_j+h\bfu} - \mathbf{b}_{\bfs_i+h\bfu}\T \widetilde{\mathbf{C}}_{\mathcal{N}(\bfs_i+h\bfu),\bfs_j}}{h}\right) \\&\quad-\frac{1}{h}\left( \frac{\widetilde{\mathbf{C}}_{\bfs_i,\mathcal{N}(\bfs_j+h\bfu)}\mathbf{b}_{\bfs_j+h\bfu} - \widetilde{\mathbf{C}}_{\bfs_i,\bfs_j})}{h}\right) \\
%& = \frac{1}{h}\left(\mathbf{b}_{\bfs_i+h\bfu}\T\widetilde{\mathbf{C}}_{\mathcal{N}(\bfs_i+h\bfu),\mathcal{N}(\bfs_j)}\mathbf{C}_{\mathcal{N}_0(\bfs_j)}^{-1}(\mathbf{C}_{\mathcal{N}_0(\bfs_j),\mathcal{N}_\Delta(\bfs_j)}-\mathbf{C}_{\mathcal{N}_0(\bfs_j)})\mathbf{e}_1- \right) \\
&= \frac{1}{h^2}(\mathbf{b}_{\bfs_i+h\bfu}\T \widetilde{\mathbf{C}}_{\mathcal{N}_0(\bfs_i),\mathcal{N}_0(\bfs_j)}-\widetilde{\mathbf{C}}_{\bfs_i,\mathcal{N}_0(\bfs_j)})\mathbf{C}_{\mathcal{N}_0(\bfs_j)}^{-1}(\mathbf{C}_{\mathcal{N}_0(\bfs_j),\mathcal{N}_\Delta(\bfs_j)}-\mathbf{C}_{\mathcal{N}_0(\bfs_j)})\mathbf{e}_1 \\
&= \frac{1}{h^2}\mathbf{e}_1\T (\mathbf{C}_{\mathcal{N}_\Delta(\bfs_i),\mathcal{N}_0(\bfs_i)}-\mathbf{C}_{\mathcal{N}_0(\bfs_i)})\mathbf{C}_{\mathcal{N}_0(\bfs_i)}^{-1}\widetilde{\mathbf{C}}_{\mathcal{N}_0(\bfs_i),\mathcal{N}_0(\bfs_j)}\mathbf{C}_{\mathcal{N}_0(\bfs_j)}^{-1}(\mathbf{C}_{\mathcal{N}_0(\bfs_j),\mathcal{N}_\Delta(\bfs_j)}-\mathbf{C}_{\mathcal{N}_0(\bfs_j)})\mathbf{e}_1 \\
& \rightarrow D_{\bfu} \mathbf{C}(\bfs_i,\mathcal{N}_0(\bfs_i))\mathbf{C}_{\mathcal{N}_0(\bfs_i)}^{-1}\widetilde{\mathbf{C}}_{\mathcal{N}_0(\bfs_i),\mathcal{N}_0(\bfs_j)}\mathbf{C}_{\mathcal{N}_0(\bfs_j)}^{-1} D_{\bfu} \mathbf{C}(\mathcal{N}_0(\bfs_j),\bfs_j).
\end{align*}

\item $\mathbf{v}_1,\mathbf{v}_2 \in \mathcal{S}$, $\mathbf{v}_1= \mathbf{v}_2=\bfs_i, \mathbf{v}_1+h\bfu=\mathbf{v}_2+h\bfu$.

Let $\partial^2 \widetilde{\mathbf{C}}_{\bfs_i} = D_{\bfu} \mathbf{C}(\bfs_i,\mathcal{N}_0(\bfs_i))\mathbf{C}_{\mathcal{N}_0(\bfs_i)}^{-1}\widetilde{\mathbf{C}}_{\mathcal{N}_0(\bfs_i),\mathcal{N}_0(\bfs_i)}\mathbf{C}_{\mathcal{N}_0(\bfs_i)}^{-1} D_{\bfu} \mathbf{C}(\mathcal{N}_0(\bfs_i),\bfs_i)$,
\begin{align*}
&\lim_{h\to 0}\frac{1}{h^2}(\widetilde{\mathbf{C}}(\mathbf{v}_1+h\bfu,\mathbf{v}_2+h\bfu)-\widetilde{\mathbf{C}}(\mathbf{v}_1+h\bfu,\mathbf{v}_2)-\widetilde{\mathbf{C}}(\mathbf{v}_1,\mathbf{v}_2+h\bfu)+\widetilde{\mathbf{C}}(\mathbf{v}_1,\mathbf{v}_2))  \\
& = \lim_{h\to 0}\frac{1}{h}\left(\frac{(\mathbf{b}_{\bfs_i+h\bfu}\T \widetilde{\mathbf{C}}_{\mathcal{N}(\bfs_i+h\bfu),\mathcal{N}(\bfs_i+h\bfu)} \mathbf{b}_{\bfs_i+h\bfu} - \mathbf{b}_{\bfs_i+h\bfu}\T \widetilde{\mathbf{C}}_{\mathcal{N}(\bfs_i+h\bfu),\bfs_i}}{h}\right) \\&\quad-\lim_{h\to 0}\frac{1}{h}\left( \frac{\widetilde{\mathbf{C}}_{\bfs_i,\mathcal{N}(\bfs_i+h\bfu)}\mathbf{b}_{\bfs_i+h\bfu} - \widetilde{\mathbf{C}}_{\bfs_i,\bfs_i})}{h}\right) + \lim_{h\to 0}\frac{1}{h^2}\mathbf{F}_{\bfs_i+h\bfu} \\
&= \partial^2 \widetilde{\mathbf{C}}_{\bfs_i} +  \lim_{h\to 0}\frac{1}{h^2}(\mathbf{C}_{\bfs_i+h\bfu,\bfs_i+h\bfu}-\mathbf{C}_{\bfs_i+h\bfu,\mathcal{N}(\bfs_i+h\bfu)}\mathbf{C}_{\mathcal{N}(\bfs_i+h\bfu)}^{-1}\mathbf{C}_{\mathcal{N}(\bfs_i+h\bfu),\bfs_i+h\bfu})\\
&= \partial^2 \widetilde{\mathbf{C}}_{\bfs_i} +\lim_{h\to 0}\frac{1}{h^2}(\mathbf{C}_{\bfs_i+h\bfu,\bfs_i+h\bfu} - \mathbf{C}_{\bfs_i+h\bfu,\bfs_i} + \mathbf{C}_{\bfs_i+h\bfu,\bfs_i} \\&\quad - \mathbf{C}_{\bfs_i+h\bfu,\mathcal{N}(\bfs_i+h\bfu)}\mathbf{C}_{\mathcal{N}(\bfs_i+h\bfu)}^{-1}\mathbf{C}_{\mathcal{N}(\bfs_i+h\bfu),\bfs_i+h\bfu}) \\
&= \partial^2 \widetilde{\mathbf{C}}_{\bfs_i} + \lim_{h\to 0}\frac{1}{h} D_{\bfu} \mathbf{C} (\bfs_i+h\bfu,\bfs_i) + \lim_{h\to 0}\frac{1}{h^2} ( \mathbf{C}_{\bfs_i+h\bfu,\mathcal{N}_0(\bfs_i)}\mathbf{e}_1 - \mathbf{C}_{\bfs_i+h\bfu,\mathcal{N}_0(\bfs_i)}\mathbf{C}_{\mathcal{N}_0(\bfs_i)}^{-1}\mathbf{C}_{\mathcal{N}_0(\bfs_i),\mathcal{N}_\Delta(\bfs_i)}\mathbf{e}_1) \\
&=\partial^2 \widetilde{\mathbf{C}}_{\bfs_i} + \lim_{h\to 0}\frac{1}{h} D_{\bfu} \mathbf{C} (\bfs_i+h\bfu,\bfs_i) - \lim_{h\to 0}\frac{1}{h} \mathbf{C}_{\bfs_i+h\bfu,\mathcal{N}_0(\bfs_i)}\mathbf{C}_{\mathcal{N}_0(\bfs_i)}^{-1}D_{\bfu} \mathbf{C}(\mathcal{N}_0(\bfs_i),\bfs_i) \\
&=\partial^2 \widetilde{\mathbf{C}}_{\bfs_i} + \lim_{h\to 0}\frac{1}{h}(D_{\bfu} \mathbf{C} (\bfs_i+h\bfu,\bfs_i) - D_{\bfu} \mathbf{C} (\bfs_i,\bfs_i)) \\&\quad+ \lim_{h\to 0}\frac{1}{h} (D_{\bfu} \mathbf{C} (\bfs_i,\bfs_i) -\mathbf{C}_{\bfs_i+h\bfu,\mathcal{N}_0(\bfs_i)}\mathbf{C}_{\mathcal{N}_0(\bfs_i)}^{-1}D_{\bfu} \mathbf{C}(\mathcal{N}_0(\bfs_i),\bfs_i)) \\
&=D^{(2)}_{\bfu}\mathbf{C}(\bfs_i,\bfs_i) + D_{\bfu} \mathbf{C}(\bfs_i,\mathcal{N}_0(\bfs_i))\mathbf{C}_{\mathcal{N}_0(\bfs_i)}^{-1}\widetilde{\mathbf{C}}_{\mathcal{N}_0(\bfs_i)}\mathbf{C}_{\mathcal{N}_0(\bfs_i)}^{-1} D_{\bfu} \mathbf{C}(\mathcal{N}_0(\bfs_i),\bfs_i)  \\
&\quad - D_{\bfu} \mathbf{C}(\bfs_i,\mathcal{N}_0(\bfs_i))\mathbf{C}_{\mathcal{N}_0(\bfs_i)}^{-1}D_{\bfu} \mathbf{C}(\mathcal{N}_0(\bfs_i),\bfs_i).
\end{align*}

\item $\mathbf{v}_1=\bfs_i,~\mathbf{v}_2 \notin \mathcal{S}$.
\begin{align*}
&\frac{1}{h^2}(\widetilde{\mathbf{C}}(\mathbf{v}_1+h\bfu,\mathbf{v}_2+h\bfu)-\widetilde{\mathbf{C}}(\mathbf{v}_1+h\bfu,\mathbf{v}_2)-\widetilde{\mathbf{C}}(\mathbf{v}_1,\mathbf{v}_2+h\bfu)+\widetilde{\mathbf{C}}(\mathbf{v}_1,\mathbf{v}_2)) \\
&= \frac{1}{h^2}(\mathbf{b}_{\bfs_i+h\bfu}\T \widetilde{\mathbf{C}}_{\mathcal{N}(\bfs_i+h\bfu),\mathcal{N}(\bfv_2+h\bfu)} \mathbf{b}_{\mathbf{v}_2+h\bfu} - \mathbf{b}_{\bfs_i+h\bfu}\T \widetilde{\mathbf{C}}_{\mathcal{N}(\bfs_i+h\bfu),\mathcal{N}(\bfv_2)} \mathbf{b}_{\mathbf{v}_2}\\&\quad-  \widetilde{\mathbf{C}}_{\bfs_i,\mathcal{N}(\bfv_2+h\bfu)}\mathbf{b}_{\mathbf{v}_2+h\bfu} + \widetilde{\mathbf{C}}_{\bfs_i,\mathcal{N}(\bfv_2)}\mathbf{b}_{\mathbf{v}_2}) \\
&= \frac{1}{h^2}(\mathbf{b}_{\bfs_i+h\bfu}\T \widetilde{\mathbf{C}}_{\mathcal{N}_0(\bfs_i),\mathcal{N}(\bfv_2)}-\widetilde{\mathbf{C}}_{\bfs_i,\mathcal{N}(\bfv_2)})\mathbf{C}_{\mathcal{N}(\bfv_2)}^{-1}(\mathbf{C}_{\mathcal{N}(\bfv_2),\mathbf{v}_2+h\bfu}-\mathbf{C}_{\mathcal{N}(\bfv_2),\mathbf{v}_2})\T \\
&= \frac{1}{h^2}\mathbf{e}_1\T (\mathbf{C}_{\mathcal{N}_\Delta(\bfs_i),\mathcal{N}_0(\bfs_i)}-\mathbf{C}_{\mathcal{N}_0(\bfs_i)})\mathbf{C}_{\mathcal{N}_0(\bfs_i)}^{-1}\widetilde{\mathbf{C}}_{\mathcal{N}_0(\bfs_i),\mathcal{N}(\bfv_2)}\mathbf{C}_{\mathcal{N}(\bfv_2)}^{-1}(\mathbf{C}_{\mathcal{N}(\bfv_2),\mathbf{v}_2+h\bfu}-\mathbf{C}_{\mathcal{N}(\bfv_2),\mathbf{v}_2})\T \\
& \rightarrow D_{\bfu} \mathbf{C}(\bfs_i,\mathcal{N}_0(\bfs_i))\mathbf{C}_{\mathcal{N}_0(\bfs_i)}^{-1}\widetilde{\mathbf{C}}_{\mathcal{N}_0(\bfs_i),\mathcal{N}(\bfv_2)}\mathbf{C}_{\mathcal{N}(\bfv_2)}^{-1} D_{\bfu} \mathbf{C}(\mathcal{N}(\bfv_2),\mathbf{v}_2).
\end{align*}

\item $\mathbf{v}_1\notin \mathcal{S},\mathbf{v}_2=\bfs_j.$
\begin{align*}
&\frac{1}{h^2}(\widetilde{\mathbf{C}}(\mathbf{v}_1+h\bfu,\mathbf{v}_2+h\bfu)-\widetilde{\mathbf{C}}(\mathbf{v}_1+h\bfu,\mathbf{v}_2)-\widetilde{\mathbf{C}}(\mathbf{v}_1,\mathbf{v}_2+h\bfu)+\widetilde{\mathbf{C}}(\mathbf{v}_1,\mathbf{v}_2)) \\
&=\frac{1}{h^2}(\mathbf{b}_{\mathbf{v}_1+h\bfu}\T \widetilde{\mathbf{C}}_{\mathcal{N}(\bfv_1+h\bfu),\mathcal{N}(\bfs_j+h\bfu)} \mathbf{b}_{\bfs_j+h\bfu} - \mathbf{b}_{\mathbf{v}_1+h\bfu}\T \widetilde{\mathbf{C}}_{\mathcal{N}(\bfv_1+h\bfu),\bfs_j} \\&\quad-   \mathbf{b}_{\mathbf{v}_1}\T\widetilde{\mathbf{C}}_{\mathcal{N}(\bfv_1),\mathcal{N}(\bfs_j+h\bfu)}\mathbf{b}_{\bfs_j+h\bfu} + \mathbf{b}_{\mathbf{v}_1}\T\widetilde{\mathbf{C}}_{\mathcal{N}(\bfv_1),\bfs_j}) \\
&= \frac{1}{h^2}(\mathbf{b}_{\mathbf{v}_1+h\bfu}\T \widetilde{\mathbf{C}}_{\mathcal{N}(\bfv_1),\mathcal{N}_0(\bfs_j)}-\mathbf{b}_{\mathbf{v}_1}\T \widetilde{\mathbf{C}}_{\mathcal{N}(\bfv_1),\mathcal{N}_0(\bfs_j)})\mathbf{C}_{\mathcal{N}_0(\bfs_j)}^{-1}(\mathbf{C}_{\mathcal{N}_0(\bfs_j),\mathcal{N}_\Delta(\bfs_j)}-\mathbf{C}_{\mathcal{N}_0(\bfs_j)})\mathbf{e}_1 \\
& \rightarrow D_{\bfu} \mathbf{C}(\mathbf{v}_1,\mathcal{N}(\bfv_1))\mathbf{C}_{\mathcal{N}(\bfv_1)}^{-1}\widetilde{\mathbf{C}}_{\mathcal{N}(\bfv_1),\mathcal{N}_0(\bfs_j)}\mathbf{C}_{\mathcal{N}_0(\bfs_j)}^{-1} D_{\bfu} \mathbf{C}(\mathcal{N}_0(\bfs_j),\bfs_j).
\end{align*}

\item $\mathbf{v}_1,\mathbf{v}_2\notin \mathcal{S}$, $\mathbf{v}_1 \neq \mathbf{v}_2$, $\mathbf{v}_1+h\bfu \neq \mathbf{v}_2$. 
\begin{align*}
&\frac{1}{h^2}(\widetilde{\mathbf{C}}(\mathbf{v}_1+h\bfu,\mathbf{v}_2+h\bfu)-\widetilde{\mathbf{C}}(\mathbf{v}_1+h\bfu,\mathbf{v}_2)-\widetilde{\mathbf{C}}(\mathbf{v}_1,\mathbf{v}_2+h\bfu)+\widetilde{\mathbf{C}}(\mathbf{v}_1,\mathbf{v}_2)) \\
&= \frac{1}{h^2}(\mathbf{b}_{\mathbf{v}_1+h\bfu}\T \widetilde{\mathbf{C}}_{\mathcal{N}(\bfv_1+h\bfu),\mathcal{N}(\bfv_2+h\bfu)} \mathbf{b}_{\mathbf{v}_2+h\bfu} - \mathbf{b}_{\mathbf{v}_1+h\bfu}\T \widetilde{\mathbf{C}}_{\mathcal{N}(\bfv_1+h\bfu),\mathcal{N}(\bfv_2)} \mathbf{b}_{\mathbf{v}_2}-  \\
&\mathbf{b}_{\mathbf{v}_1}\T\widetilde{\mathbf{C}}_{\mathcal{N}(\bfv_1),\mathcal{N}(\bfv_2+h\bfu)}\mathbf{b}_{\mathbf{v}_2+h\bfu} + \mathbf{b}_{\mathbf{v}_1}\T\widetilde{\mathbf{C}}_{\mathcal{N}(\bfv_1),\mathcal{N}(\bfv_2)}\mathbf{b}_{\mathbf{v}_2}) \\
&= \frac{1}{h^2}(\mathbf{b}_{\mathbf{v}_1+h\bfu}\T-\mathbf{b}_{\mathbf{v}_1}\T) \widetilde{\mathbf{C}}_{\mathcal{N}(\bfv_1),\mathcal{N}(\bfv_2)}(\mathbf{b}_{\mathbf{v}_2+h\bfu}\T-\mathbf{b}_{\mathbf{v}_2}\T)\T \\
& \rightarrow D_{\bfu} \mathbf{C}(\mathbf{v}_1,\mathcal{N}(\mathbf{v}_1))\mathbf{C}_{\mathcal{N}(\bfv_1)}^{-1}\widetilde{\mathbf{C}}_{\mathcal{N}(\bfv_1),\mathcal{N}(\bfv_2)}\mathbf{C}_{\mathcal{N}(\bfv_2)}^{-1} D_{\bfu} \mathbf{C}(\mathcal{N}(\mathbf{v}_2),\mathbf{v}_2).
\end{align*}

\item $\mathbf{v}_1,\mathbf{v}_2 \notin \mathcal{S}$, $\mathbf{v}_1= \mathbf{v}_2, \mathbf{v}_1+h\bfu=\mathbf{v}_2+h\bfu$.

Let $\partial^2 \widetilde{\mathbf{C}}_{\bfv_i} = D_{\bfu} \mathbf{C}(\bfv_1,\mathcal{N}(\bfv_1))\mathbf{C}_{\mathcal{N}(\bfv_1)}^{-1}\widetilde{\mathbf{C}}_{\mathcal{N}(\bfv_1),\mathcal{N}(\bfv_1)}\mathbf{C}_{\mathcal{N}(\bfv_1)}^{-1} D_{\bfu} \mathbf{C}(\mathcal{N}(\bfv_1),\bfv_1)$,
\begin{align*}
&\lim_{h\to 0}\frac{1}{h^2}(\widetilde{\mathbf{C}}(\mathbf{v}_1+h\bfu,\mathbf{v}_2+h\bfu)-\widetilde{\mathbf{C}}(\mathbf{v}_1+h\bfu,\mathbf{v}_2)-\widetilde{\mathbf{C}}(\mathbf{v}_1,\mathbf{v}_2+h\bfu)+\widetilde{\mathbf{C}}(\mathbf{v}_1,\mathbf{v}_2))  \\
& = \lim_{h\to 0}\frac{1}{h}\left(\frac{(\mathbf{b}_{\bfv_1+h\bfu}\T \widetilde{\mathbf{C}}_{\mathcal{N}(\bfv_1+h\bfu),\mathcal{N}(\bfv_1+h\bfu)} \mathbf{b}_{\bfv_1+h\bfu} - \mathbf{b}_{\bfv_1+h\bfu}\T \widetilde{\mathbf{C}}_{\mathcal{N}(\bfv_1+h\bfu),\bfv_1}}{h}\right) \\&\quad-\lim_{h\to 0}\frac{1}{h}\left( \frac{\widetilde{\mathbf{C}}_{\bfv_1,\mathcal{N}(\bfv_1+h\bfu)}\mathbf{b}_{\bfv_1+h\bfu} - \widetilde{\mathbf{C}}_{\bfv_1,\bfv_1})}{h}\right) + \lim_{h\to 0}\frac{1}{h^2}(\mathbf{F}_{\bfv_1+h\bfu}-\mathbf{F}_{\bfv_1}) \\
&= \partial^2 \widetilde{\mathbf{C}}_{\bfv_1} +  \lim_{h\to 0}\frac{1}{h^2}(\mathbf{C}_{\bfv_1+h\bfu,\bfv_1+h\bfu}-\mathbf{C}_{\bfv_1+h\bfu,\mathcal{N}(\bfv_1+h\bfu)}\mathbf{C}_{\mathcal{N}(\bfv_1+h\bfu)}^{-1}\mathbf{C}_{\mathcal{N}(\bfv_1+h\bfu),\bfv_1+h\bfu}\\&\quad - \mathbf{C}_{\bfv_1,\bfv_1}+\mathbf{C}_{\bfv_1,\mathcal{N}(\bfv_1)}\mathbf{C}_{\mathcal{N}(\bfv_1)}^{-1}\mathbf{C}_{\mathcal{N}(\bfv_1),\bfv_1})\\
&= \partial^2 \widetilde{\mathbf{C}}_{\bfv_1} +\lim_{h\to 0}\frac{1}{h^2}(\mathbf{C}_{\bfv_1+h\bfu,\bfv_1+h\bfu} - \mathbf{C}_{\bfv_1+h\bfu,\bfv_1} + \mathbf{C}_{\bfv_1+h\bfu,\bfv_1} -\mathbf{C}_{\bfv_1,\bfv_1}) \\&\quad - D_{\bfu} \mathbf{C}(\bfv_1,\mathcal{N}(\bfv_1))\mathbf{C}_{\mathcal{N}(\bfv_1)}^{-1}D_{\bfu} \mathbf{C}(\mathcal{N}(\bfv_1),\bfv_1) \\
&=D^{(2)}_{\bfu}\mathbf{C}(\bfv_1,\bfv_1) + D_{\bfu} \mathbf{C}(\bfv_1,\mathcal{N}(\bfv_1))\mathbf{C}_{\mathcal{N}(\bfv_1)}^{-1}\widetilde{\mathbf{C}}_{\mathcal{N}(\bfv_1)}\mathbf{C}_{\mathcal{N}(\bfv_1)}^{-1} D_{\bfu} \mathbf{C}(\mathcal{N}(\bfv_1),\bfv_1)  \\
&\quad - D_{\bfu} \mathbf{C}(\bfv_1,\mathcal{N}(\bfv_1))\mathbf{C}_{\mathcal{N}(\bfv_1)}^{-1}D_{\bfu} \mathbf{C}(\mathcal{N}(\bfv_1),\bfv_1).
\end{align*}
\end{enumerate}

Next, we prove the continuity of $\cov(D_\bfu \widetilde{\omega}(\bfv_1),D_\bfu \widetilde{\omega}(\bfv_2))$, that is
$$\cov(D_\bfu \widetilde{\omega}(\bfv_1+\bfu_1),D_\bfu \widetilde{\omega}(\bfv_2+\bfu_2))\xrightarrow{\bfu_1,\bfu_2\to0}\cov(D_\bfu \widetilde{\omega}(\bfv_1),D_\bfu \widetilde{\omega}(\bfv_2))$$

The limit of $\mathcal{N}(\bfv_1+\bfu_1)$, $\mathcal{N}_0(\bfv_1+\bfu_1)$, $\mathcal{N}(\bfv_2+\bfu_2)$, $\mathcal{N}_0(\bfv_2+\bfu_2)$ holds same as stated in the proof of continuity of $\cov(D_\bfu \widetilde{\omega}(\bfv_1),\widetilde{\omega}(\bfv_2))$. We consider the following four cases:

\begin{enumerate}[label=\roman*.]
\item $\mathbf{v}_1=\bfs_i,~\mathbf{v}_2=\bfs_j \in \mathcal{S}$, $\mathbf{v}_1\neq \mathbf{v}_2$.
\begin{align*}
&\cov(D_\bfu \widetilde{\omega}(\bfv_1+\bfu_1),D_\bfu \widetilde{\omega}(\bfv_2+\bfu_2)) \\
& = D_{\bfu} C(\bfv_1+\bfu_1,\mathcal{N}_0(\bfv_1+\bfu_1))\mathbf{C}_{\mathcal{N}(\bfv_1+\bfu_1)}^{-1}\widetilde{\mathbf{C}}_{\mathcal{N}(\bfv_1+\bfu_1),\mathcal{N}(\bfv_2+\bfu_2)}
\\&\quad \times \mathbf{C}_{\mathcal{N}(\bfv_2+\bfu_2)}^{-1} D_{\bfu} C(\mathcal{N}_0(\bfv_2+\bfu_2),\bfv_2+\bfu_2) \\
& \rightarrow D_{\bfu} C(\bfs_i,\mathcal{N}_0(\bfs_i))\mathbf{C}_{\mathcal{N}_0(\bfs_i)}^{-1}\widetilde{\mathbf{C}}_{\mathcal{N}_0(\bfs_i),\mathcal{N}_0(\bfs_j)}\mathbf{C}_{\mathcal{N}_0(\bfs_j)}^{-1} D_{\bfu} C(\mathcal{N}_0(\bfs_j),\bfs_j) \\
& = \cov(D_\bfu \widetilde{\omega}(\bfv_1),D_\bfu \widetilde{\omega}(\bfv_2)).
\end{align*}

\item $\mathbf{v}_1=\bfs_i,~\mathbf{v}_2 \notin \mathcal{S}$.
\begin{align*}
&\cov(D_\bfu \widetilde{\omega}(\bfv_1+\bfu_1),D_\bfu \widetilde{\omega}(\bfv_2+\bfu_2)) \\
& = D_{\bfu} C(\bfv_1+\bfu_1,\mathcal{N}(\bfv_1+\bfu_1))\mathbf{C}_{\mathcal{N}(\bfv_1+\bfu_1)}^{-1}\widetilde{\mathbf{C}}_{\mathcal{N}(\bfv_1+\bfu_1),\mathcal{N}(\bfv_2+\bfu_2)}
\\&\quad \times \mathbf{C}_{\mathcal{N}(\bfv_2+\bfu_2)}^{-1} D_{\bfu} C(\mathcal{N}(\bfv_2+\bfu_2),\bfv_2+\bfu_2) \\
& \rightarrow D_{\bfu} C(\bfs_i,\mathcal{N}_0(\bfs_i))\mathbf{C}_{\mathcal{N}_0(\bfs_i)}^{-1}\widetilde{\mathbf{C}}_{\mathcal{N}_0(\bfs_i),\mathcal{N}(\bfv_2)}\mathbf{C}_{\mathcal{N}(\bfv_2)}^{-1} D_{\bfu} C(\mathcal{N}(\bfv_2),\mathbf{v}_2) \\
& = \cov(D_\bfu \widetilde{\omega}(\bfv_1),D_\bfu \widetilde{\omega}(\bfv_2)).
\end{align*}
\item $\mathbf{v}_1\notin \mathcal{S},\mathbf{v}_2=\bfs_j.$ The proof is same as the previous case.

\item $\mathbf{v}_1,\mathbf{v}_2 \notin \mathcal{S},\mathbf{v}_1\neq\bfv_2.$ 
\begin{align*}
&\cov(D_\bfu \widetilde{\omega}(\bfv_1+\bfu_1),D_\bfu \widetilde{\omega}(\bfv_2+\bfu_2)) \\
& = D_{\bfu} C(\bfv_1+\bfu_1,\mathcal{N}(\bfv_1+\bfu_1))\mathbf{C}_{\mathcal{N}(\bfv_1+\bfu_1)}^{-1}\widetilde{\mathbf{C}}_{\mathcal{N}(\bfv_1+\bfu_1),\mathcal{N}(\bfv_2+\bfu_2)}
\\&\quad \times \mathbf{C}_{\mathcal{N}(\bfv_2+\bfu_2)}^{-1} D_{\bfu} C(\mathcal{N}(\bfv_2+\bfu_2),\bfv_2+\bfu_2) \\
& \rightarrow D_{\bfu} C(\bfv_1,\mathcal{N}(\bfv_1))\mathbf{C}_{\mathcal{N}(\bfv_1)}^{-1}\widetilde{\mathbf{C}}_{\mathcal{N}(\bfv_1),\mathcal{N}(\bfv_2)}\mathbf{C}_{\mathcal{N}(\bfv_2)}^{-1} D_{\bfu} C(\mathcal{N}(\bfv_2),\mathbf{v}_2) \\
& = \cov(D_\bfu \widetilde{\omega}(\bfv_1),D_\bfu \widetilde{\omega}(\bfv_2)).
\end{align*}

\iffalse
\item $\mathbf{v}_1,\mathbf{v}_2 \notin \mathcal{S},\mathbf{v}_1=\bfv_2.$ 
\begin{align*}
&\cov(D_\bfu \widetilde{\omega}(\bfv_1+\bfu_1),D_\bfu \widetilde{\omega}(\bfv_2+\bfu_2)) \\
& = D_{\bfu} C(\bfv_1+\bfu_1,\mathcal{N}(\bfv_1+\bfu_1))\mathbf{C}_{\mathcal{N}(\bfv_1+\bfu_1)}^{-1}\widetilde{\mathbf{C}}_{\mathcal{N}(\bfv_1+\bfu_1),\mathcal{N}(\bfv_2+\bfu_2)}
\\&\quad \times \mathbf{C}_{\mathcal{N}(\bfv_2+\bfu_2)}^{-1} D_{\bfu} C(\mathcal{N}(\bfv_2+\bfu_2),\bfv_2+\bfu_2) \\
& \rightarrow D_{\bfu} C(\bfv_1,\mathcal{N}(\bfv_1))\mathbf{C}_{\mathcal{N}(\bfv_1)}^{-1}\widetilde{\mathbf{C}}_{\mathcal{N}(\bfv_1),\mathcal{N}(\bfv_2)}\mathbf{C}_{\mathcal{N}(\bfv_2)}^{-1} D_{\bfu} C(\mathcal{N}(\bfv_2),\mathbf{v}_2) \\
& \neq \cov(D_\bfu \widetilde{\omega}(\bfv_1),D_\bfu \widetilde{\omega}(\bfv_2)).
\end{align*}
Thus, $\cov(D_\bfu \widetilde{\omega}(\bfv_1),D_\bfu \widetilde{\omega}(\bfv_2))$ is not continuous at $\mathbf{v}_1,\mathbf{v}_2 \notin \mathcal{S},\mathbf{v}_1=\bfv_2.$
\fi
\end{enumerate}
\end{proof}

\section{Proof of \Cref{thm:NNDP_proper}: The joint density of NNDP}

%\begin{theorem}
    %Graph of NNDP is still a directed acyclic graph (DAG), so the joint density derived from NNDP is still a proper joint density.
%\end{theorem}

\begin{proof}[Proof of \Cref{thm:NNDP_proper}]
We first prove that the graph of NNDP is still a directed acyclic graph (DAG), so the joint density derived from NNDP is still a proper joint density. Given $\mathcal{G} = \{\mathcal{S},\mathcal{N}_\mathcal{S}\}$, the DAG of NNGP, let $\mathcal{G}_{NNDP} = \{(\mathcal{S},\mathcal{S}^d),(\mathcal{N}(\mathcal{S}),\mathcal{N}(\mathcal{S}^d))\}$ be the graph of NNDP. Here we use $\mathcal{S}^d$ to denote a copy of $\mathcal{S}$ for the NNDP, that is, $\bfs^d_i = \bfs_i$. 
    Consider $\bfs_i\in \mathcal{S}$,
    $$\widetilde{p}(D_\bfu \widetilde{\omega}(\bfs_i),\boldsymbol{\widetilde{\omega}_\mathcal{S}}) = \widetilde{p}(D_\bfu \widetilde{\omega}(\bfs_i)\given \boldsymbol{\widetilde{\omega}_\mathcal{S}}) \widetilde{p}(\boldsymbol{\widetilde{\omega}_\mathcal{S}}).$$
According to \Cref{thm:C_duC_cov} and \Cref{thm:duC_var}, the matrix needed in the distribution are $\mathbf{C}_{\mathcal{N}_0(\bfs_i)}^{-1}$, $D_\bfu C(\bfs_i,\mathcal{N}_0(\bfs_i))$, $\widetilde{\mathbf{C}}_{\mathcal{N}_0(\bfs_i)}$, $\widetilde{\mathbf{C}}_{\mathcal{N}_0(\bfs_i),s_j}$. In addition to the matrix calculated on $\boldsymbol{\omega_\mathcal{S}}$, the only information needed is $D_\bfu C(\bfs_i,\mathcal{N}_0(\bfs_i))$, which means the neighbor set of $\bfs^d_i$ is $\mathcal{S}_Q \bigcup \mathcal{N}_0(\bfs^d_i)\setminus \{s^d_i\}$, where $\mathcal{S}_Q$ is some subset of $\mathcal{S}$. This ensures that $\mathcal{G}_{NNDP}$ is still a DAG.

Then, following Supplement A1 in \cite{datta2016hierarchical}, the joint density of NNDP is still a proper joint density.

Now we can write the joint density of NNDP as 
\begin{align*}
p(\boldsymbol{\omega_\mathcal{S}},D_\bfu \boldsymbol{\omega_\mathcal{S}}) &=p(\omega(\bfs_1))\prod_{i=2}^{k}p(\omega(\bfs_i)\given \omega(\bfs_{i-1}),\cdots, \omega(\bfs_1))  \\
&~~ \cdot p(D_\bfu \omega(\bfs_1)\given\boldsymbol{\omega_\mathcal{S}})\prod_{i=2}^{k}p(D_\bfu\omega(\bfs_i)\given D_\bfu\omega(\bfs_{i-1}),\cdots, D_\bfu\omega(\bfs_1),\boldsymbol{\omega_\mathcal{S}})\\
& = p(\omega(\bfs_1))\prod_{i=2}^k p(\omega(\bfs_i)\given\boldsymbol{\omega_{\mathcal{N}(\bfs_i)}}) p(D_\bfu\omega( \bfs_1)\given \boldsymbol{\omega_\mathcal{S}}) \prod_{i=2} p(D_\bfu \omega(\bfs_i)\given D_\bfu\boldsymbol{\omega_{\mathcal{N}_0(\bfs_i)\setminus \{s_i\}}}, \boldsymbol{\omega_\mathcal{S}}).
\end{align*} 

%\end{proof}

%\section{Kolmogorov Consistency for NNDP}
The proof of Kolmogorov consistentcy for NNDP shares the same road map as in NNGP~\citep{datta2016hierarchical}.

For $\mathcal{V} = \{\bfv_1,\bfv_2,\dots,\bfv_n\}$ in $\mathcal{D}$, and for every permutation $\pi(1),\pi(2),\dots,\pi(n)$, since the joint density of $\omega(\bfv)$ and $D_\bfu \omega(\bfv)$ only depends on the neighbor sets $\mathcal{N}(\bfv)$, thus, we still have
\begin{align*}
&\widetilde{p}(\omega(\bfv_1),\omega(\bfv_2),\dots,\omega(\bfv_n),D_\bfu \omega(\bfv_1), D_\bfu \omega(\bfv_2),\dots,D_\bfu \omega(\bfv_n))  \\
&=\widetilde{p}(\omega(\bfv_\pi(1)),\omega(\bfv_\pi(2)),\dots,\omega(\bfv_\pi(n)),D_\bfu \omega(\bfv_\pi(1)), D_\bfu \omega(\bfv_\pi(2)),\dots,D_\bfu \omega(\bfv_\pi(n))).
\end{align*}

We next prove that for every location $\bfv_0\in \mathcal{D}$, we have 
\[\widetilde{p}(\boldsymbol{\omega_\mathcal{V}},D_\bfu \boldsymbol{\omega_\mathcal{V}}) = \int \widetilde{p}(\boldsymbol{\omega_{\mathcal{V}\bigcup \{\bfv_0\}}},D_\bfu \boldsymbol{\omega_{\mathcal{V}\bigcup \{\bfv_0\}}})d(\omega(\bfv_0))d(D_\bfu \omega(\bfv_0)).\]

If $\bfv_0\in \mathcal{S}$, let $\mathcal{V}_1 = \mathcal{V}\bigcup \{\bfv_0\}$
\(\mathcal{V}_1 \setminus \mathcal{S} = \mathcal{V} \setminus \mathcal{S} = \mathcal{U}\), we obtain
\begin{align*}
&\int \tilde{p}(\boldsymbol{\omega_{\mathcal{V}_1}},D_\bfu \boldsymbol{\omega_{\mathcal{V}_1}}) d(\omega(\mathbf{v}_0))d(D_\bfu \omega(\bfv_0)) \\
&= \int \tilde{p}(\boldsymbol{\omega_{\mathcal{S}}}, D_\bfu \boldsymbol{\omega_{\mathcal{S}}}) \tilde{p}(\boldsymbol{\omega_{\mathcal{V}_1 \setminus \mathcal{S}}}, D_\bfu \boldsymbol{\omega_{\mathcal{V}_1 \setminus \mathcal{S}}}| \boldsymbol{\omega_{\mathcal{S}}}, D_\bfu \boldsymbol{\omega_\mathcal{S}}) \prod_{s_i \in \mathcal{S} \setminus \mathcal{V}_1} d(\omega(s_i)) d(D_\bfu \omega(s_i)) d(\omega(\mathbf{v}_0)) d(D_\bfu \omega(\mathbf{v}_0)) \\
&= \int \tilde{p}(\boldsymbol{\omega_{\mathcal{S}}},D_\bfu \boldsymbol{\omega_{\mathcal{S}}}) \tilde{p}(\boldsymbol{\omega_{\mathcal{V} \setminus \mathcal{S}}}, D_\bfu \boldsymbol{\omega_{\mathcal{V} \setminus \mathcal{S}}} | \boldsymbol{\omega_{\mathcal{S}}}, D_\bfu \boldsymbol{\omega_{\mathcal{S}}}) \prod_{s_i \in \mathcal{S} \setminus \mathcal{V}} d(\omega(s_i)) d(D_\bfu \omega(s_i)) \\
&= \tilde{p}(\boldsymbol{\omega_{\mathcal{U}}}, D_\bfu \boldsymbol{\omega_{\mathcal{U}}}).
\end{align*}

If \( \bfv_0 \notin \mathcal{S} \), then \( \omega(\bfv_0) \) does not appear in the neighborhood set of any other term. So, \( p(\omega(\bfv_0) D_\bfu \omega(\bfv_0) \mid \boldsymbol{\omega_{\mathcal{S}}}, D_\bfu \boldsymbol{\omega_{\mathcal{S}}}) \) integrates to one with respect to \( d(\omega(\bfv_0)) d(D_\bfu \omega(\bfv_0)) \). The result now follows from
\[
\int p(\boldsymbol{\omega_{\mathcal{V}_1}} D_\bfu \boldsymbol{\omega_{\mathcal{V}_1}} \given \boldsymbol{\omega_{\mathcal{S}}}, D_\bfu \boldsymbol{\omega_{\mathcal{S}}}) d(\omega(\bfv_0)) d(D_\bfu \omega(\bfv_0)) = p(\boldsymbol{\omega_{\mathcal{V}}} \given \boldsymbol{\omega_{\mathcal{S}}}, D_\bfu \boldsymbol{\omega_{\mathcal{S}}}).
\]
\end{proof}

\section{Proof of \Cref{thm:NNGP_GP_equal}: The relationship between $\mathbf{C}$ and $\widetilde{\mathbf{C}}$}

%Proof of \Cref{thm:NNGP_GP_equal}:
\begin{proof}[Proof of \Cref{thm:NNGP_GP_equal}]
Let $\mathbf{e}_i = (0,0,\dots, 1,\dots,0)$ be the i-th canonical basis. Let $\mathbf{e}_{\mathcal{N}(\bfs_i)} = (\mathbf{e}_{\mathcal{N}(\bfs_i)_1}\T,\dots,\mathbf{e}_{\mathcal{N}(\bfs_i)_{m_i}}\T)\T$, then
\begin{align*}
    \widetilde{C}(\bfs_i,\mathcal{N}(\bfs_i)) &= \widetilde{\EE}(\widetilde{\cov}(\bfs_i,\mathcal{N}(\bfs_i))\given \mathcal{N}(\bfs_i)) + \widetilde{\cov}(\widetilde{\EE}(\bfs_i),\widetilde{\EE}(\mathcal{N}(\bfs_i))\given \mathcal{N}(\bfs_i)) \\
    &= 0 + \widetilde{\cov}(\mathbf{b}_{\bfs_i}\T\mathcal{N}(\bfs_i),\mathcal{N}(\bfs_i)) \\
    &= \mathbf{C}_{\bfs_i,\mathcal{N}(\bfs_i)}\mathbf{C}^{-1}_{\mathcal{N}(\bfs_i)}\widetilde{\mathbf{C}}_{\mathcal{N}(\bfs_i)} \\
    &= \mathbf{C}_{\bfs_i,\mathcal{N}(\bfs_i)}.
\end{align*}
Similarly,
\begin{align*}
    \widetilde{C}(\bfs_i,\bfs_i) &= \widetilde{\EE}(\widetilde{\cov}(\bfs_i)\given \mathcal{N}(\bfs_i)) + \widetilde{\cov}(\widetilde{\EE}(\bfs_i),\widetilde{\EE}(\bfs_i)\given \mathcal{N}(\bfs_i)) \\
    &= \widetilde{\EE}(C(\bfs_i,\bfs_i)-\mathbf{C}_{\bfs_i,\mathcal{N}(\bfs_i)}\mathbf{C}^{-1}_{\mathcal{N}(\bfs_i)}\mathbf{C}_{\mathcal{N}(\bfs_i),\bfs_i}) + \widetilde{\cov}(\mathbf{b}_{\bfs_i}\T\mathcal{N}(\bfs_i),\mathbf{b}_{\bfs_i}\T\mathcal{N}(\bfs_i)) \\
    &= C(\bfs_i,\bfs_i)-\mathbf{C}_{\bfs_i,\mathcal{N}(\bfs_i)}\mathbf{C}^{-1}_{\mathcal{N}(\bfs_i)} + \mathbf{C}_{\bfs_i,\mathcal{N}(\bfs_i)}\mathbf{C}^{-1}_{\mathcal{N}(\bfs_i)}\widetilde{\mathbf{C}}_{\mathcal{N}(\bfs_i)}\mathbf{C}^{-1}_{\mathcal{N}(\bfs_i)}\mathbf{C}_{\mathcal{N}(\bfs_i),\bfs_i} \\
    & = C(\bfs_i,\bfs_i).
\end{align*}
\end{proof}

\section{Simplified version of prediction on new locations} \label{apdx:simplified}

Consider $\bfv_1\notin \mathcal{S},\bfv_2=\bfs_j $, 
$$\cov(D_\bfu \widetilde{\omega}(\bfv_1),\widetilde{\omega}(\bfv_2)) = D_\bfu C(\bfv_1,\mathcal{N}(\bfv_1))\mathbf{C}_{\mathcal{N}(\bfv_1)}^{-1} \widetilde{\mathbf{C}}_{\mathcal{N}(\bfv_1),\bfs_j},$$
\begin{align*}
\cov(D_\bfu \widetilde{\omega}(\bfv_i),D_\bfu \widetilde{\omega}(\bfv_j)) &= D_{\bfu} C(\bfv_1,\mathcal{N}(\bfv_1))\mathbf{C}_{\mathcal{N}(\bfv_1)}^{-1}\widetilde{\mathbf{C}}_{\mathcal{N}(\bfv_1)}\mathbf{C}_{\mathcal{N}(\bfv_1)}^{-1} \\
&~~+D^{(2)}_{\bfu}C(\bfv_1,\bfv_1)- D_{\bfu} C(\bfv_1,\mathcal{N}(\bfv_1))\mathbf{C}_{\mathcal{N}(\bfv_1)}^{-1}D_{\bfu} C(\mathcal{N}(\bfv_1),\bfv_1).
\end{align*}

Here $\mathcal{N}(\bfv_1) = \mathcal{N}_0(\bfs_i),~\bfs_i$ is the closed point of $\bfv_1$ in $\mathcal{S}$. Similarly to the construction of Algorithm S2, we have $\widetilde{\mathbf{C}}_{\mathcal{N}(\bfv_1),\bfs_j} = \mathbf{C}_{\mathcal{N}(\bfv_1),\bfs_j}$ for $\bfs_j \in \mathcal{N}_0(\bfs_i)$, and $\widetilde{\mathbf{C}}_{\mathcal{N}(\bfv_1),\mathcal{N}_0(\bfs_j)} = \widetilde{\mathbf{C}}_{\mathcal{N}_0(\bfs_i),\mathcal{N}_0(\bfs_j)} = \mathbf{C}_{\mathcal{N}_0(\bfs_i),\mathcal{N}_0(\bfs_j)}.$

Thus, 
\begin{align*}
\cov(D_\bfu \widetilde{\omega}(\bfv_1),\widetilde{\omega}(\bfs_j)) & = 
D_\bfu C(\bfv_1,\mathcal{N}_0(\bfs_i))\mathbf{C}_{\mathcal{N}_0(\bfs_i)}^{-1} \widetilde{\mathbf{C}}_{\mathcal{N}_0(\bfs_i),\bfs_j} \\
& = D_\bfu C(\bfv_1,\mathcal{N}_0(\bfs_i))\mathbf{C}_{\mathcal{N}_0(\bfs_i)}^{-1} \mathbf{C}_{\mathcal{N}_0(\bfs_i),\bfs_j} \\
& = D_\bfu C(\bfv_1,\bfs_j),
\end{align*}
\begin{align*}
\cov(D_\bfu \widetilde{\omega}(\bfv_i),D_\bfu \widetilde{\omega}(\bfs_j)) & = 
D_{\bfu} C(\bfv_1,\mathcal{N}_0(\bfs_i))\mathbf{C}_{\mathcal{N}_0(\bfs_i)}^{-1}\widetilde{\mathbf{C}}_{\mathcal{N}_0(\bfs_i)}\mathbf{C}_{\mathcal{N}_0(\bfs_i)}^{-1} \\
&~~+D^{(2)}_{\bfu}C(\bfv_1,\bfv_1)- D_{\bfu} C(\bfv_1,\mathcal{N}_0(\bfs_i))\mathbf{C}_{\mathcal{N}_0(\bfs_i)}^{-1}D_{\bfu} C(\mathcal{N}_0(\bfs_i),\bfv_1) \\
& = D^{(2)}_{\bfu}C(\bfv_1,\bfv_1).
\end{align*}

\section{NNDP algorithm}

\begin{algorithm}[H]
\caption{Nearest-Neighbor Derivative Process (NNDP)}\label{alg:NNDP}
\begin{algorithmic}
\State {\bfseries Input:} Observed data $\{(\bfs_i, y_i)\}_{i=1}^n$, where $\bfs_i \in \mathbb{R}^d$ and $y_i\in\RR$ is the observed outcome at location $\bfs_i$.
\State \textbf{Model Specification}: Assume a Gaussian process prior for $\omega(\bfs)$ with mean zero and covariance function $C(\bfs, \bfs'; \theta)$, where $\theta$ represents the set of covariance parameters. The observations are modeled as $y_i = \omega(\bfs_i)$.
\State \textbf{Fitting NNGP}:
Fit a NNGP model to obtain $R$ MCMC samples of the parameters $\theta^{(k)} = \{\sigma^2, \phi, \ldots\}$ for $k = 1, \ldots, R$. The referense set $\mathcal{S}$ is the observation set $\mathcal{T}$.

\State \textbf{Gradient Estimation}:
\State \textbf{    Compute Conditional Distribution}:
For each location $\bfs$, compute the conditional posterior distribution of $\nabla \widetilde{\omega}(\bfs)$ given $\widetilde{\omega}^{(k)}$ and $\theta^{(k)}$:
\[
\nabla \widetilde{\omega}(\bfs) \given \widetilde{\omega}^{(k)}, \theta^{(k)} \sim N\left( \mu_{\nabla \widetilde{\omega}(\bfs)}, \Sigma_{\nabla \widetilde{\omega}(\bfs)} \right),
\]
where
\[
\mu_{\nabla \omega(s)} = \nabla \widetilde{\mathbf{C}}_\bfs^\top \widetilde{\mathbf{C}}_\bfs^{-1} \omega(\bfs), \quad \Sigma_{\nabla \omega(\bfs)} = \nabla^2 \widetilde{\mathbf{C}}_\bfs - \nabla \widetilde{\mathbf{C}}_\bfs^\top \widetilde{\mathbf{C}}_\bfs^{-1} \nabla \widetilde{\mathbf{C}}_\bfs,
\]
with $\widetilde{\mathbf{C}}_\mathcal{S} = [\widetilde{C}(\bfs, \bfs_i)]_{i=1}^n$ computed from \Cref{eqn:C_tilde}, $\nabla \widetilde{\mathbf{C}}_\bfs = [\nabla \widetilde{C}(\bfs, \bfs_i)]_{i=1}^n$ computed from \Cref{eqn:C_duC_cov}, and $\nabla^2 \widetilde{\mathbf{C}}_\bfs$ computed from \Cref{eqn:duC_var}.
\State \textbf{   Sample Gradients}:
Sample $\nabla \omega^{(k)}(\bfs)$ from the conditional distribution:
\[
\nabla \omega^{(k)}(\bfs) \sim N\left( \mu_{\nabla \omega(\bfs)}, \Sigma_{\nabla \omega(\bfs)} \right).
\]
\State \textbf{Output}:
Posterior samples $\{\nabla \omega^{(k)}(\bfs)\}_{k=1}^R$ for gradient estimation at location $\bfs$.
\end{algorithmic}
\end{algorithm}

\begin{algorithm}[H]
\caption{Nearest-Neighbor Derivative Process (NNDP): fast version}\label{alg:NNDP_NN}
\begin{algorithmic}
\State {\bfseries Input:} Observed data $\{(\bfs_i, y_i)\}_{i=1}^n$, where $\bfs_i \in \mathbb{R}^d$ and $y_i\in\RR$ is the observed outcome at location $\bfs_i$.
\State \textbf{Model Specification}: Assume a Gaussian process prior for $\omega(\bfs)$ with mean zero and covariance function $C(\bfs, \bfs'; \theta)$, where $\theta$ represents the set of covariance parameters. The observations are modeled as $y_i = \omega(\bfs_i)$.

\State \textbf{Fitting NNGP}:
Fit a NNGP model to obtain $R$ MCMC samples of the parameters $\theta^{(k)} = \{\sigma^2, \phi, \ldots\}$ for $k = 1, \ldots, R$. The referense set $\mathcal{S}$ is the observation set $\mathcal{T}$.
    
\State \textbf{Gradient Estimation}:
\State \textbf{    Compute Conditional Distribution}:
 For each location $\bfs$, find $\mathcal{N}(\bfs)$ - the set of $m$ nearest neighbors of $\bfs$ from $\mathcal{S}$, which should be same as the neighbors used in NNGP model. Compute the conditional posterior distribution of $\nabla \widetilde{\omega}(\bfs)$ given $\widetilde{\omega}^{(k)}$ and $\theta^{(k)}$:
\[
\nabla \widetilde{\omega}(\bfs) \given \widetilde{\omega}^{(k)}, \theta^{(k)} \sim N\left( \mu_{\nabla \widetilde{\omega}(\bfs)}, \Sigma_{\nabla \widetilde{\omega}(\bfs)} \right),
\]
where
\[
\mu_{\nabla \omega(\bfs)} = \nabla \widetilde{\mathbf{C}}_\bfs^\top \widetilde{\mathbf{C}}_\bfs^{-1} \omega(\bfs), \quad \Sigma_{\nabla \omega(\bfs)} = \nabla^2 \widetilde{\mathbf{C}}_\bfs - \nabla \widetilde{\mathbf{C}}_\bfs^\top \widetilde{\mathbf{C}}_\bfs^{-1} \nabla \widetilde{\mathbf{C}}_\bfs,
\]
with $\widetilde{\mathbf{C}}_\mathcal{S} = [C(\bfs, \bfs_i)]_{i=1}^m$, $\bfs_i \in \mathcal{N}(\bfs)$, $\nabla \widetilde{\mathbf{C}}_\bfs = [\nabla \widetilde{C}(\bfs, \bfs_i)]_{i=1}^m$, $\bfs_i \in \mathcal{N}(\bfs)$, computed from \Cref{eqn:dC_NN}, and $\nabla^2 \widetilde{\mathbf{C}}_\bfs$ computed from \Cref{eqn:varC_NN}.

\State \textbf{    Sample Gradients}:
 Sample $\nabla \omega^{(k)}(\bfs)$ from the conditional distribution:
\[
\nabla \omega^{(k)}(\bfs) \sim N\left( \mu_{\nabla \omega(\bfs)}, \Sigma_{\nabla \omega(\bfs)} \right).
\]
\State \textbf{Output}: Posterior samples $\{\nabla \omega^{(k)}(\bfs)\}_{k=1}^R$ for gradient estimation at location $\bfs$.
\end{algorithmic}
\end{algorithm}

\iffalse
\section{Posterior sampling details}

We use $\phi \sim \text{unif}(0.01,300)$, $\sigma^2 \sim IG(0.01,10)$ for priors. Gradient estimates are the batch median of the posterior samples. Batch size is set as $50$ for the simulation of pattern 2 equal spacing case (\Cref{tab:pattern2_scale}, \Cref{fig:pattern2_scale1}), and $100$ for all other experiments.
\fi

\section{VisiumHD experimental details}

The mouse brain VisiumHD data is available at \textit{https://www.10xgenomics.com/datasets/visium-hd-cytassist-gene-expression-libraries-of-mouse-brain-he}. We used the $16 \mu m \times 16 \mu m$ bin size version.

\section{Air temperature experimental details}

We obtained air temperature data at 2 meters above ground level, recorded on June 05, 2019 from \textit{https://github.com/WentaoZhan1998/geospaNN}.

\end{document}